\documentclass[prb,aps,showpacs,twocolumn,longbibliography]{revtex4-1}

\usepackage{graphicx}
\usepackage{amsfonts}
\usepackage{amssymb}
\usepackage{amsmath}
\usepackage{mathtools}
\usepackage{float}
\usepackage{epsfig}
\usepackage{color}
\usepackage{gensymb}
\usepackage{multirow}
\usepackage[mathscr]{euscript}

\newcommand{\Skip}[1]{}

\begin{document}

\title{Basic Understanding of Condensed Phases of Matter via Packing Models}

\author{S. Torquato}

\email{torquato@princeton.edu}

\affiliation{\emph{Department of Chemistry,  Department of Physics,
Princeton Center for Theoretical Science,Princeton Institute for the Science and Technology of Materials, and 
Program in Applied and Computational Mathematics, Princeton University}, Princeton
New Jersey, 08544 USA}

\begin{abstract}

Packing problems have  been a source of fascination for millenia and 
their study  has produced a  rich literature that spans numerous disciplines.
Investigations  of hard-particle packing models have provided basic 
insights into the structure and bulk properties of condensed phases of matter,
including low-temperature states (e.g., molecular and colloidal liquids, crystals and glasses),  
multiphase heterogeneous media, granular media, and  biological systems. The densest packings are of great interest
in pure mathematics, including discrete geometry and number theory.
This perspective reviews pertinent theoretical and computational literature concerning the  equilibrium, metastable and nonequilibrium
packings of hard-particle packings in various Euclidean space dimensions.
In the case of jammed packings, emphasis will be placed on the  ``geometric-structure" approach,
 which provides a powerful and unified means to quantitatively
characterize individual packings  via jamming categories
and ``order" maps. It incorporates extremal jammed states, including the densest packings,
maximally random jammed states, and lowest-density jammed structures. Packings of identical spheres, spheres with a size distribution, and nonspherical particles
are also surveyed. We close this review by identifying challenges and open questions
for future research.

\end{abstract}

\maketitle

\section{Introduction}

We will call a {\it packing} a large collection of nonoverlapping (i.e., hard) particles
in either a finite-sized container or in $d$-dimensional Euclidean
space $\mathbb{R}^d$. Exclusion-volume effects that often arise
in dense many-particle systems in physical and biological contexts are
naturally captured  by simple packing models.
 They have been studied to  help understand the 
symmetry, structure and physical properties of condensed matter phases, including
 liquids, glasses and crystals, as well as the associated phase transitions.~\cite{Ma40,Be60,Ber65,St64b,St69,We71,As76,Wo81,Ha86,Sp94,Di94,Chaik95,Ri96b}
Packings also serve as excellent models  of the structure of
multiphase heterogeneous materials,~\cite{Fe82b,To02a,Zo06,Li14} colloids,~\cite{Ru89,Chaik95,To09a} suspensions,\cite{Ki91a,Sp14} and granular media,~\cite{Ed94,Zo04} which enables predictions of their effective transport, mechanical and electromagnetic properties.\cite{To02a}

Packing phenomena are also ubiquitous in  biological
contexts and occur in systems across a  wide spectrum of length scales.
This includes DNA packing within the nucleus of a cell, \cite{Cr01}``crowding" of macromolecules within living cells,
~\cite{El01} the  packing of cells to form tissue,~\cite{To02a,Ge08a,Ji14,Ch16} the fascinating spiral patterns seen in plant shoots  and
flowers (phyllotaxis),~\cite{Pr90,Ni10}  and the competitive settlement of territories.~\cite{Tan80,To02a}

Understanding the symmetries and other mathematical properties
of the densest sphere packings in various spaces and dimensions 
is a challenging area of long-standing interest in discrete geometry and number theory~\cite{Co93,Co03,Ro64} as well as coding theory.~\cite{Sh48,Co93,Co07} Packing problems are mathematically easy to pose, but they are notoriously difficult
to solve rigorously. For example, in 1611, Kepler was asked: What is the densest way to stack 
equal-sized cannon balls? His solution, known as ``Kepler's conjecture," 
was the face-centered-cubic (fcc) arrangement (the way your green grocer stacks oranges). 
Gauss~\cite{Ga31}  proved that this is the densest Bravais lattice packing (defined below). 
Remarkably, nearly four centuries passed before Hales proved the general conjecture that there 
is no other sphere packing  in three-dimensional (3D) Euclidean space whose density can exceed
that of the fcc packing.\cite{Ha05} Even the proof
that the densest packing of congruent (identical) circles in the plane
is the triangular-lattice  packing appeared only 80 years ago;
see Refs.  \onlinecite{Ro64,Co93} for the history of this problem.



One objective of this perspective  is  to survey recent developments
concerning the simplest but venerable packing model: identical  frictionless spheres in the absence
of gravity subject only to a nonoverlap constraint, i.e., the spheres do not interact
for any nonoverlapping configuration. This ``stripped-down" hard-sphere model
can be viewed as the particle-system analog of the idealized Ising model for magnetic systems,\cite{To10c}
which is regarded  as one of the pillars of statistical mechanics. \cite{On44,Do60,Ga99}
More complex packing models can include interactions that extend 
beyond hard-core distances, but our main concern here is the aforementioned classic version.
This model enables one to uncover unifying principles that describe a broad range of phenomena,
including the nature of equilibrium states of matter (e.g., liquids and crystals),
metastable and nonequilibrium states of matter (e.g., supercooled
liquids and structural glasses), and jamming phenomena.

Jammed sphere packings represent an important subset of  hard-sphere
configurations and have attracted great attention because they capture salient
characteristics of crystals, glasses, granular media and biological systems
\cite{Ed94,To00b,Oh03,Pa10,To10c,Ji14,Jaeg15} and naturally arise in pure mathematical
contexts.\cite{Co93,Mar03,Cohn11} Jammed packings are those in
which each particle has a requisite number contacting particles
in order for the packing to achieve a certain level of mechanical stability.
This review focuses on the so-called ``geometric-structure" approach to jammed particle packings,
which provides a powerful and unified means to quantitatively understand such
many-particle systems. It incorporates not only the maximally dense
packings  and disordered jammed packings, but a
myriad of other significant jammed states, including maximally random jammed
states,\cite{To00b,To10c} which can be regarded to be prototypical structural glasses, as well
as the lowest-density jammed packings.\cite{To07}

Importantly, the simplified hard-sphere model embodies the primary structural attributes 
of dense many-particle systems in which steep repulsive interparticle interactions
are dominant. For example, the densest sphere packings are intimately related
to the (zero-temperature) ground states of such molecular systems\cite{St01}
and high-pressure crystal phases. 
Indeed, the equilibrium hard-sphere model~\cite{Ha86} also serves as a natural reference system
in the thermodynamic perturbation theory of liquids characterized
by  steep isotropic repulsive  interparticle interactions at short distances
as well short-range attractive interactions.~\cite{We71,Ve72}
Moreover, the classic hard-sphere model
provides a good description of certain classes of colloidal systems.~\cite{Ru89,Ru96,Di98,Di99,Di14}
Note that the hard-core constraint does not uniquely
specify the hard-sphere model; there are an infinite number of nonequilibrium
hard-sphere ensembles, some of which will be surveyed.


We will also review work that describes  generalizations
of this simplified hard-sphere model to include its natural extension
to packings of hard spheres with a size distribution. This topic has relevance
to understanding  dispersions of technological importance,
(e.g., solid propellants combustion\cite{Ke87} flow in packed beds, \cite{Sc74} 
and sintering of powders\cite{Ra95}), packings of biological cells, 
and phase behavior of various molecular systems. For example,
the densest packings of hard-sphere mixtures are intimately related
to high-pressure phases of compounds for a range of temperatures.\cite{Dem06,Ca09}
Another natural extension of the hard-sphere model that will be surveyed are
hard {\it nonspherical} particles in two and three dimensions. Asphericity in particle shape is capable
of capturing salient features of phases of molecular systems with anisotropic pair interactions
(e.g., liquid crystals) and is also  a more realistic characteristic of real granular media.  
In addition, we will review aspects
of dense sphere packings in high-dimensional
Euclidean spaces, which provide  useful physical insights and are relevant to error-correcting codes and information
theory.\cite{Sh48,Co93}

We begin this perspective by introducing relevant definitions and background (Sec. \ref{defs}).
This is followed by a survey of work on equilibrium, metastable and nonequilibrium
packings of identical spheres and polydisperse spheres in one, two and three
dimensions (Sec. \ref{lessons}). The geometric-structure approach to jammed and unjammed packings,
including their classification via order maps, is emphasized.
Subsequently, corresponding results for sphere packings in high Euclidean and non-Euclidean space dimensions,  (Sec. \ref{high})
and packings of nonspherical particles in low-dimensional Euclidean spaces (Sec. \ref{nonspherical})
are reviewed. Finally, we describe some 
challenges and open questions for future research (Sec. \ref{open}).

\section{Definitions and Background}
\label{defs}

A {\it packing} $P$ is a collection of nonoverlapping solid objects or particles 
of general shape in $d$-dimensional
Euclidean space $\mathbb{R}^d$. Packings can be defined in other spaces
(e.g., hyperbolic spaces and compact spaces, such as the surface
of a $d$-dimensional sphere), but our primary focus in this review
is $\mathbb{R}^d$. A {\it saturated} packing
is one in which there is no space available to add another particle 
of the same kind to the packing. The {\it packing fraction}  $\phi$ is the fraction of
space $\mathbb{R}^d$ covered by the particles. 

A $d$-dimensional particle is {\it centrally symmetric} if it has a center $C$ that
bisects every chord through $C$ connecting any two boundary points
of the particle, i.e., the center is a point of inversion symmetry.
 Examples of centrally symmetric particles in  $\mathbb{R}^d$ include spheres, ellipsoids and
superballs (defined in Section \ref{nonspherical}). A triangle and tetrahedron are
examples of non-centrally symmetric 2D  and 3D particles, respectively.
A $d$-dimensional centrally symmetric particle for $d \ge 2$
is said to possess
$d$ equivalent principal axes (orthogonal directions) associated
with the moment of inertia tensor if those
directions are two-fold rotational symmetry axes.
Whereas a $d$-dimensional superball has $d$
equivalent directions, a $d$-dimensional ellipsoid generally does not.
The reader is referred to Ref. \onlinecite{To10c} for further discussion
concerning particle symmetries.

A {\it lattice} $\Lambda$ in $\mathbb{R}^d$ is a subgroup
consisting of the integer linear combinations of vectors that constitute
a basis for $\mathbb{R}^d$. In the physical sciences, this is referred to as a {\it Bravais}
lattice. The term ``lattice" will refer here
to a Bravais lattice only. Every lattice has a dual (or reciprocal) lattice $\Lambda^*$
in which the lattice sites are specified by the dual (reciprocal) lattice vectors. \cite{Co93}
A {\it lattice packing} $P_{L}$ is one in which  the centroids of the nonoverlapping identical
particles are located at the points of $\Lambda$, and all particles have a common orientation.
The set of lattice packings is a subset of all possible packings in $\mathbb{R}^d$.
In a lattice packing, the space $\mathbb{R}^d$ can be geometrically divided into identical
regions $F$ called {\it fundamental cells}, each of which has volume $\mbox{Vol}(F)$ and contains
the centroid of just one particle of volume $v_1$. Thus, the lattice packing fraction is
\begin{equation}
\phi= \frac{v_1}{\mbox{Vol}(F)}.
\end{equation}
Common $d$-dimensional lattices include the {\it hypercubic} {$\mathbb{Z}^d$,
}{\it checkerboard} {$D_d$, and }{\it root} {$A_d$ lattices; see  Ref. \onlinecite{Co93}.
Following Conway and Sloane\cite{Co93}, we say two lattices are }{\it {equivalent}} {or }{\it {similar}} {if one becomes identical
to the other possibly by a rotation, reflection, and change of scale,
for which we use the symbol $\equiv$.
The $A_d$ and $D_d$ lattices are $d$-dimensional generalizations
of the face-centered-cubic (FCC) lattice defined by $A_3 \equiv D_3$; however, for $d\ge 4$, they are no longer
equivalent. In two dimensions, $A_2 \equiv A_2^*$ is the triangular lattice.
In three dimensions, $A_3^* \equiv D_3^*$ is the body-centered-cubic (BCC)
lattice. In four dimensions, the checkerboard lattice and its dual are equivalent,
i.e., $D_4\equiv D_4^*$. The hypercubic lattice $\mathbb{Z}^d\equiv \mathbb{Z}^d_*$ and its dual
lattice are equivalent for all $d$.

A {\it periodic} packing of congruent particles
is obtained by placing a fixed configuration of $N$ particles (where $N\ge 1$)
with {\it arbitrary orientations} subject to the nonoverlapping condition in one 
fundamental cell of a lattice $\Lambda$, which is then periodically replicated.
 Thus, the packing is still
periodic under translations by $\Lambda$, but the $N$ particles can occur
anywhere in the chosen fundamental cell subject to the overall nonoverlap condition.
The packing fraction of a  periodic packing is given by
\begin{equation}
\phi=\frac{N v_1}{\mbox{Vol}(F)}=\rho v_1,
\end{equation}
where $\rho=N/\mbox{Vol}(F)$ is the number density, i.e., the number of particles per unit volume.

For simplicity, consider a packing of $N$ identical $d$-dimensional
spheres of diameter $D$ centered at the positions ${\bf r}^N \equiv \{{\bf r}_1, {\bf r}_2, \ldots, {\bf r}_N$\} in
a region of volume $V$ in $d$-dimensional Euclidean space $\mathbb{R}^d$.
Ultimately, we will pass to the {\it thermodynamic limit}, i.e.,
$N \rightarrow \infty$, $V \rightarrow \infty$ such that 
the {\it number density} $\rho=N/V$ is a fixed positive constant
and its corresponding packing fraction is given by
\begin{equation}
\phi=\rho v_1(D/2),
\label{phi}
\end{equation}
where $\rho$ is the number density,
\begin{equation}
v_1(R) = \frac{\pi^{d/2}}{\Gamma(1+d/2)} R^d
\label{v(R)}
\end{equation}
is the volume of a $d$-dimensional sphere of radius $R$, and
$\Gamma(x)$ is the gamma function. For an individual sphere, the {\it kissing} or {\it contact} number $Z$ 
is the number of spheres that may simultaneously touch this sphere. In a sphere packing, the {\it mean} kissing or contact
number per particle, $\overline Z$, is the average of $Z$ over all spheres. For a lattice packings, $Z=\overline Z$. 
For statistically homogeneous sphere packings in $\mathbb{R}^d$, the quantity
$\rho^n g_{n}({\bf r}^n)$  is proportional to
the probability density associated with simultaneously finding $n$ sphere centers at
locations ${\bf r}^n\equiv \{{\bf r}_1,{\bf r}_2,\dots,{\bf r}_n$\} in $\mathbb{R}^d$; see Ref. \onlinecite{Ha86}
and references therein.
 With this convention, each {\it $n$-particle correlation function} $g_n$ approaches
unity when all particles become widely separated from one another for any system
without long-range order.
Statistical homogeneity implies that $g_n$ is translationally
invariant and therefore only depends on the relative displacements 
of the positions with respect to some arbitrarily chosen origin of the system, {\it i.e.},
$g_n=g_n({\bf r}_{12}, {\bf r}_{13}, \ldots, {\bf r}_{1n})$,
where ${\bf r}_{ij}={\bf r}_j - {\bf r}_i$.

The {\it pair correlation} function $g_2({\bf r})$ is 
of basic interest in this review. If the
system is also rotationally invariant (statistically
isotropic), then $g_2$ depends on the radial distance $r \equiv |{\bf r}|$ only, {\it i.e.},
$g_2({\bf r}) = g_2(r)$. The {\it total correlation} function is defined by $h({\bf r})\equiv g_2({\bf r})-1$.
Importantly, we focus in this review on  {\it disordered} packings in which
$h({\bf r})$ decays to zero for large $|{\bf r}|$ sufficiently rapidly
so that its volume integral over all space exists.\cite{To06b}

As usual, we define the nonnegative {\it structure factor} $S({\bf k})$ for
a statistically homogeneous packing as
\begin{equation}
S({\bf k})=1+\rho {\tilde h}({\bf k}),
\label{factor}
\end{equation}
where ${\tilde h}({\bf k})$ is the Fourier transform of 
$h({\bf r})$ and $\bf k$ is the wave vector.
The nonnegativity of $S(\bf k)$ for all $\bf k$ follows physically from the fact
that it is proportional to the intensity of the scattering of incident radiation
on a many-particle system. \cite{Ha86}
The structure factor $S({\bf k})$ provides a measure of the density fluctuations in particle configurations
at a particular wave vector $\bf k$. For any point configuration
in which the minimal pair distance is some positive number, such as a sphere packing, 
$g_2({\bf r=0})=0$ and $h({\bf r=0})=-1$, the structure factor obeys the following sum rule:\cite{To18a}
\begin{equation}
\frac{1}{(2\pi)^{d}}\int_{\mathbb{R}^d} [S({\bf k})-1]\, d{\bf k}=-\rho.
\label{SUM}
\end{equation}
For a single-component system in thermal equilibrium at number density $\rho$
and absolute temperature $T$, the structure factor at the origin
is directly related to the isothermal compressibility $\kappa_T$
via the relation \cite{Ha86}
\begin{equation}
\rho k_B T \kappa_T  =S(0),
\label{comp}
\end{equation}
where $k_B$ is the Boltzmann constant.

It is well known from Fourier transform theory that
if a real-space radial function $f(r)$ in $\mathbb{R}^d$ decreases sufficiently rapidly
to zero for large $r$ such that
all of its even moments exist, then its Fourier transform ${\tilde f}(k)$, where
$k\equiv |\bf k|$ is a wavenumber, is an even and analytic  function
at $k=0$. Hence,
$S(k)$, defined by  (\ref{factor}), has an expansion about $k=0$ in any space dimension $d$  of the general form
\begin{equation}
S(k)= S_0 + S_2 k^2 + {\cal O}(k^4),
\label{low-k}
\end{equation}
where $S_0$ and $S_2$  are the $d$-dependent constants defined by
\begin{equation}
S_0=1+  d \rho s_1(1) \int_0^\infty r^{d-1} h(r) dr \ge 0
\end{equation}
and
\begin{equation}
S_2=-\frac{\rho s_1(1)}{2d} \int_0^\infty r^{d+1} h(r) dr,
\end{equation}
where 
\begin{equation}
s_1(R)= =\frac{d\,\pi^{d/2} R^{d-1}}{\Gamma(d/2+1)}
\label{surface}
\end{equation}
is the surface area of $d$-dimensional sphere of radius $R$.
This behavior  is to be contrasted with that of 
maximally random jammed sphere packings,\cite{To00b}  which possess a structure factor
that is nonanalytic at $k=0$ (Ref. \onlinecite{Do05d}), as discussed
in greater detail in Sec. \ref{mrj}.

A hyperuniform \cite{To03a}  point configuration in $\mathbb{R}^d$  is one in which the structure factor $S({\bf k})$ tends to zero
as the wavenumber  tends to zero, i.e.,
\begin{equation}
\lim_{|{\bf k}| \rightarrow 0} S({\bf k}) = 0,
\label{hyper}
\end{equation}
implying that single scattering of incident radiation at infinite wavelengths is completely suppressed.
Equivalently, a hyperuniform system is one in which the number variance $\sigma^2_{_N}(R) \equiv \langle N(R)^2 \rangle -\langle N(R) \rangle^2$ of particles within a spherical observation window of radius $R$ grows more slowly than the window volume, i.e., $R^d$, in the large-$R$ limit.
This class of point configurations
includes perfect crystals,  many perfect quasicrystals 
and special disordered many-particle systems. \cite{To03a,Za09,Og17} Note that structure-factor definition (\ref{factor})
and the hyperuniformity requirement (\ref{hyper}) dictate that the  
following  {\it sum rule} involving $h({\bf r})$ that a hyperuniform point process must obey:
\begin{equation}
\rho \int_{\mathbb{R}^d} h({\bf r}) d{\bf r}=-1.
\label{sum-1}
\end{equation}
This sum rules implies that $h({\bf r})$ must exhibit negative correlations, i.e., {\it anticorrelations}, for some values of $\bf r$.

The hyperuniformity concept was generalized to incorporate two-phase heterogeneous
media  (e.g., composites, porous media and dispersions). \cite{Za09,To16b}
Here the phase volume fraction
fluctuates within a spherical window of radius $R$ and hence can be characterized by the
volume-fraction variance $\sigma_{_V}^2(R)$. For typical disordered two-phase media, the variance  $\sigma_{_V}^2(R)$ for large $R$ goes to zero
like $R^{-d}$. However, for hyperuniform disordered two-phase media, $\sigma_{_V}^2(R)$  goes to zero asymptotically more
rapidly than the inverse of the window volume, i.e., faster than $R^{-d}$, which is equivalent to the following condition
on the {\it spectral density} $\tilde{\chi}_{_V}(\mathbf{k})$:\cite{To16b}
\begin{eqnarray}
\lim_{|\mathbf{k}|\rightarrow 0}\tilde{\chi}_{_V}(\mathbf{k}) = 0.
\label{hyper-2}
\end{eqnarray}
The spectral density is proportional to scattered intensity associated with ``mass" (volume) 
content of the phases.\cite{De57}

\section{Sphere Packings in Low Dimensions}
\label{lessons}

The classical statistical mechanics of hard-sphere systems 
has generated a huge  collection of scientific publications, dating back at least to 
Boltzmann; \cite{Bo64} see also Refs. \onlinecite{Al57,Re59,Sa62,Ha86}.
Here we focus on packings of frictionless, congruent
spheres of diameter $D$ in one, two and three dimensions in the absence of gravity.

It is important to observe that the impenetrability
constraint alone of this idealized hard-sphere model does not uniquely specify the statistical ensemble. Hard-sphere
systems can be in thermal equilibrium (as discussed in Sec. \ref{phase}) 
or derived from one of an infinite number of  nonequilibrium
ensembles \cite{To02a} (see Sec. \ref{noneq}  for examples).

\subsection{Equilibrium and Metastable Phase Behavior}
\label{phase}

The phase behavior of hard spheres provides
powerful insights into the nature of liquid, crystal and  metastable
states as well as the associated phase transitions in molecular 
and colloidal systems.
It is well known that the pressure $p$ of a stable thermodynamic
phase in $\mathbb{R}^d$ at packing fraction $\phi$ 
and temperature $T$ is  simply related to the
contact value of the pair correlation function, $g_2(D^+)$: \cite{To02a}
\begin{equation}
\frac{p}{\rho k_B T}=1 +2^{d-1}\phi g_2(D^+).
\end{equation}
Away from jammed states, it has been proved that the  
mean nearest-neighbor distance between spheres, $\lambda$,   is bounded from above
by the pressure,\cite{To95a} namely, $\lambda \le 1+1/[2d(p/(\rho k_B T)-1]$.

 Figure \ref{hsphere} schematically shows the 3D phase behavior
in the $\phi$-$p$ plane.
At sufficiently low densities,  an infinitesimally slow 
compression of the system at constant temperature defines a  thermodynamically stable liquid branch for packing
fractions up to the ``freezing" point ($\phi \approx 0.49$). Increasing the density
beyond the freezing point putatively results in an entropy-driven first-order phase transition \cite{Al57,Frenk99}
to a crystal branch that begins at the melting point ($\phi \approx 0.55$).
While there is no rigorous proof that such a first-order freezing transition
occurs in three dimensions, there is overwhelming simulational evidence for its existence, beginning with the pioneering work
of Alder and Wainwright.\cite{Al57} Slow compression of the system along the crystal branch must end at one of the optimal (maximally dense)
sphere packings with  $\phi = \pi/\sqrt{18}=0.74048 \ldots$, 
each of which is a  jammed packing in which each particle contacts 12 others
(see Sec. \ref{map}).
This equilibrium state has an infinite pressure and is
putatively entropically favored by the fcc packing.\cite{Mau99}

\begin{figure}[bthp]
\centerline{\includegraphics[width=3in,keepaspectratio,clip=]{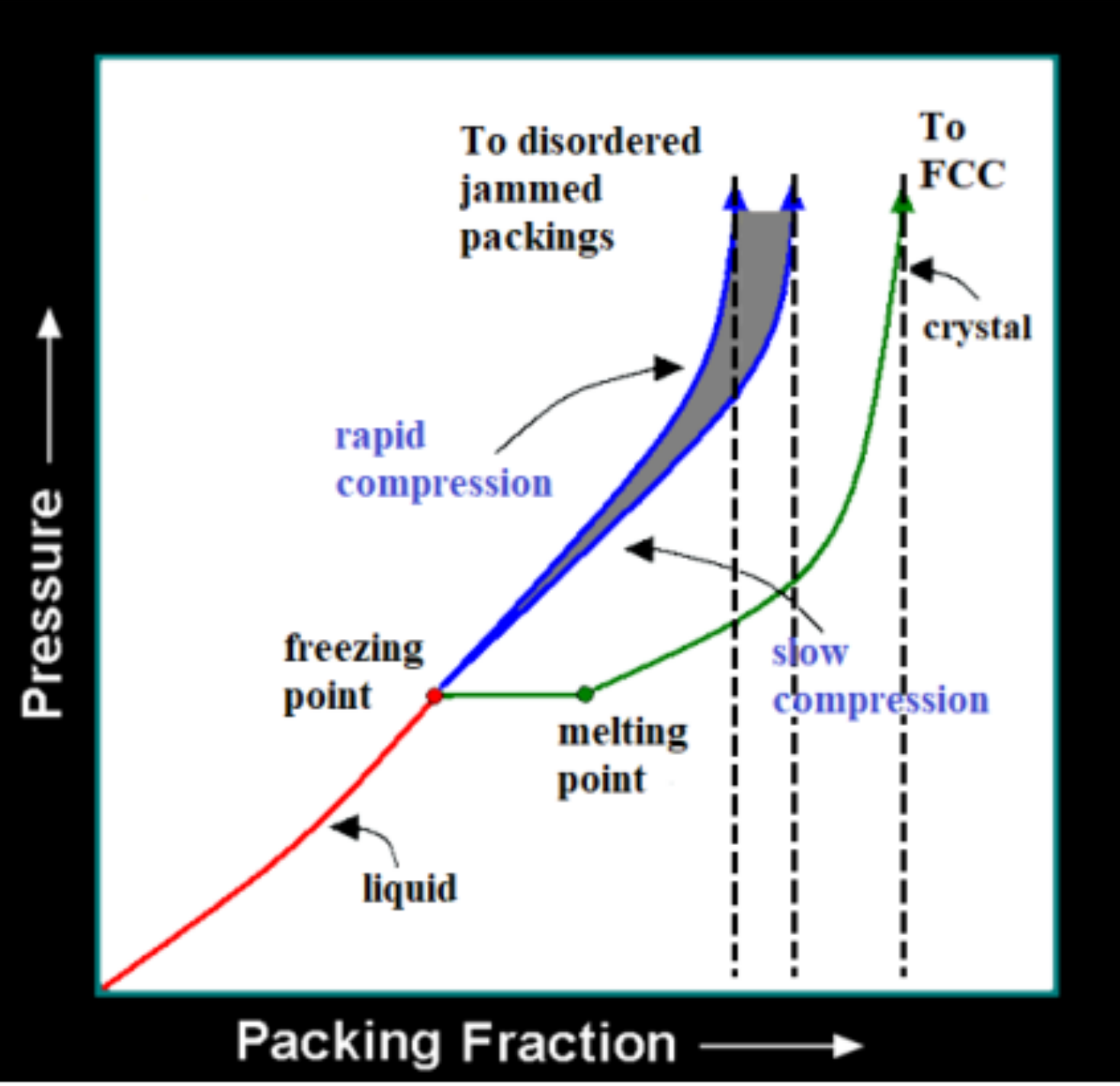}}
\caption{\footnotesize The isothermal phase behavior of the 3D hard-sphere model in the pressure-packing fraction plane, adapted from 
Ref. \onlinecite{To02a}.
 Three different isothermal 
densification paths by which a hard-sphere liquid 
may jam are shown.  An infinitesimal compression rate of the liquid traces out the thermodynamic equilibrium path (shown in green), 
including a discontinuity resulting from the first-order freezing transition
to a crystal branch that ends at a maximally dense, infinite-pressure
jammed state.   Rapid compressions of the liquid while suppressing 
some degree of local order (blue curves) can avoid 
crystal nucleation (on short time scales) and produce a range
of amorphous metastable extensions of the liquid branch that jam
only at the their density maxima. }
\label{hsphere}
\end{figure}

However, compressing a hard-sphere liquid rapidly, under the
constraint that significant crystal nucleation is suppressed,
can produce a range of metastable branches whose density
end points are disordered ``jammed" packings,\cite{Ri96b,To02a,Ja13} which can be regarded to be glasses.
A rapid compression leads to a lower random jammed
density than that for a slow compression. 
The most rapid compression ending in  mechanically stable packing is presumably
 the maximally random jammed (MRJ) state\cite{To00b} with $\phi \approx 0.64$.
Accurate approximate formulas 
for the pressure along such metastable extensions up to the jamming points have been obtained. \cite{To95b,To02a}
This ideal amorphous state is described in greater detail in Sec. \ref{mrj}. Note that rapid compression
a hard-sphere system is analogous  to supercooling a molecular liquid.

Pair statistics are exactly known only in the case of 1D ``hard-rod" systems.\cite{Ze27}
For $d \ge 2$, approximate formulas for $g_2(r)$ are known along liquid branches.\cite{Ha86}
Approximate closures of the Ornstein-Zernike integral equation linking the direct 
correlation function $c(r)$ to the total correlation function $h(r)$,~\cite{Or14}
such as the Percus-Yevick (PY) and hypernetted chain schemes,\cite{Pe58,St63,Ha86}  provide
reasonably accurate estimates of  $g_2(r)$ for hard-spheres liquids. 
Because $g_2(r)$ decays to unity exponentially fast for liquid states, we can conclude
that it must have a corresponding structure factor $S(k)$ that is an even function and analytic at $k=0$;
see Eq. (\ref{low-k}). Also, since the leading-order term $S_0$ must be positive
because the isothermal compressibility is positive [cf. (\ref{comp})], hard-sphere liquids are not hyperuniform. 
Figure \ref{g2-S-equi} shows $g_2(r)$ and the corresponding structure factor $S(k)$
in three dimensions at $\phi=0.35$ as obtained from the PY approximation.

\begin{figure}
\begin{center}
{\includegraphics[width=2.8in,keepaspectratio,clip=]{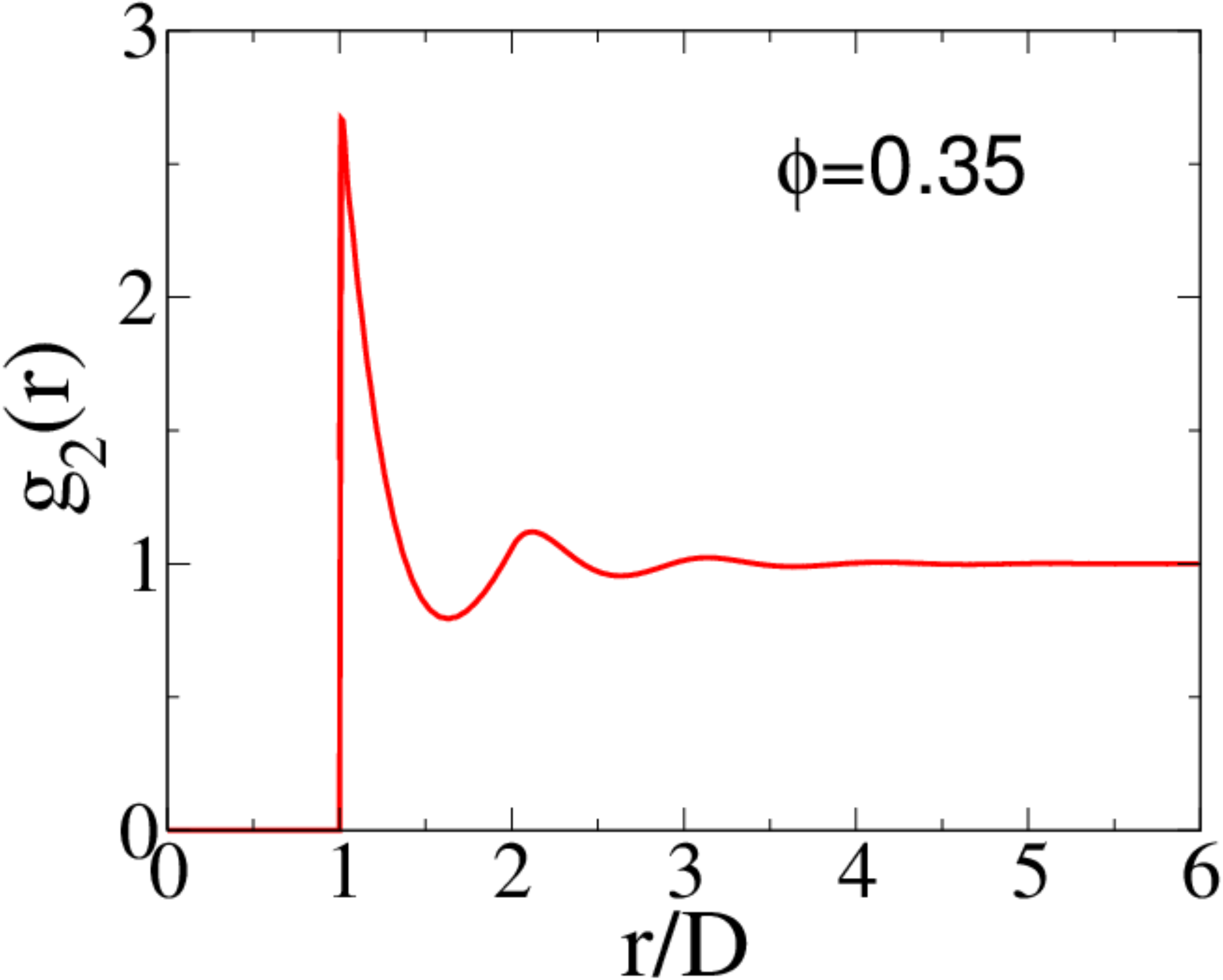}\hspace{0.25in}
\includegraphics[width=2.7in,keepaspectratio,clip=]{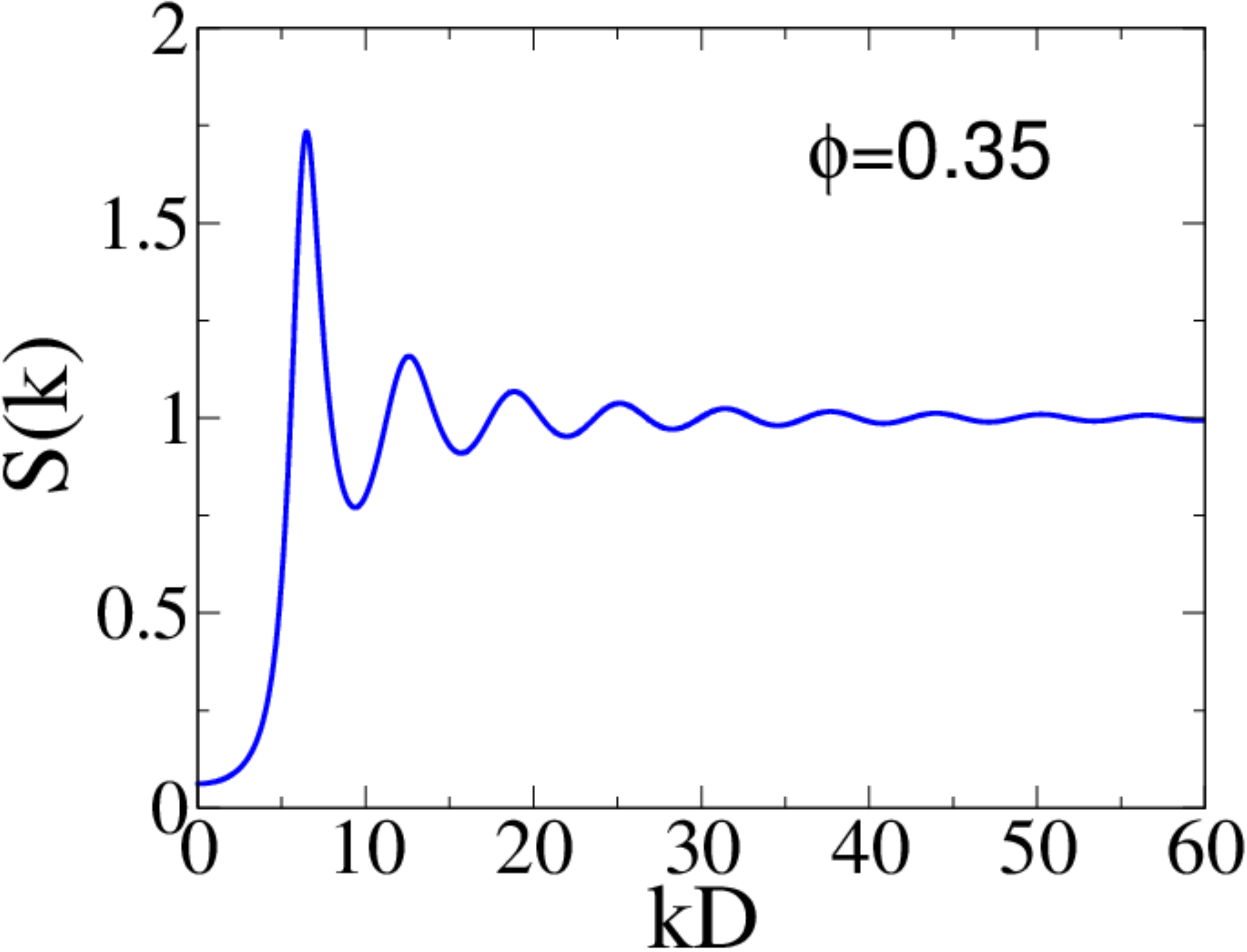}}
\end{center}
\caption{\footnotesize Pair statistics for an equilibrium hard-sphere fluid 
in three dimensions at $\phi=0.35$ as obtained from the PY approximation. \cite{Pe58,Ha86}
Top panel:  Pair correlation function  $g_{2}(r)$ versus $r/D$, where $D$ is the sphere diameter.
Bottom panel: The corresponding structure factor $S(k)$ as a function
of the dimensionless wavenumber $kD$, which clearly shows nonhyperuniformity.}
\label{g2-S-equi}
\end{figure}

\subsection{Nonequilibrium Disordered Packings}
\label{noneq}

Here we briefly discuss three different nonequilibrium 
sphere packings: random sequential addition, ``ghost" random 
sequential addition, and random close packing.

\subsubsection{Random Sequential Addition}

Perhaps one of the most well-known nonequilibrium packing models is the random sequential addition (RSA)
packing procedure, which  is a time-dependent process that generates  disordered sphere packings in $\mathbb{R}^d$; see Refs.
\onlinecite{Re63,Wi66,Fi79,Fe80,Sw81,Co88,Ta91,To06d,Zh13b}. The RSA packing process in the first three space dimensions has been used to model 
a variety of different condensed phases, including  protein adsorption, \cite{Fe80} polymer oxidation, \cite{Fl39}
particles in cell membranes \cite{Fi79} and ion implantation in semiconductors. \cite{Ro83} 

In its simplest rendition, a RSA sphere packing  is produced by randomly,
irreversibly, and sequentially placing nonoverlapping objects
into a large volume in $\mathbb{R}^d$ that at some initial time is empty of
spheres. If an attempt to add a sphere at some time $t$ results in an
overlap with an existing sphere in the packing, the attempt is
rejected and further attempts are made until it can be added
without overlapping existing spheres. 
One can stop the addition process at any finite time $t$, obtaining an RSA configuration with a 
time-dependent packing fraction $\phi(t)$ but this value cannot exceed  the maximal {\it saturation} 
packing fraction $\phi_s=\phi(\infty)$ that occurs in the infinite-time and thermodynamic limits. 
Even at the saturation state, the spheres are never in contact and hence are not {\it jammed}
in the sense described in Sec. \ref{jamming}; moreover, the pair correlation function $g_2(r)$ decays to unity
for large $r$ super-exponentially fast.~\cite{Bo94} The latter property
implies that  the structure factor $S(k)$ of RSA  packings 
in $\mathbb{R}^d$ must be an even and analytic function at $k=0$,~\cite{To06d} as indicated
in Eq. (\ref{low-k}). It is notable that saturated RSA packings are not
hyperuniform for any finite space dimension.~\cite{To06d,Zh13b} 

In the one-dimensional case, also known as the ``car-parking'' problem, 
the saturation packing fraction was obtained analytically: $\phi_s=0.7475979202...$.\cite{Re63}
However, for $d \ge 2$, $\phi_s$ in the thermodynamic limit has only been estimated via  numerical simulations.
The most precise numerical study to date~\cite{Zh13b} has yielded  $\phi_s = 0.5470735 \pm 0.0000028$ for $d=2$  
and $\phi_s =  0.3841307 \pm 0.0000021$ for $d=3$.  
Estimates of  $\phi_s$ in higher dimensions are discussed in Sec. \ref{high}.
RSA packings have also been examined for spheres with a size distribution
 \cite{Ad97} 
and other particle shapes, including squares,~\cite{Fe80,Bro91,Vi92} rectangles,~\cite{Vi89} ellipses,~\cite{Vi92}
superdisks,~\cite{Gr09} and polygons,~\cite{Ci14,Zhang18} in $\mathbb{R}^2$ and spheroids,~\cite{Sh97} spherocylinders~\cite{Vi92} 
and cubes~\cite{Bo01,Ci18} in $\mathbb{R}^3$.

\begin{figure}[bthp]
\centerline{\includegraphics[height=1.5in,keepaspectratio,clip=]{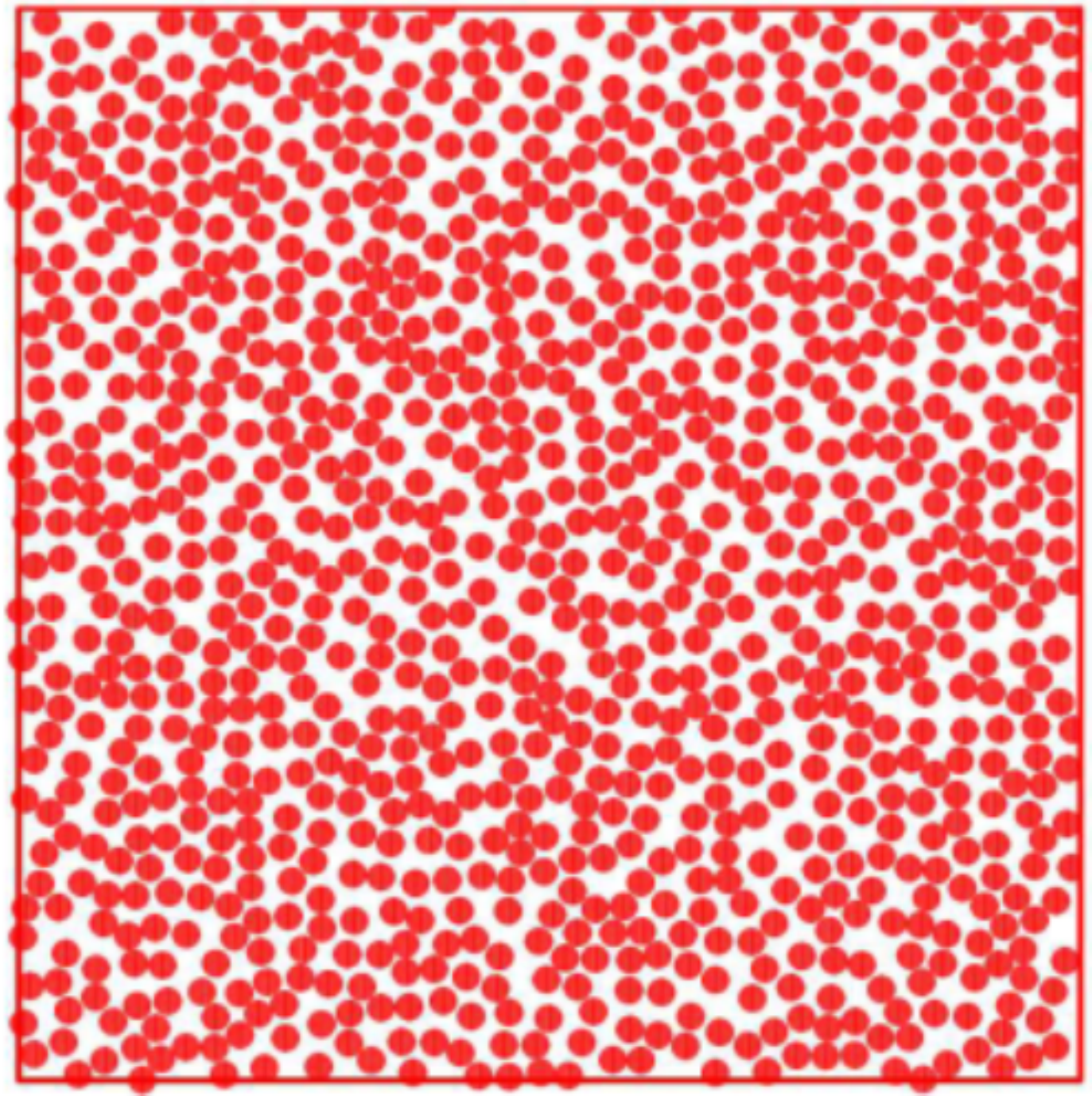} \hspace{0.25in}\includegraphics[height=1.5in,keepaspectratio,clip=]{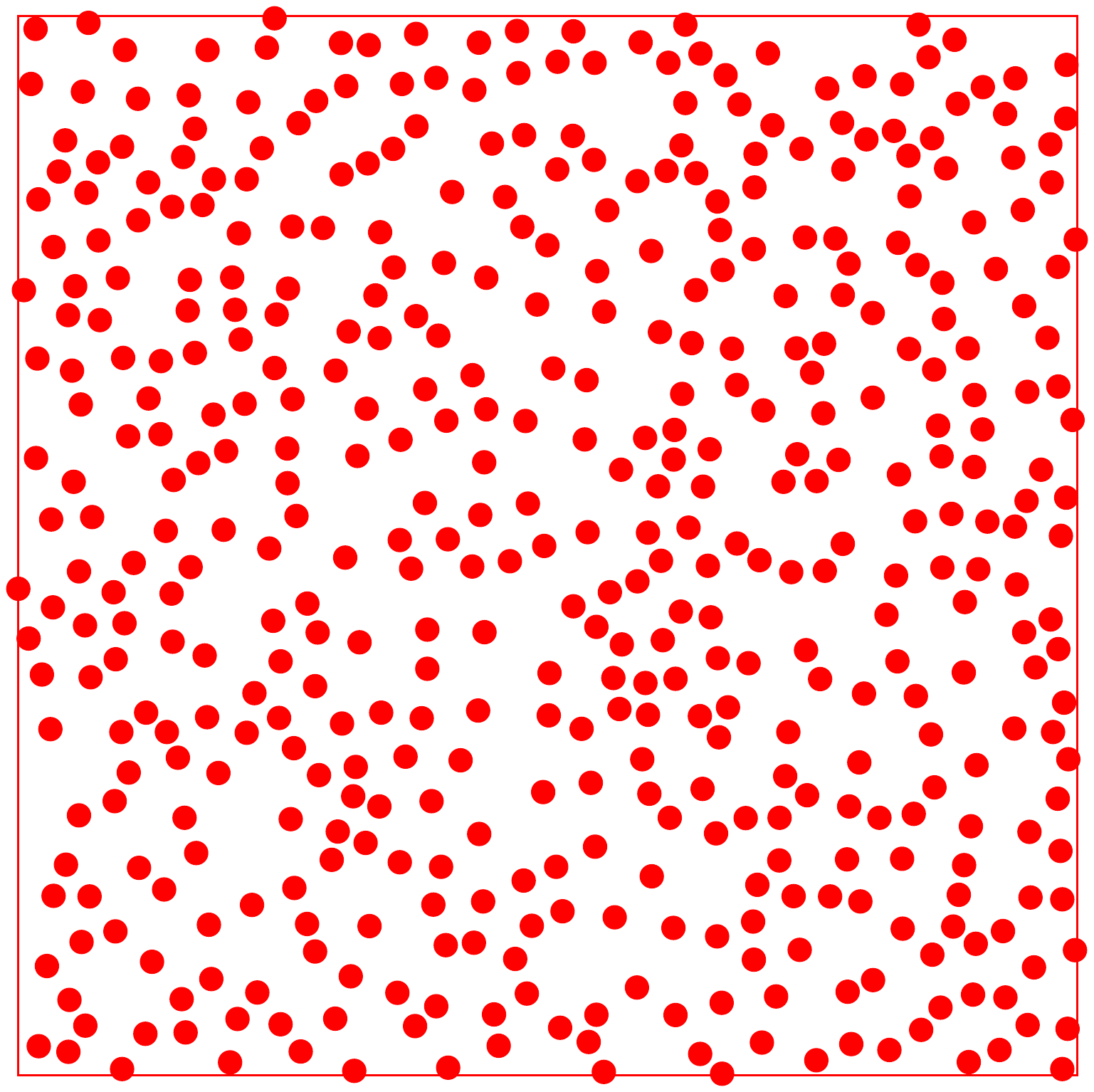}}
\caption{(Color online) \footnotesize Examples of two nonequilibrium  
packing models in two dimensions under periodic boundary conditions. Left panel: 
A configuration of the standard RSA packing at saturation with $\phi_s \approx 0.5$.
Right panel: A configuration of a ghost RSA packing at a packing 
fraction $\phi$ very near its maximal value of 0.25.
Note that the packing is clearly unsaturated (as defined in Sec. \ref{defs}) and there are no contacting particles.}
\label{grsa}
\end{figure}

\subsubsection{``Ghost" Random Sequential Addition}
\label{ghost}

There is a generalization of the aforementioned standard RSA process, which
can be viewed as a ``thinning" process of a Poisson distribution of sphere and is parameterized
by the positive constant  $\kappa$ that lies in the closed interval $[0,1]$; see Ref. \onlinecite{To06a}.
When $\kappa=0$, one recovers the standard RSA process and when $\kappa=1$,
one obtains the ``ghost" random sequential addition (RSA) process
that enables one  to obtain exactly not only the time-dependent packing
fractions but all of the $n$-particle correlation functions $g_n$ for
any  $n$ and $d$. The reader is referred to Ref. \onlinecite{To06a} for details
about the general model. 

In the ghost RSA process, one attempts to place spheres into an 
initially empty region of space randomly, irreversibly and sequentially. However, here 
one keeps track of any rejected sphere, which is called  a ``ghost" sphere.
No additional spheres can be added whenever they overlap an existing sphere 
or a ghost sphere. The packing fraction at time $t$ for spheres of unit diameter is 
given by $\phi(t)=[1-\exp(-v_1(1) t)]/2^d$, where $v_1(R)$ is the volume
of sphere of radius $R$. Thus, we see that as $t \rightarrow +\infty$, $\phi = 2^{-d}$,
which  is appreciably smaller than the RSA saturation packing fraction $\phi_s$ in low dimensions; see Fig. \ref{grsa} for 2D examples. Nonetheless, 
it is notable that the ghost RSA process is the only hard-sphere packing model that is exactly
solvable for any dimension $d$ and all realizable densities, which
has implications for high-dimensional packings,
as discussed in Sec. \ref{high}.

\subsubsection{Random Close Packing}

The ``random close packing" (RCP) notion was pioneered by Bernal\cite{Be60,Ber65} to model 
the structure of liquids, and 
has been one of the more persistent themes with a venerable history. \cite{Sc69,An72,Go74b,Be83,Jo85,Zi94,Ju97,Po97}
Two decades ago, the prevailing notion of the RCP state was that it 
is the {\it maximum} density that a large, random collection of congruent (identical) spheres can attain and that this density is a well-defined  quantity.
This traditional view has been summarized as
follows: ``Ball bearings and similar objects have been shaken,
settled in oil, stuck with paint, kneaded inside rubber
balloons--and all with no better result than (a packing fraction of) $\ldots$ $0.636$''; see Ref. \onlinecite{An72}.

Torquato, Truskett and Debenedetti \cite{To00b} have argued that this 
RCP-state concept is actually ill-defined because ``randomness" and ``closed-packed" 
were never defined  and, even if they were, 
are at odds with one another. Using the Lubachevsky-Stillinger (LS) \cite{Lu90} molecular-dynamics growth 
algorithm to generate jammed packings, it was shown \cite{To00b} that fastest particle growth rates
generated the most disordered sphere packings with $\phi \approx 0.64$, but that by
slowing the growth rates larger packing fractions could
be continuously achieved up to the densest value 
$\phi_{\mbox{\scriptsize max}} = 0.74048\ldots$ 
such that the degree of order increased monotonically with $\phi$. These results demonstrated 
that once can increase the packing fraction
by an arbitrarily small amount at the 
expense of correspondingly small increases in order and thus the notion of RCP is ill-defined 
as the highest possible density that a random sphere packing can attain.
 To remedy these  flaws, Torquato, Truskett and Debenedetti \cite{To00b} replaced
the notion of ``close packing" with
``jamming" categories (defined precisely in Sec. \ref{jamming}) and introduced the notion
of an ``order metric" to quantify the degree
of order (or disorder) of a single packing configuration.
 This led them to supplant the concept of RCP with the maximally random jammed (MRJ) state,
which is defined, roughly speaking,  to be that jammed state 
with a minimal value of an order metric (see Sec. \ref{order} for details).
This work pointed the way toward a quantitative means
of characterizing all packings, namely, the geometric-structure approach.

\begin{figure}[bthp]
\centerline{\includegraphics[height=1.5in,keepaspectratio,clip=]{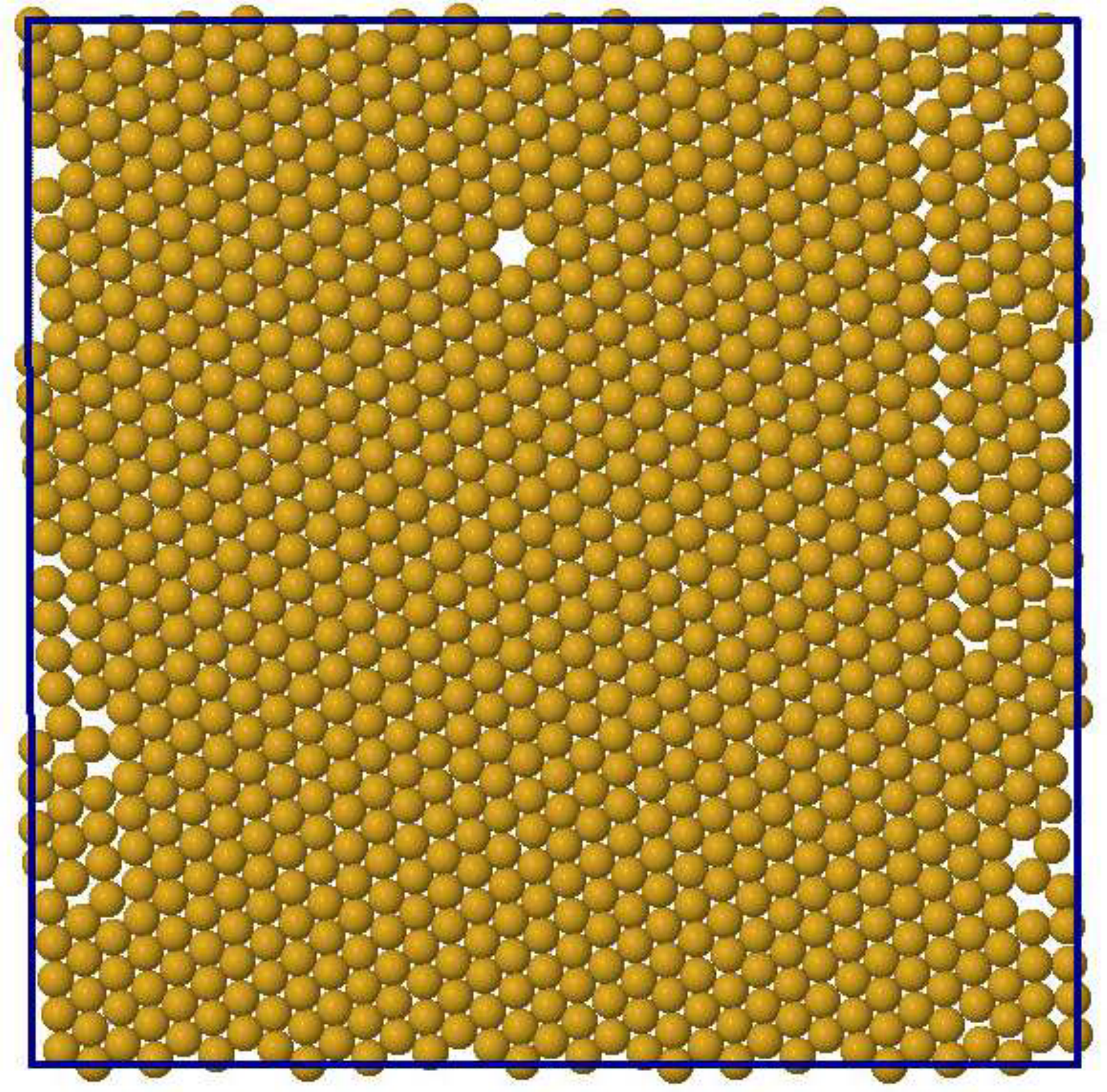}
\hspace{0.25in}
\includegraphics[height=1.7in,keepaspectratio,clip=]{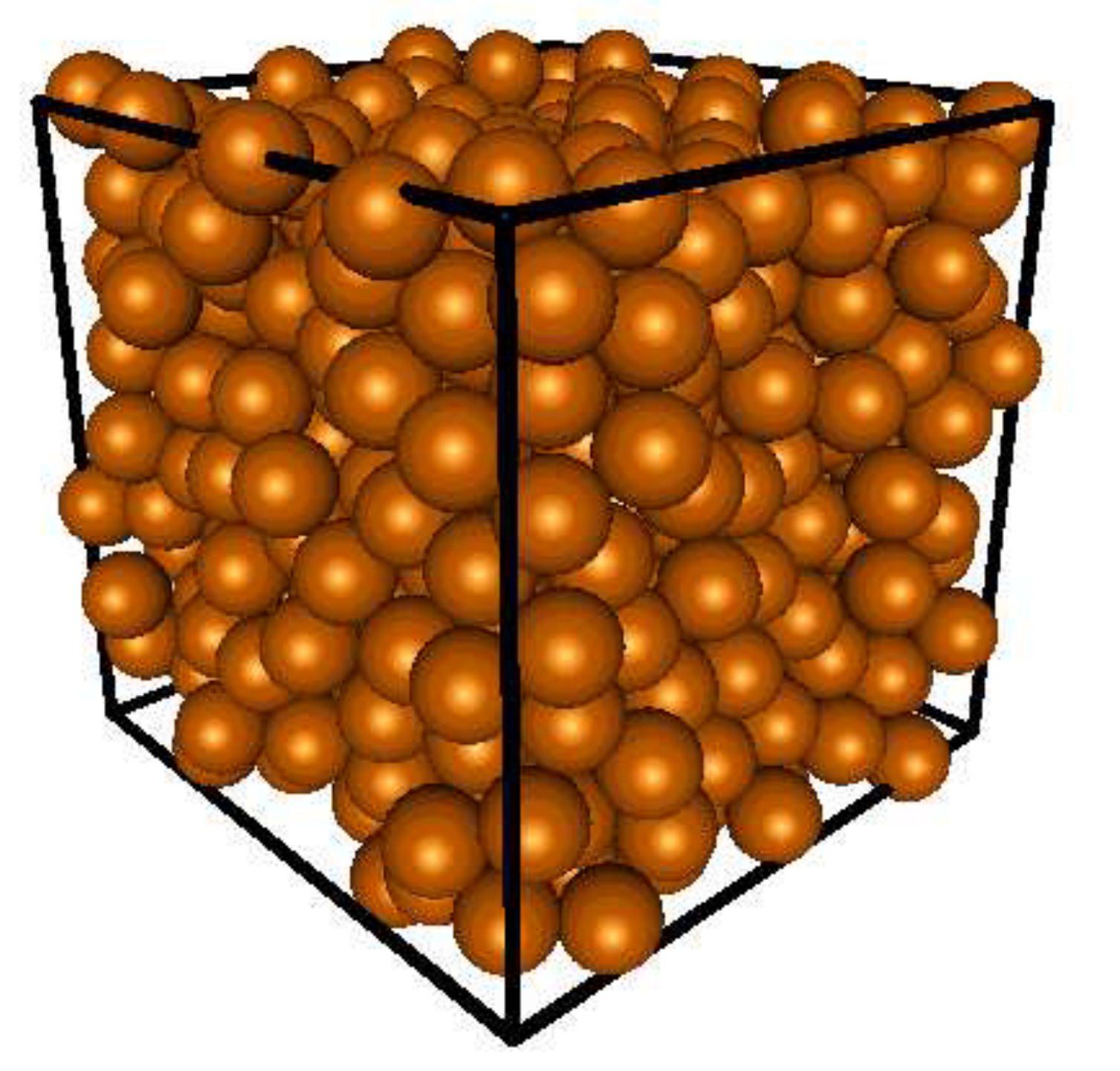}
 }
\caption{\footnotesize Typical protocols, used to generate disordered sphere packings
in three dimensions, produce highly crystalline packings in two dimensions. 
Left panel: A highly ordered collectively jammed configuration 
(Sec. \ref{jamming-cat}) of 1000 disks with $\phi \approx 0.88$ produced
using the Lubachevsky-Stillinger (LS) algorithm \cite{Lu90} with a fast expansion rate. \cite{Do04a}
Right panel: A 3D MRJ-like configuration
of 500 spheres with $\phi \approx 0.64$ produced
using the  LS algorithm with a fast expansion rate. \cite{To00b}
}
\label{MRJ}
\end{figure}


We note that whereas the  LS packing protocol with a fast growth rate typically leads to disordered jammed states in
three dimensions, it invariably produces  highly crystalline ``collectively" jammed packings in two
dimensions. Figure \ref{MRJ}  vividly illustrates the differences between the textures produced in
three and in two dimensions (see Section VII for further remarks).

\subsection{Geometric-Structure Approach to Jammed Packings}
\label{jamming}

A ``jammed" packing is one  in which each particle
is in contact with its nearest neighbors  such  that
mechanical stability of a specific type is conferred to the packing, as detailed below.
Two conceptual approaches for their study  have emerged. One is  
the ``ensemble" approach, \cite{Be60,Ber65,Ed94,Ed01,Li98,Ma02,Si02,Oh03,Wy05,Si05,Ga06,Ma08,Pa10} which for
a given packing protocol  aims to understand typical
configurations and their frequency of occurrence. The other, more recently, is the
``geometric-structure" approach, \cite{To00b,To01b,Ka02d,To03c,Do04a,Do04b,Do07a,To07} which emphasizes quantitative characterization of single-packing configurations without regard
to their occurrence frequency in the protocol used to produce them.
Here we focus on the latter approach, which enables one 
to enumerate and classify packings with a diversity
of order/disorder and packings fractions, including extremal
packings, such as the densest sphere packing  and MRJ packings.
 
\subsubsection{Jamming Categories}
\label{jamming-cat}
 
Three broad and mathematically precise
``jamming" categories of sphere packings can be distinguished depending on the nature of their 
mechanical stability. \cite{To01b,To03a}  In order of increasing stringency (stability), for
a finite sphere packing, these are the following:
\begin{enumerate}
\item {\it Local~jamming}: Each particle in the packing is locally trapped by
its neighbors (at least $d+1$ contacting particles,
not all in the same hemisphere), i.e., it cannot be translated while fixing the positions
of all other particles;
\item {\it Collective~jamming:} Any locally jammed configuration is collectively jammed if no
subset of particles can be simultaneously displaced so that its members
move out of contact with one another and with the remainder set; and
\item {\it Strict~jamming:} Any collectively jammed configuration that disallows
all  uniform volume-nonincreasing strains of the system
boundary is strictly jammed.
\end{enumerate}

\noindent{These hierarchical jamming categories are closely related to the concepts of ``rigid" and ``stable"
packings found in the mathematics literature, \cite{Co98} 
and imply that there can be no ``rattlers" (i.e., movable but caged particles) in the packing.
The jamming category of a given packing depends on the boundary conditions employed; see Refs. \onlinecite{To01b,To10c}
for specific examples.
Rigorous and efficient linear-programming (LP) algorithms have been devised
to assess whether a particular sphere packing is locally, collectively, 
or strictly jammed.~\cite{Do04a,Do04c} These jamming categories can
now be ascertained in real-system experiments by applying the LP tests
to configurational coordinates of a packing determined via a variety of 
imaging techniques, including tomography \cite{As06}, confocal microscopy  \cite{Bru03} and magnetic resonance imaging. 
~\cite{Ma05}}
     
\subsubsection{Polytope Picture and Pressure in Jamming Limit}
\label{polytope}

A  packing of $N$ hard spheres of diameter $D$ in a
jammed framework in $\mathbb{R}^d$ is
specified by a $Nd$-dimensional configurational
position vector $\bf{R}={\bf r}^N \equiv \{{\bf r}_{1},\ldots,{\bf r}_{N}\}$.
\emph{Isostatic} jammed packings  possess the minimal
number of contacts for a jamming category and boundary conditions. \cite{Do05c}
The relative differences between isostatic collective and strict jammed packings
diminish as $N$ becomes large, and since the number
of degrees of freedom is essentially equal to $N d$, 
an isostatic packing has a {\it mean contact number per particle}, $\overline Z$,
equal to  $2d$ in the large-$N$ limit. Packings having more and fewer contacts than
isostatic ones are \emph{hyperstatic} and \emph{hypostatic}, respectively.
Sphere packings that are hypostatic cannot be collectively or strictly jammed.\cite{Do07a}

Consider decreasing the density slightly 
in a sphere packing that is at least collectively 
jammed by reducing the particle diameter
by $\Delta{D}$,  so that the packing fraction is
lowered to $\phi=\phi_{J}\left(1-\delta\right)^{d}$, where $\delta=\Delta{D}/D\ll1$
is called  the \emph{jamming gap} or distance to jamming.
There is a sufficiently small $\delta$ that
does not destroy the jamming confinement property.
For fixed $N$ and sufficiently small $\delta$, it can be shown asymptotically, 
through first order in $\delta$),  that  the set of displacements accessible to the packing approaches
a convex \emph{limiting polytope} (high-dimensional polyhedron) $\mathcal{P}$.
\cite{Sa62,St69} This polytope $\mathcal{P}$
is determined from the linearized impenetrability equations \cite{Do04a,Do04c}
and is necessarily bounded for a jammed
configuration.

Now consider adding thermal kinetic energy to a nearly jammed
sphere packing in the {\it absence of rattlers}. While the system will not be 
globally ergodic over the full system configuration space and thus
not in thermodynamic equilibrium,  one can still define a 
macroscopic pressure $p$ for the trapped but locally ergodic system
by considering time averages as the system
executes a tightly confined motion around the \emph{particular} jammed configuration
$\bf{R}_{J}$. The probability distribution $P_{f}(f)$ of the {\it time-averaged} interparticle forces $f$ has been rigorously linked to the contact 
value $r=D$ of the pair correlation function $g_2(r)$. \cite{Do05c}
Moreover, the available (free)
configuration volume scales with the jamming
gap $\delta$ such that  the reduced pressure is asymptotically given by
the free-volume equation of state:~\cite{Sa62,St69,Do05c}
\begin{equation}
\frac{p}{\rho k_{B}T} \sim \frac{1}{\delta}=\frac{d}{1-\phi/\phi_{J}},
\label{free}
\end{equation}
where $T$ is the absolute temperature and $\rho$ is the number density.
 Relation (\ref{free}) is remarkable, since it enables
one to determine  accurately the true jamming density of a given packing,
even if the actual jamming point has not quite yet been reached, just by
measuring the pressure and extrapolating to $p=+\infty$.
This free-volume form has been used to estimate
the equation of state along ``metastable" extensions of the hard-sphere
fluid up to the  infinite-pressure endpoint, 
assumed to be  disordered jammed states (see Fig. \ref{hsphere} and Sec. \ref{phase}). \cite{To95b,To02a} 
Kamien and Liu \cite{Ka07} assumed the same free-volume 
form to fit data for the pressure of ``metastable" hard-sphere states.

\subsubsection{Hard-Particle Jamming Algorithms}

For many years, the Lubachevsky-Stillinger (LS) algorithm~\cite{Lu90} has served as the premier 
numerical scheme to generate a wide spectrum of dense jammed hard-sphere packings with
variable disorder in both two and three dimensions.~\cite{To00b,Tr00,Ka02d}
This is an event-driven or collision-driven  molecular-dynamics algorithm: an initial configuration of
spheres of a given size within a periodic cell are given initial
random velocities and the motion of the spheres are followed
as they collide elastically and also grow uniformly until the
spheres can no longer expand. The final density is generally inversely related
to the particle growth rate. This algorithm has been generalized by 
Donev, Torquato, and Stillinger~\cite{Do05a,Do05b} to generate
jammed packings of smoothly shaped nonspherical particles,
including ellipsoids,~\cite{Do05b,Do07a} superdisks,~\cite{Ji08b}, and superballs.~\cite{Ji09a}
Event-driven packing protocols with
growing particles, however, do not guarantee jamming of the final
packing configuration, since jamming is not explicitly incorporated
as a termination criterion.

The task of generating dense packings of
particles of general shape within an adaptive periodic
fundamental cell 
has been posed by Torquato and Jiao~\cite{To09b,To09c} as an optimization problem called the
{\it adaptive-shrinking cell} (ASC) scheme. The negative of the packing fraction, $-\phi$
(which can be viewed as an ``energy")
is to be minimized subject to constraints, and 
the design variable are the shape and size of the deformable periodic fundamental cell 
as well as the centroids and orientations of the particles.
The so-called Torquato-Jiao (TJ) sphere-packing
algorithm \cite{To10e} is a sequential linear-programming (SLP) solution
of the ASC optimization problem for the special case
of packings of spheres with a size distribution for which
linearization of the design variables is {\it exact}.
The  deterministic SLP solution in principle always leads to
strictly jammed packings up to a high numerical tolerance with a
wider range of densities and degree of disorder than packings
produced by the LS algorithm.  From an initial configuration,
the TJ algorithm   leads to a mechanically stable local ``energy" minimum
(local density maximum), which in principle is the inherent
structure associated with the starting initial many-particle
configuration; see Fig. \ref{inherent}.   A broad range
of inherent structures can be obtained across space dimensions, including locally maximally dense and mechanically
stable packings, such as MRJ states,\cite{To10e,At13,At14,At16a} disordered hyperstatic packings~\cite{Ji11a} and  the globally maximally dense inherent
structures,\cite{To10e,Ho11b,Ho12a,Ma13c,Ka13} with very small computational cost, provided that
the system sizes are not too large. 

It is notable   the TJ algorithm creates disordered jammed sphere packings that are closer to the ideal MRJ state
than previous algorithms.\cite{At13} It was shown that the rattler concentration
 of these packings converges towards 1.5\% in the infinite-system
limit, which is markedly lower than previous estimates for the MRJ state using the LS
protocol (with about 3\% rattlers).

\begin{figure}[htp]
\begin{center}
$\begin{array}{c}
\includegraphics[height=7.5cm, keepaspectratio,clip]{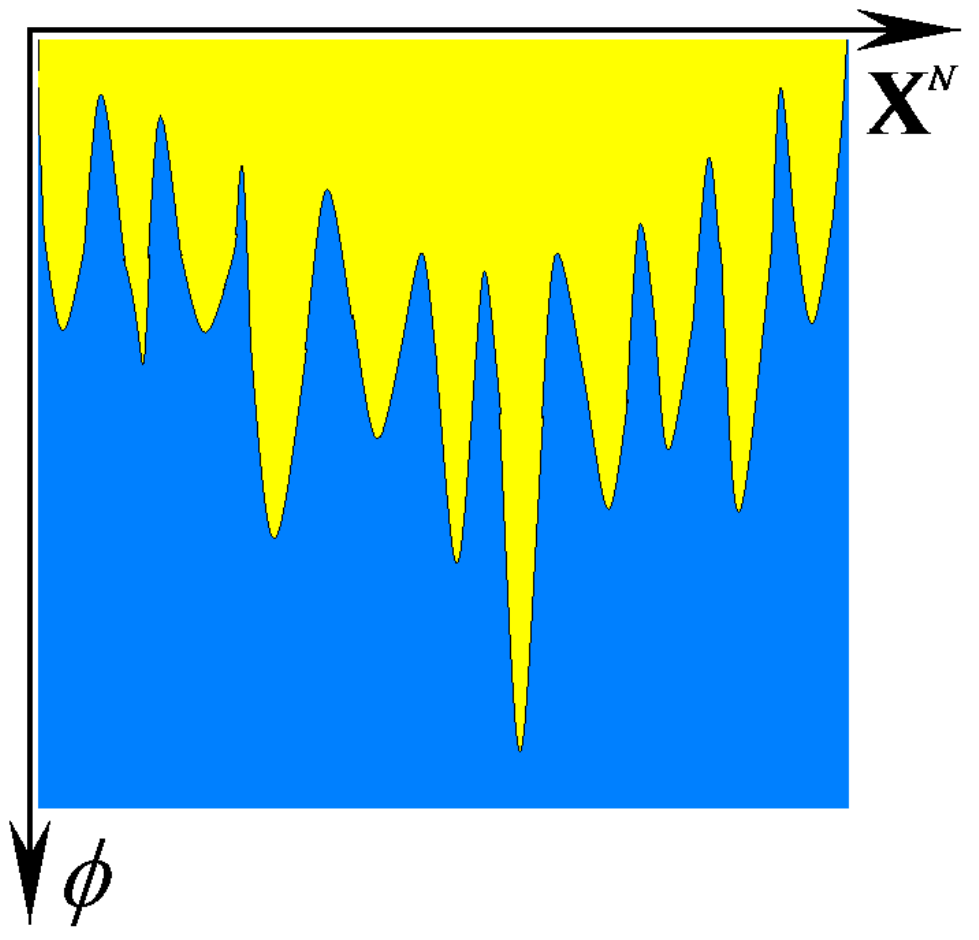} \\
\end{array}$
\end{center}
\caption{(Color online).  A schematic diagram of inherent structures (local density maxima) for  
sphere packings of $N$ spheres, as taken from Ref. \onlinecite{To10e}. The
horizontal axis labeled ${\bf X}^N$ stands for the entire set
of centroid positions, and  $-\phi$ (``energy") decreases downward.
The jagged curve is the boundary between accessible 
configurations (yellow)  and inaccessible ones (blue).  The deepest point of the
accessible configurations corresponds to the maximal density
packings of hard spheres.} \label{inherent}
\end{figure}

\subsubsection{Order Metrics}
\label{order}

The enumeration and classification of both ordered and disordered 
sphere packings is an outstanding problem.
Since the difficulty of the complete enumeration of packing configurations rises
exponentially with the number of particles,  it is 
desirable to devise a small set of intensive parameters that can characterize packings well. 
One  obvious property of a sphere packing is the packing fraction $\phi$.
Another important characteristic of a packing is some
measure of its ``randomness" or degree of disorder.
Devising such  measures  is a highly  nontrivial challenge, but even the tentative solutions
that have been put forth during the last two decades
have been fruitfully applied   to  characterize not only
sphere packings \cite{To00b,Tr00,Ka02d,To03a} but also  simple liquids,\cite{Ri96a,Ri96b,Tr00,Er03} glasses,\cite{Tr00,Zh16a} water,~\cite{Er01,Er02} disordered ground states,~\cite{To15} random media,~\cite{Di18}
and biological systems.\cite{Ji14}

It is quite reasonable to consider ``entropic" measures to characterize the
randomness of packings.
However, as pointed out by Kansal {\it et al.}~\cite{Ka02d}, a substantial hurdle to
overcome in implementing such an order metric is the necessity to generate all possible jammed
states or, at least, a representative sample of such states in an
unbiased fashion using a ``universal" protocol in the large-system
limit, each of which is an intractable problem. Even if such a universal protocol could be developed,
however, the issue of what weights to assign the resulting
configurations remains. Moreover, there are other basic
problems with the use of entropic measures as order metrics, as we will discuss in Sec.~\ref{mrj},
including its significance for certain 2D hard-disk packings.

We know that a many-body system of $N$ particles is completely characterized statistically 
by its $N$-body probability density function $P({\bf R};t)$ that is associated
with finding the $N$-particle system with configuration ${\bf R}$ at
some time $t$. Such complete
information is virtually never  available for large $N$ and, in practice, one
must settle for reduced information, such
as a scalar order metric $\psi$.
Any order metric $\psi$ conventionally possesses the following
three properties: (1) it is a well-defined scalar function of
a configuration ${\bf R}$; (2) it is subject 
typically to the normalization $0\le \psi \le 1$;
and, (3) for any two configurations ${\bf R}_A$ and ${\bf R}_B$, $\psi({\bf R}_A) > \psi({\bf R}_B)$ implies that 
configuration  ${\bf R}_A$ is to be
considered as more ordered than configuration ${\bf R}_B$. 
The set of order metrics that one selects is unavoidably subjective, given that there 
appears to be no single universally applicable scalar metric capable of describing order across
all lengths scales. However, one can construct order metrics that lead to consistent
results, as we will be discussed below.

Many relevant order metrics have been devised. While a comprehensive discussion of this topic
is beyond the scope of this article, it is useful to here to note some order metrics
that have been identified, including bond-orientational
order metrics in two \cite{Ka00} and three dimensions, \cite{St83}
 translational order metrics,  \cite{To00b,Tr00,To15} and hyperuniformity
order metrics. \cite{To03a,Za09} These specific order metrics have both strengths and weaknesses.
This raises the question of what are the characteristics
of a good order metric?  It has been suggested that
a  good scalar order metric should have the following additional properties: \cite{Ka02d,To10c}
(1) sensitivity to any type of
ordering without  bias toward any reference system; (2) ability to reflect the hierarchy of
ordering between prototypical systems given by common
physical intuition (e.g., perfect crystals with high symmetry
should be highly ordered, followed by quasicrystals, correlated disordered packings
without long-range order, and finally spatially uncorrelated or Poisson
distributed particles); (3) incorporation of both the variety of local coordination patterns 
as well as the spatial distribution of such patterns should be included;
and  (4) the capacity to detect translational and orientational order at any length scale.   
Moreover, any useful set of order metrics
should consistently produce results that are positively correlated with one another.
\cite{To00b,To02a} The development of improved order
metrics deserves continued research attention.

Order metrics and maps have been fruitfully extended
to characterize the degree of structural order in condensed phases 
in which the constituent particles (jammed or not) possess both attractive and repulsive interactions. 
This includes the determination of the order maps of models of simple liquids, glasses and crystals with isotropic interactions, \cite{Ri96a,Ri96b,Tr00,Er03}
models of water, \cite{Er01,Er02} disordered ground states of
long-ranged isotropic pair potentials,\cite{To15} and models of amorphous polymers. \cite{Stach03}

\subsection{Order Maps and Extremal Packings}
\label{map}

The geometric-structure classification naturally emphasizes that  
there is a great diversity in the types of attainable
packings with varying magnitudes of overall order, packing fraction $\phi$, mean contact number per particle, 
$\overline Z$, among many other intensive parameters.
Any attainable hard-sphere configuration, jammed or not, can be mapped to a point
in this multidimensional parameter space that we call  an {\it order map}. 
The use
    of ``order maps" in combination with the mathematically precise ``jamming
    categories" enable one to view and characterize individual packings, including the densest sphere packing (e.g., Kepler's conjecture in 3D) and 
MRJ packings as extremal states in the order map for a
    given jamming category. Indeed, this picture encompasses not only these
    special jammed states, but an uncountably infinite number of other
    packings, some of which have only recently been identified as physically
    significant, e.g., the jamming-threshold states~\cite{To07} (least dense 
jammed packings) as well as states between these and MRJ.

\begin{figure}[bthp]
\centerline{
\includegraphics[height=1.8in,keepaspectratio,clip=]{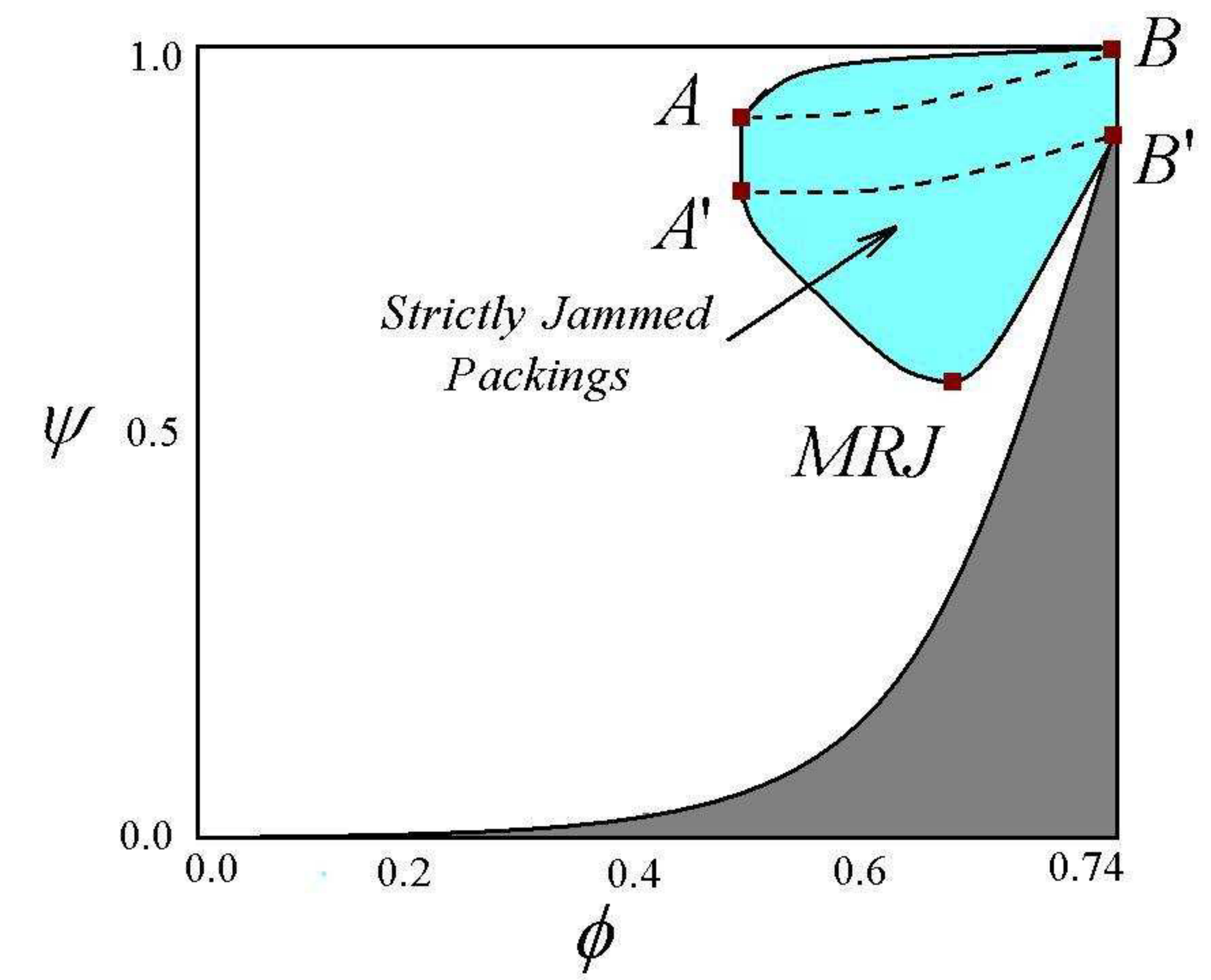}}
\caption{\footnotesize Schematic {\it order map} of sphere packings  in $\mathbb{R}^3$ in the density-order
($\phi$-$\psi$) plane. White and blue regions contain the attainable
packings, blue regions represent the jammed subspaces, and  dark shaded regions
contain no packings. The boundaries of the jammed region delineate {\it extremal} structures.
The locus of points $A$-$A^{\prime}$ correspond to the lowest-density jammed
packings. The locus of points $B$-$B^\prime$ correspond
to the densest jammed packings. Points MRJ represent
the maximally random jammed states, i.e., the most disordered
states subject to the jamming constraint.}
\label{maps}
\end{figure}

\begin{figure}
\centerline{\includegraphics[height=1.in,keepaspectratio,clip=]{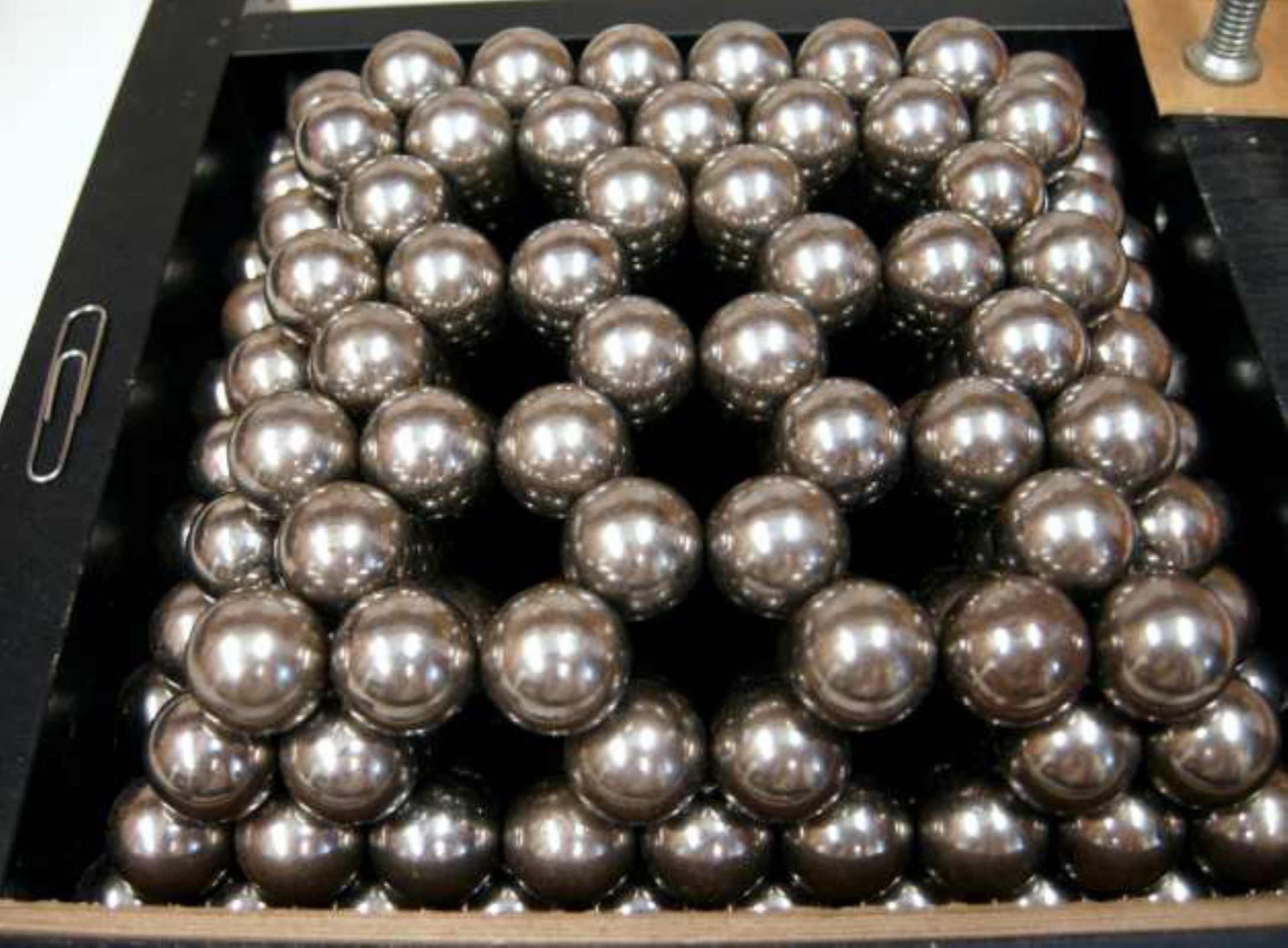}
\hspace{0.1in}\includegraphics[height=1.1in,keepaspectratio,clip=]{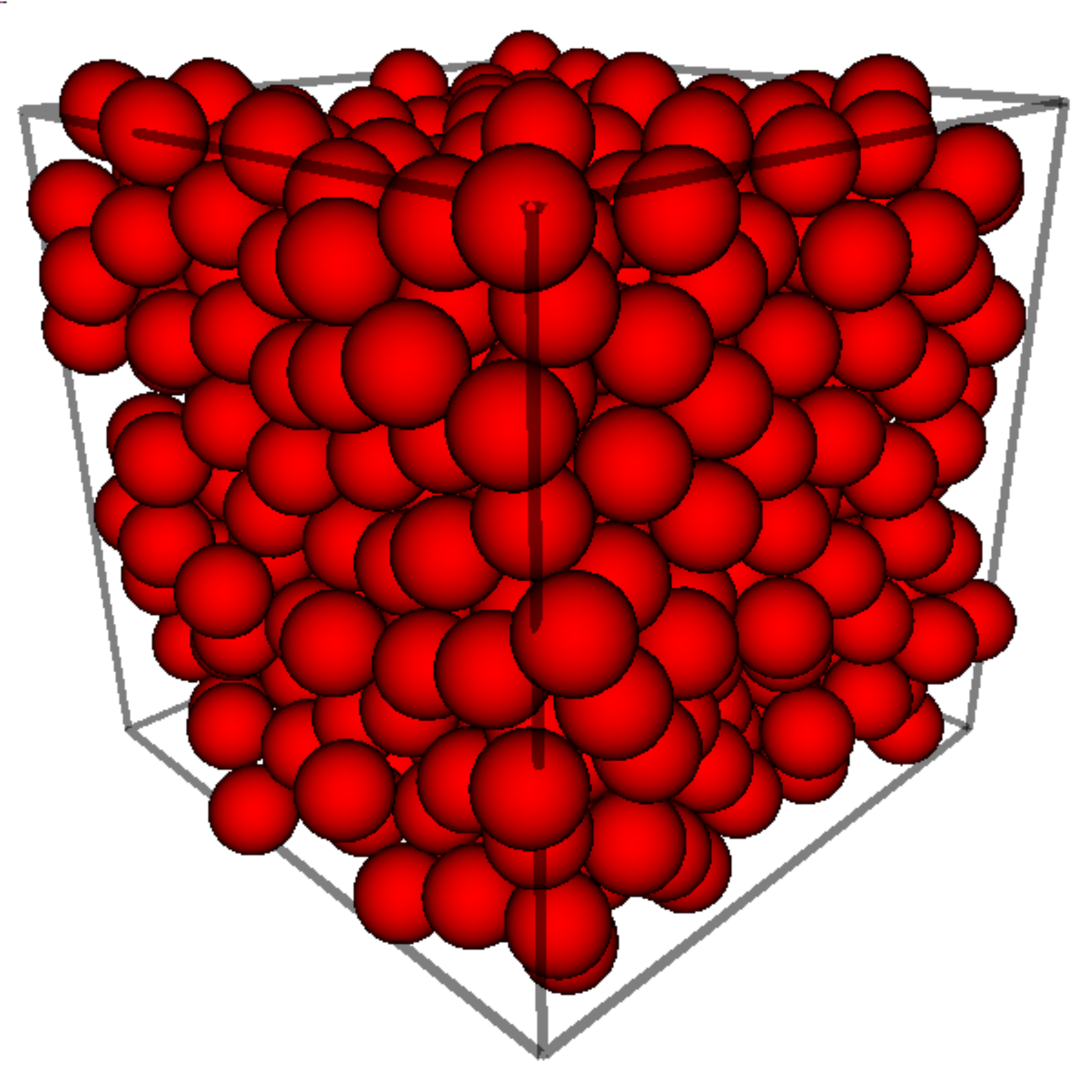}
\hspace{0.1in}\includegraphics[height=1.in,keepaspectratio,clip=]{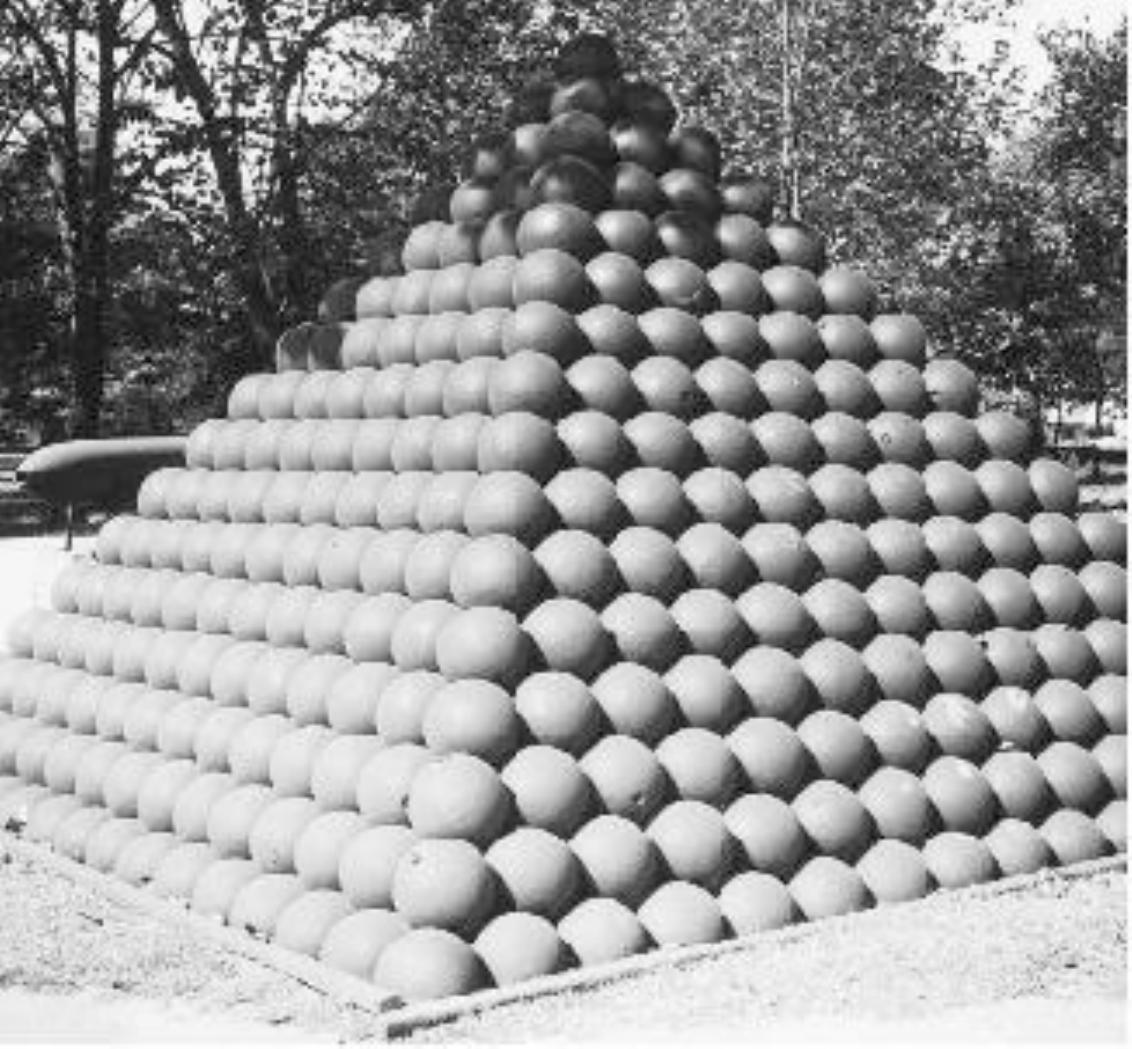}}
\caption{Three different  extremal strictly jammed packings in $\mathbb{R}^3$ identified in  Fig. \ref{maps},
as taken from Ref. \onlinecite{To10c}.
Left panel: (A) A tunneled crystal with $Z=7$ . Middle panel: MRJ packing with $Z=6$ (isostatic). 
Right panel: (B) FCC packing with $Z=12$.}
\label{extremal}
\end{figure}

The simplest order map is the one that maps any hard-sphere configuration
to a point in the $\phi$-$\psi$ plane. 
This two-parameter description is but a very small subset
of the relevant parameters that are necessary to fully characterize a 
configuration, but it  nonetheless enables one to draw important
conclusions.  Figure \ref{maps} 
shows such an order map that delineates the set of 
strictly jammed packings \cite{To00b,Ka02d,Do04a,To07} from non-jammed packings
in three dimensions.
The boundaries of the jammed region delineate extremal structures (see Fig. \ref{extremal}).
The densest sphere packings,~\cite{Ha05} which
lie along the locus $B$-$B^\prime$ with $\phi_{\mathrm{max}}=\pi/\sqrt{18}\approx 
0.74$, are strictly jammed. \cite{To01b,Do04a}
Point B corresponds to the face-centered-cubic (fcc) packing, i.e., it is 
the {\it most ordered} and {\it symmetric} densest packing, implying that their
shear moduli are infinitely large.\cite{To03c} The most disordered subset of the stacking variants
of the fcc packing is denoted by point $B^{\prime}$. 
In two dimensions, the strictly jammed triangular lattice is the unique
densest packing \cite{Fe64} and so the line $B$-$B^\prime$ in $\mathbb{R}^3$ collapses to a single point $B$
in $\mathbb{R}^2$.

The least dense strictly jammed packings are conjectured to be the  ``tunneled crystals'' in $\mathbb{R}^3$ with
$\phi_{\mathrm{min}}=2\phi_{\mathrm{max}}/3 = 0.49365\ldots $, corresponding to the  locus of points $A$-$A^{\prime}$.\cite{To07}  These infinitely-degenerate
sparse structures were analytically determined by appropriate stackings of planar ``honeycomb"
layers of spheres. These constitute a set of zero measure among the possible packings
with the same density and thus are virtually impossible to discover using
packing algorithms, illustrating the importance of analyzing individual packings, regardless of their frequency of occurrence in 
the space of jammed configurations, as advocated by the geometric-structure approach. The tunneled crystals are subsets of the densest packings
but are permeated with infinitely long chains of particle vacancies. \cite{To07}    Every sphere in a tunneled crystal 
contacts 7 immediate neighbors. Interestingly, the tunneled crystals exist at the edge of mechanical stability, since removal of any one sphere from the interior would cause the entire packing to collapse.  The tunneled crystals are magnetically frustrated
structures and Burnell and Sondhi\cite{Bu08} showed that their
underlying topologies greatly simplify the determination
of their antiferromagnetic properties.
In $\mathbb{R}^2$, the ``reinforced" kagom{\' e} packing with
exactly four contacts per particle  (isostatic value)
is evidently the lowest density strictly jammed packing \cite{Do04a}
with $\phi_{\mathrm{min}}=\sqrt{3}\pi/8=0.68017\ldots$ and so the line $A$-$A^\prime$ in $\mathbb{R}^3$ collapses to a 
single point $A$ in $\mathbb{R}^2$.

\subsection{Maximally Random Jammed States}
\label{mrj}

Among all strictly jammed sphere packings  in  $\mathbb{R}^d$,
the one that exhibits maximal disorder (minimizes some given order metric $\psi$)
is of special interest. This  is called the maximally random jammed (MRJ) state \cite{To00b}; see Fig.~\ref{maps}.
The MRJ state is automatically compromised
by passing either to the maximal packing fraction (fcc and its stacking variants) or the minimal
possible density for strict jamming (tunneled crystals), thereby
causing any reasonable order metric to rise on either side. A variety of
sensible order metrics produce a 3D MRJ state with a packing
fraction $\phi_{\mbox{\scriptsize MRJ}} \approx 0.64$  (see Ref. \onlinecite{Ka02d}), 
close to the traditionally  advocated density of the RCP state, and with
an isostatic mean contact number $\overline Z=6$ (see Ref. \onlinecite{Do05c}). This consistency among
the different order metrics speaks to their utility, even if a perfect order metric has not yet been identified.
However, the density of the MRJ state is not sufficient to fully
specify it.  It is possible to have a rather
ordered strictly jammed packing at this very same density, \cite{Ka02d} as indicated in 
Fig. \ref{maps}. The  MRJ state refers to a single packing
that is maximally disordered subject to the strict jamming constraint, regardless of its probability of occurrence
in some packing protocol.  Thus, the MRJ state is conceptually and quantitatively
different from random close packed (RCP) packings \cite{Be60,Ber65}, which, more recently, have  been suggested to be the most probable jammed configurations within an ensemble.\cite{Oh03} The differences between these states are even starker
in two dimensions, e.g.,   MRJ packings of identical circular disks in $\mathbb{R}^2$ have been shown to be dramatically
different from their RCP counterparts, including their respective densities,
average contact numbers, and degree of order, \cite{At14} as detailed below.

\begin{figure}
\begin{center}
{\includegraphics[width=2.8in,keepaspectratio,clip=]{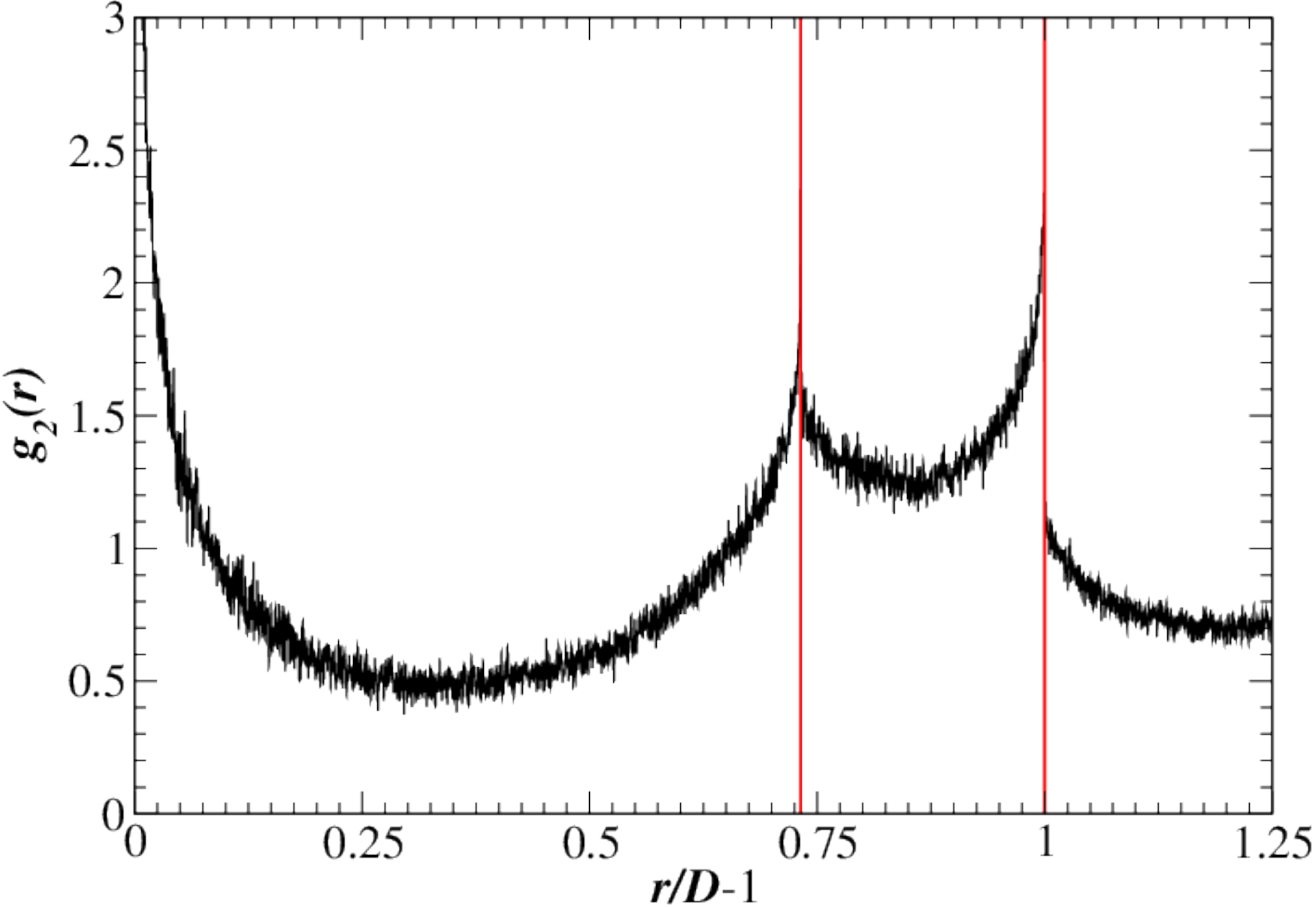}\hspace{0.25in}
\includegraphics[width=2.7in,keepaspectratio,clip=]{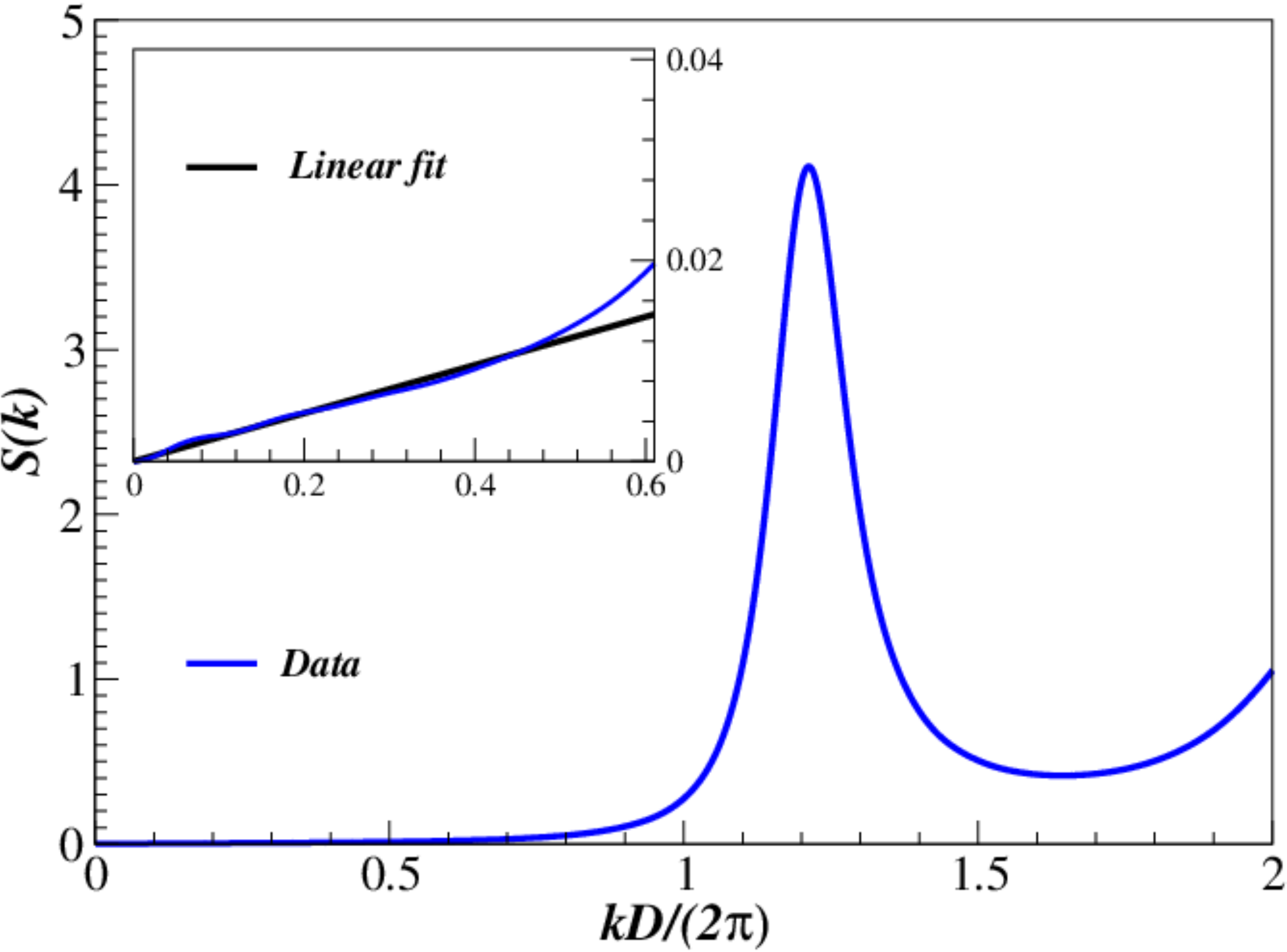}}
\end{center}
\caption{\footnotesize Pair statistics for packings of spheres of diameter $D$ in the immediate neighborhood of the 
3D MRJ state with $\phi \approx 0.64$. Top panel: The pair correlation function  $g_{2}(r)$ versus $r/D-1$. \cite{Do05c}
 The split second peak, the discontinuity at twice the sphere diameter, and the divergence near contact
are clearly visible. Bottom panel: The corresponding structure factor $S(k)$ as a function
of the dimensionless wavenumber $kD/(2\pi)$ for a million-particle packing.\cite{Do05d} 
The inset shows that $S(k)$ tends to zero (within numerical error) linearly in $k$
hence an MRJ packing is effectively hyperuniform. }
\label{g2-S}
\end{figure}

The MRJ state under the strict-jamming constraint is a prototypical glass
\cite{To07} in that it is maximally disordered (according to a variety of order metrics) without any  long-range order (Bragg peaks)
and perfectly rigid (i.e., the elastic moduli are  unbounded. \cite{To03c,To10c})
Its pair correlation function  can be decomposed into a Dirac-delta-function contribution
from particle contacts and a continuous-function contribution $g_2^{cont}(r)$:
\begin{equation}
g_2(r) = \frac{Z}{\rho s_1(D)} \delta(r-D) +  g_2^{cont}(r),
\end{equation}
where $s_1(R)$ is given by (\ref{surface}),
$\delta(r)$ is a radial Dirac delta function and $Z=2d$.
The corresponding structure factor in the long-wavenumber limit is 
\begin{equation}
S(k) \sim 1 + \frac{2 Z}{s_1(1)} \left(\frac{2\pi}{k D}\right)^{\frac{d-1}{2}} \cos[kD -(d-1)\pi/4] \quad (k \to \infty).
\end{equation}
For $d=3$, $g_2^{cont}(r)$ possesses the well-known feature of a split
second peak,\cite{Za83} with a prominent discontinuity at twice the sphere diameter, 
as shown in  Fig. \ref{g2-S}.  Interestingly, an integrable power-law divergence 
($1/r^{\alpha}$ with $\alpha \approx 0.4$) exists for near contacts. \cite{Do05c,Do05d} 
Moreover, an  MRJ packing in  $\mathbb{R}^d$ has a structure factor $S({\bf k})$ that  
tends to zero linearly in $|\bf k|$ (within numerical error)
in the limit $|{\bf k}| \rightarrow 0$ and hence is hyperuniform (see Fig. \ref{g2-S}),
belonging to the same class as Fermi-sphere point processes \cite{To08b} and
superfluid helium in its ground state. \cite{Fe56,Re67} This {\it nonanalytic} behavior
at $|\bf k|=0$ implies that MRJ packings
are characterized by  negative ``quasi-long-range" (QLR) pair correlations
in which the total correlation function $h({\bf r})$ decays to zero asymptotically
like $-1/|{\bf r}|^{d+1}$; see Refs. \onlinecite{Do05d,Sk06,Ho12b,At16a,At16b,To18a}. 
The QLR hyperuniform behavior distinguishes the
MRJ state strongly from that of the equilibrium hard-sphere
fluid \cite{Ha13}, which possesses a structure factor that is analytic at 
${\bf k = 0}$ [cf. (\ref{low-k})], and thus its $h(r)$ decays to zero
exponentially fast  for large $r$; see  Ref. \onlinecite{To10c}.  Interestingly,
the hyperuniformity concept enables one to identify a static nonequilibrium growing length scale
in overcompressed (rapidly compressed) hard-sphere systems as a function of $\phi$ up to the jammed glassy state.\cite{Ho12b,At16b}
This led to the identification of static nonequilibrium  growing length scales in supercooled liquids
on approaching the glass transition.\cite{Ma13a}

The following is a general conjecture due to Torquato and Stillinger \cite{To03a} 
concerning the conditions under which certain  strictly jammed packings are hyperuniform:
\smallskip

\noindent{\bf Conjecture}: {\sl Any strictly jammed saturated infinite packing of identical spheres in $\mathbb{R}^d$ is hyperuniform.}\\

\noindent{To date, there is no rigorously known counterexample to this conjecture. Its justification
and rigorously known examples are discussed in Refs.  \onlinecite{To18a} and \onlinecite{At16a}.}

A recent numerical study of necessarily finite disordered packings calls into question the connection between jamming and perfect hyperuniformity.\cite{Ik15} It is problematic to try to test
this conjecture using current packing protocols to construct  supposedly disordered jammed
 states for a variety of reasons. First, we stress that the conjecture eliminates packings that may have a rigid backbone but possess ``rattlers'' - -- a subtle point that has not been fully appreciated in numerical investigations. \cite{Ik15,Wu15,Ik17}
Current numerically generated disordered packings that are putatively jammed tend to contain a small concentration of rattlers;\cite{Lub91,Oh03,Do05d,To10e,At13} because of
these movable particles, the entire packing cannot be deemed to be ``jammed." 
Moreover, it  has been recently shown that various standard packing protocols struggle to reliably create packings that are jammed for even modest system sizes. \cite{At16a} Although these packings {\it appear} to be jammed by conventional tests,
 rigorous linear-programming jamming tests \cite{Do04a,Do04c} reveal that they are not.
Recent evidence has emerged that suggest that deviations from hyperuniformity in putative MRJ packings also can in part be explained by a shortcoming of the numerical protocols to generate exactly-jammed configurations as a result of a type of  ``critical slowing down'' \cite{At16a} as the packing's collective rearrangements in configuration space become locally confined by high-dimensional ``bottlenecks'' through which escape is a rare event.
Thus, a critical slowing down implies that it becomes increasingly difficult numerically to drive the value of $S(0)$ exactly down to its minimum
value of zero if a true jammed critical state could  be attained; typically,\cite{Do05d} $S(0) \approx 10^{-4}$.
 Moreover, the inevitable presence of even a small fraction of
rattlers generated by current packing algorithms destroys perfect hyperuniformity.
In summary, the difficulty of ensuring jamming in disordered packings
as the particle number becomes sufficiently large to access the small-wavenumber regime in the structure factor as well as the presence of rattlers that degrade hyperuniformity makes it extremely challenging
impossible to test the Torquato-Stillinger jamming-hyperuniformity conjecture on disordered jammed
packings  via current
numerical protocols. This raises the possibility that the ideal
MRJ packings have no rattlers, and provides a challenge to develop packing algorithms that
produce large disordered strictly jammed packings that are rattler-free.

A variety of different attributes of MRJ packings generated via the TJ packing algorithm
have been investigated in separate studies.
In the first paper of a three-part series, Klatt and Torquato \cite{Kla14} ascertained  Minkowski correlation functions associated with the Voronoi cells
of such MRJ packings and found that they exhibited even stronger anti-correlations than
those shown in the standard pair-correlation function. \cite{Kla14}
In the second paper of this series,\cite{Kl16} a variety of different correlation functions that arise in rigorous expressions for the effective physical properties of MRJ sphere packings were computed. 
In the third paper of this series,\cite{Kl18}  these structural descriptors
were used  to estimate effective transport and electromagnetic properties
of composites composed of MRJ sphere packings dispersed throughout
a matrix phase and showed, among other things, that  electromagnetic waves in the long-wavelength limit 
can propagate without dissipation, as generally predicted in Ref. \onlinecite{Re08a}.
In a separate study, Ziff and
Torquato~\cite{Zi17} determined the site and bond percolation thresholds
of TJ generated MRJ packings to be $p_c = 0.3116(3)$ and $p_c = 0.2424(3)$, respectively.

Not surprisingly, ensemble methods that produce ``most probable"
configurations typically miss interesting extremal points in the order map, such
as the locus of points $A$-$A^{\prime}$ and the rest of the
jamming-region boundary. However, numerical protocols
can be devised to yield unusual extremal jammed states.
For example, disordered jammed packings can be created in the entire 
non-trivial range of packing fraction $0.6 < \phi < 0.74048\ldots$ \cite{To00b,Ka02d,Ji11a}
Thus, in Fig. \ref{maps},  the locus of points along the boundary of the jammed set to the left
and right of the MRJ state are achievable. 
The  TJ algorithm \cite{To10e} 
was applied  to yield disordered strictly jammed  packings\cite{Ji11a} with $\phi$ as low as $0.60$, which
are overconstrained with $Z \approx 6.4$, and hence are more ordered than the MRJ state.

It is well known that lack of ``frustration" \cite{Ju97,To02a} in 2D
analogs of 3D computational and experimental
protocols that lead to putative RCP states result in packings of identical disks
that are highly crystalline, forming rather large triangular
coordination domains (grains). Such a 1000-particle  packing with $\phi \approx 0.88$ is depicted 
 in the right panel of Fig. \ref{MRJ}, and is only collectively
jammed at this high density.  Because such highly ordered
packings are the most probable outcomes for these typical protocols, 
``entropic measures" of disorder would identify these as the most disordered,
a misleading conclusion.  Recently, Atkinson {\it et al.}\cite{At14} applied the TJ algorithm  to generate relatively large
MRJ disk packings  with exactly isostatic ($\overline Z=4$)
jammed backbones and a packing fraction (including rattlers) of
$\phi_{\mbox{\scriptsize MRJ}} \approx 0.827$. Uncovering such disordered jammed packings of identical hard disks
challenges the traditional notion that the most probable distribution
is necessarily correlated with randomness and hence the RCP state.
An analytical formula was derived for  MRJ packing fractions of more
general 2D packings,\cite{Ti15} yielding the prediction $\phi_{\mbox{\scriptsize MRJ}} =0.834$
in the monodisperse-disk limit,
which is in very good agreement with the aforementioned recent numerical estimate.\cite{At14}

\subsection{Packings of Spheres With a Size Distribution}
\label{poly}

Sphere packings with a size distribution (polydispersity), sometimes called hard-sphere mixtures,   exhibit
intriguing structural features, some of which are only beginning to be understood.  Our knowledge of sphere packings with a size distribution
is very limited due in part to the infinite-dimensional  parameter space, i.e., 
all particle size ratios and relative concentrations.
It is known, for example, that a relatively small degree of polydispersity
can  suppress the disorder-order phase
transition seen in monodisperse hard-sphere systems (see Fig. \ref{hsphere}).\cite{Hen96}  Size polydispersity 
constitutes a fundamental feature of the microstructure
of a wide class of dispersions of technological importance,
including those involved in composite solid propellant combustion,
\cite{Ke87} flow in packed beds, \cite{Sc74} sintering of powders,\cite{Ra95} colloids, \cite{Ru89} transport and mechanical
properties of particulate composite materials. \cite{To02a} Packings of biological cells
in tissue are better modeled by assuming that the spheres have a size
distribution.\cite{Ji14,Ch16}

Generally, we allow the spheres to possess a continuous distribution in radius $R$, which is
characterized by a probability density function $f(R)$.
The average of any function $w(R)$ is defined by
$\langle {w(R)} \rangle = \int_{0}^{\infty} w(R) f(R) dR$.
The overall packing fraction $\phi$ is then defined as
\begin{equation}
\phi=\rho \langle v_1(R) \rangle,
\label{eta-poly}
\end{equation}
where $\rho$ is the total number density, $v_1(R)$ is given by (\ref{v(R)}). Two continuous probability densities
that have been widely used  are the Schulz \cite{Sc39} and log-normal \cite{Cr54}
 distributions. One can obtain corresponding results for  spheres
with $M$ discrete different sizes from Eq. (\ref{eta-poly})  by letting
\begin{equation}
f(R) = \sum_{i=1}^{M} \frac{\rho_{i}}{\rho} \delta (R -
R_{i}), 
\label{f-discrete}
\end{equation}
where $\rho_{i}$ and $R_{i}$ are number density and radius
of type-$i$ particles, respectively, and $\rho$ is the {\it total number
density}. Therefore, the overall volume fraction using (\ref{eta-poly})
is given by  $\phi= \sum_{i=1}^M \phi^{(i)}$, 
where $\phi^{(i)}=\rho_i v_1(R_i)$
is the packing fraction of the $i$th component.

\subsubsection{Equilibrium and Metastable Behavior}

The problem of determining the equilibrium phase behavior
of hard sphere mixtures is substantially more challenging and richer than that for identical hard spheres, including the possibilities of metastable or stable
fluid-fluid and/or solid-solid phase transitions (apart from stable fluid or crystal
phases). While much research remains to be done, considerable progress has been made
over the years,\cite{Le64,Wi70,Le71,Man71,Sa82,Ni84,Ros89,Di95,Sh96,Lo96,Sa99,Di99,Go03,Ro10,Di14} which we only briefly touch upon here
for both additive and nonadditive cases. In {\it additive} hard-sphere mixtures,  the distance of
closest approach  between the centers of any two spheres
is the arithmetic mean of their diameters. By contrast, in {\it nonadditive} hard-sphere  mixtures, the distance of closest
approach between any two spheres is generally no longer
the arithmetic mean. 

Additive mixtures have been studied both theoretically and computationally.
Lebowitz exactly obtained the pair correlation functions of such systems
with $M$ components within the Percus-Yevick approximation.~\cite{Le64}
Accurate approximate equations of state under liquid conditions have been found for both
discrete~\cite{Man71,Sa99} and continuous~\cite{Sa82,Sa17} size distributions with additive
hard cores. Fundamental-measure theory 
can provide useful insights about the phase behavior of  hard-sphere mixtures.\cite{Ros89,Ro10}
This theory predicts the existence of stable fluid-fluid coexistence
for sufficiently large size ratios in additive binary
hard-sphere systems, while numerical simulations \cite{Di99} indicate 
that such phase separation is metastable with respect to fluid-crystal
coexistence and also shows stable solid-solid coexistence. Nonetheless, it  remains an open 
theoretical question whether a fluid-fluid demixing transition 
exists in a binary mixture of additive hard spheres for any
size ratio and relative composition. In order to quantify 
fluid-crystal phase coexistence, one must know the densest
crystal structure, which is highly nontrivial, especially
for large size ratios, although recent
progress has been made; see Sec. \ref{max-mix}.

The Widom-Rowlinson model is an extreme case of nonadditivity
in which like species do not interact and
unlike species interact via a hard-core repulsion.\cite{Wi70} As the density is increased,
this model exhibits a fluid-fluid demixing transition in low dimensions and possesses
a critical point that is in the Ising universality class.\cite{Di95,Sh96}
More general nonadditive hard-sphere mixtures in which all
spheres interact have been studied.  Mixtures of  hard spheres
with positive nonadditivity (unlike-particle 
diameters greater than the arithmetic mean of the corresponding 
like-particle diameters) can exhibit fluid-fluid demixing transition \cite{Lo96,Di99} that belongs to the Ising universality class,\cite{Go03}
while those with negative nonadditivity have tendencies to form alloyed (mixed) fluid phases.\cite{Ni84}
For certain binary mixtures of hard sphere fluids with nonadditive diameters,
the pair correlation function has been determined in the Percus-Yevick
approximation.\cite{Le71,Ni84}

Fluid phases of hard sphere mixtures, like their monodisperse counterparts,
are not hyperuniform. However, multicomponent equilibrium plasmas consisting of 
nonadditive hard spheres with Coulombic interactions
enable one to generate a very wide class of disordered hyperuniform as well as ``multihyperuniform" \cite{Ji14}
systems  at positive temperatures. \cite{Lo17,Lo18a}

\subsubsection{Maximally Random Jammed States}
\label{max-mix}

The study of dense disordered packings of 3D polydisperse additive spheres
has received considerable attention. \cite{Vi72,Cl87,Ya96,Ric98,San02,De05}
However, these investigations did not consider their mechanical
stability via our modern understanding of jamming.
Not surprisingly, the determination of the MRJ  state
for an arbitrary polydisperse sphere packing is a challenging
open question, but some progress has been made recently, as described below.

Jammed states of polydisperse spheres, whether disordered or ordered, will generally
have higher packing fractions when ``alloyed" than their monodisperse counterparts.
The first investigation that attempted to  generate 3D
amorphous jammed sphere packings  with a polydispersity in size 
was carried out by Kansal {\it et al.}  \cite{Ka02c} using the LS packing algorithm.
It was applied to show that disordered binary jammed packings with a small-to-large size ratio $\alpha$
and relative concentration $x$ can be obtained whose packing fractions exceeds 0.64
and indeed can attain $\phi=0.79$ for $\alpha=0.1$ (the smallest
value considered). Chaudhuri {\it et al.}  \cite{Cha10}
 numerically generated amorphous 50-50 binary packings
with packing fractions in the range $0.648 \le \phi \le 0.662$
for $\alpha=1.4$. It is notable Clusel {\it et al.} \cite{Cl09} carried out a series of
experiments to visualize and characterize 3D random packings of frictionless polydisperse emulsion
droplets  using confocal microscopy.

Until recently, packing protocols that have attempted
to produce disordered binary sphere packings have been limited in producing mechanically
stable, isostatic packings across a broad spectrum of packing fractions.
Many previous simulation studies of
disordered binary sphere packings have produced packings
that are not mechanically stable \cite{Cl87,Sa02,De05} and report coordination numbers
as opposed to contact numbers. Whereas a coordination
number indicates only proximity of two spheres,  a
contact number reflects mechanical stability and is derived
from a jammed network.\cite{Do04c,To10c} 

Using an the TJ
linear programming packing algorithm,\cite{To10e} Hopkins {\it et al.} \cite{Ho13}
were recently able to generate high-fidelity strictly jammed, isostatic,
disordered binary packings with an anomalously large range of  
packing fractions, $0.634 \le \phi \le  0.829$, with the size ratio 
restriction $\alpha \ge 0.1$. These packings are MRJ like due to the nature of the TJ algorithm  and
the use of RSA initial conditions. Additionally, the 
packing fractions for certain values of $\alpha$ and $x$  approach
those of the corresponding densest known ordered packings.\cite{Ho11b,Ho12a} These findings suggest that
these high-density disordered packings should be good glass formers  for entropic reasons and hence may be easy to prepare
experimentally. The identification and explicit
construction of binary packings with such high packing fractions could have important practical implications 
for the design of improved solid propellants, concrete, and ceramics.
In this connection, a recent study of MRJ binary sphere packings under {\it confinement}
is particularly relevant.\cite{Ch15}

The LS algorithm has been used successfully to generate
disordered strictly jammed packings of binary disks\cite{Do06b} with
$\phi \approx 0.84$ and $\alpha^{-1}=1.4$.  
By  explicitly constructing an exponential number of jammed packings of binary
disks with densities spanning the spectrum from the
accepted amorphous glassy state to the phase-separated crystal,
it has been argued \cite{Do06b,Do07c} that there is no
``ideal glass transition". \cite{Pa05} The existence
of an ideal glass transition remains a hotly debated
topic of research.

Simulational \cite{Za11a,Za11c,Be11} as well as experimental \cite{Dr15,Ri17} studies of 
disordered jammed spheres with a size distribution reveal
that they are effectively hyperuniform. In such cases, it has been demonstrated
that the proper means of investigating hyperuniformity 
is through the spectral density  $\tilde{\chi}_{_V}(\mathbf{k})$,\cite{Za11a,Za11c} which
is defined by condition (\ref{hyper-2}).

\subsubsection{Densest Packings}

The densest packings of spheres with a size distribution is of great
interest in crystallography, chemistry and materials science. It is notable that
the densest packings of hard-sphere mixtures are intimately related
to high-pressure phases of molecular systems, including intermetallic
compounds\cite{Dem06} and solid rare-gas compounds\cite{Ca09}
for a range of temperatures.

Except for trivial space-filling structures, there are no provably densest
packings when the spheres possess a size distribution, which is a testament
to the mathematical challenges that they present.
We begin by  noting some rigorous bounds on the maximal packing fraction
of packings of spheres with $M$ different radii $R_1,R_2, \ldots, R_M$ in $\mathbb{R}^d$.
Specifically, the overall maximal packing fraction $\phi^{(M)}_{\scriptsize \mbox{max}}$ of such
a general mixture in $\mathbb{R}^d$ [where $\phi$ is defined by
(\ref{eta-poly}) with (\ref{f-discrete})] is bounded from above and below
in terms of the maximal packing fraction $\phi^{(1)}_{\scriptsize \mbox{max}}$ for a
monodisperse packing in the infinite-volume limit:\cite{To02a}
\begin{equation}
\phi^{(1)}_{\scriptsize \mbox{max}} \le \phi^{(M)}_{\scriptsize
\mbox{max}}
\le 1- (1-\phi^{(1)}_{\scriptsize \mbox{max}})^M.
\label{upper-eta}
\end{equation}
The lower bound corresponds to the case when the $M$ components completely
demix, each at the density $\phi^{(1)}_{\scriptsize \mbox{max}}$.
The upper bound corresponds to an ideal sequential packing process for arbitrary
$M$ in which one takes the limits $R_1/R_2 \rightarrow 0$, $R_2/R_3 \rightarrow 0,
\cdots, R_{M-1}/R_M \rightarrow 0$.\cite{To02a} Specific {\it nonsequential} protocols (algorithmic or otherwise)
that can generate structures that  approach the upper bound (\ref{upper-eta}) for arbitrary
values of $M$ are currently unknown and thus the development of such protocols
is an open area of research. We see that in the limit $M\rightarrow \infty$, the
upper bound approaches unity, corresponding to space-filling polydisperse spheres with
an infinitely wide separation in sizes \cite{He90} or
a continuous size distribution with sizes ranging to the infinitesimally small.\cite{To02a}

Even the characterization of the densest jammed {\it binary} sphere packings offers great challenges
and is very far from completion. Here we briefly review some work concerning
such packings  in two and three dimensions. Ideally, it is desired to obtain 
$\phi_{\mbox{\scriptsize max}}$ as a function of $\alpha$ and $x$.
In practice, we have a sketchy understanding of the surface
defined by $\phi_{\mbox{\scriptsize max}}(\alpha,x)$.

Among the 2D and 3D cases, we know most about the determination of the maximally dense
binary packings in $\mathbb{R}^2$. 
Fejes T{\'o}th~\cite{Fe64} reported a number of candidate maximally dense 
packings
for certain values of the radii ratio in the range $\alpha \geq 0.154701 \ldots$.
Maximally dense binary disk packings have been also studied to determine the stable
crystal phase diagram of such alloys.\cite{Li93} 
The determination of $\phi_{\mbox{\scriptsize max}}$ for sufficiently small $\alpha$ amounts to finding the
optimal arrangement of the small disks within a {\it tricusp}: the nonconvex cavity between three
close-packed large disks. A particle-growth Monte Carlo algorithm was used to generate
the densest arrangements of a fixed number of small identical  disks within such a tricusp.~\cite{Uc04a}
All of these results can be compared to a relatively tight
upper bound on $\phi_{\mbox{\scriptsize max}}$ given by~\cite{Fl60}
\begin{equation}
\phi_{\mbox{\scriptsize max}} \le   \phi_U = \frac{\displaystyle \pi \alpha^2 + 2(1-\alpha^2)\sin^{-1}{\left(\frac{\alpha}{1+\alpha}\right)}}
{\displaystyle 2\alpha (1+2\alpha)^{1/2}}.
\label{supremum}
\end{equation}
 Inequality (\ref{supremum}) also applies to general
multicomponent packings, where  $\alpha$ is taken to be the ratio of
the smallest disk radius to the
largest one.

There is great interest in finding the  densest binary sphere packings in $\mathbb{R}^3$ in the physical sciences
because of their relationship to binary crystal alloys found in molecular systems; see
Refs. \onlinecite{Ho12a} and \onlinecite{Hu08} as well as  references therein
for details and some history.
Past efforts to identify such optimal packings
have employed simple crystallographic techniques (filling the interstices in uniform
3D tilings of space with spheres of different sizes) and  algorithmic methods, e.g., Monte Carlo calculations
and a genetic algorithm.\cite{Ku08,Fi09}  However, these methods
have achieved only limited success, in part due to the
infinite parameter space that is involved. When using traditional
algorithms, difficulties result from the enormous
number of steps required to escape from local minima in the
``energy" (negative of the packing fraction).
Hopkins {\it et al.} \cite{Ho11b,Ho12a} have presented the most comprehensive determination
to date of the ``phase diagram" for the densest binary sphere packings 
via the TJ
linear-programming algorithm.\cite{To10e} In Ref. \onlinecite{Ho12a}, 
19 distinct crystal alloys (compositions of large and small spheres
spatially mixed  within a fundamental cell) were identified, including 8 that were unknown at the time.
 Using the TJ algorithm, they were always able to obtain either the densest previously
known alloy or denser ones. These structures may correspond to currently unidentified stable
phases of certain binary atomic and molecular systems, particularly at high temperatures and pressures.\cite{Dem06,Ca09}
Reference \onlinecite{Ho12a} provides details about the structural
characteristics of these densest-known binary sphere packings.

\section{Packing Spheres in High Dimensions}
\label{high}

Sphere packings in four and higher-dimensional Euclidean spaces are of great interest 
in the physical and mathematical sciences; see Refs.
\onlinecite{Con95,Fr99,Pa00,El00,Pa06,Co02,Co03,To06a,To06b,To06c,Sk06,Ro07,Ad08,Sc08,Co09,Me09,Me09b,
Pa10,Lu10,To10c,Ma13c,Ka13,Ka14,An16}. Physicists have studied high-dimensional packings to gain insight
into liquid,  crystal and glassy states of matter  in lower dimensions.\cite{Fr99,Pa00,Pa06,Sk06,Ro07,Ad08,Me09,Me09b,Pa10}
Finding the densest packings
in arbitrary dimension in Euclidean and compact spaces is a problem of long-standing interest in discrete geometry.\cite{Co93,Co03,Vi17,Co17}
Remarkably, the optimal way of sending digital signals over noisy channels corresponds
to the densest sphere packing in a high-dimensional space.\cite{Sh48,Co93}
These ``error-correcting" codes underlie a variety of systems in digital
communications and storage, including compact disks, cell phones and the Internet.

\subsection{Equilibrium and Metastable Phase Behavior}

The properties of equilibrium and metastable states of hard spheres  have been studied both theoretically and computationally in spatial
dimensions greater than three. \cite{Lu82,Fr99,Sk06,Cl06,Me09,Me09b,Lu10}
This includes the evaluation of various virial coefficients across dimensions, \cite{Lu82,Cl06}
as well as  the pressure along the liquid, metastable  and crystal branches,\cite{Sk06,Ro07,Me09,Me09b,Lu10}
especially for $d=4,5,6$ and 7. Using the LS packing algorithm, Skoge {\it et al.} \cite{Sk06}
numerically estimated  the freezing and
melting packing fractions for the equilibrium hard-sphere fluid-solid
transition, $\phi_F \simeq 0.32$ and $\phi_M \simeq 0.39$,
respectively, for $d=4$, and $\phi_F \simeq 0.19$ and $\phi_M \simeq
0.24$, respectively, for $d=5$. These authors showed that nucleation
appears to be strongly suppressed with increasing dimension. The same conclusion was subsequently reached
in a study by van Meel {\it et al.}.\cite{Me09b}
Finken, Schmidt and L{\" o}wen \cite{Fi01} used a variety of approximate theoretical
methods to show that equilibrium  hard spheres have a first-order freezing transition for
dimensions as high as $d=50$.

Any disordered packing in which the pair correlation function at contact, $g_2(D^+)$,
is bounded (such as equilibrium hard spheres) induces a power-law
decay in the structure factor $S(k)$ in the limit $k \to \infty$ for any dimension $d$ 
given by
\begin{equation}
S(k) \sim 1- \frac{2^{\frac{3d+1}{2}} \Gamma(1+d/2)\phi g_2(D^+)}{\sqrt{\pi} (kD)^{\frac{d+1}{2}}} \cos[kD -(d+1)\pi/4]\;.
\end{equation}

There is a remarkable duality between the equilibrium
hard-hypersphere (hypercube) fluid system in $\mathbb{R}^d$ and the continuum percolation model of overlapping
hyperspheres (hypercubes) in $\mathbb{R}^d$. In particular, the pair connectedness function of the latter
is to a good approximation equal to the negative of the total correlation function of the former evaluated
at {\it negative} density.\cite{To12a} This mapping becomes exact for $d=1$ and
in the large-$d$ limit.

\subsection{Nonequilibrium Disordered Packings Via Sequential Addition}

In Sec. \ref{ghost}, we noted that the ``ghost" RSA packing~\cite{To06a} is a disordered but {\it unsaturated} packing construction 
whose $n$-particle correlation functions are known exactly and rigorously achieves
the infinite-time  packing fraction  $\phi=2^{-d}$ 
for any $d$; see Fig.~\ref{grsa} for a 2D realization of such a packing.

Saturated RSA sphere packings have been numerically generated and structurally characterized for
dimensions up through $d=6$ (Ref. \onlinecite{To06d}). A more efficient numerical
procedure was devised to produce such packings for dimensions  up through $d=8$ (Ref. \onlinecite{Zh13b}).
The current best estimates of the  maximal saturation packing fraction $\phi_s$
for $d=4$, $5$, $6$, $7$ and $8$ are $0.2600781 \pm 0.0000037$,  $0.1707761 \pm 0.0000046$,  
$0.109302 \pm 0.000019$,
$0.068404 \pm 0.000016$, and $0.04230 \pm 0.00021$,  respectively.\cite{Zh13b}  These are
lower than the corresponding MRJ packing fractions in those dimensions (see Sec. \ref{mrj-d}). 
The quantity $\phi_s$  apparently scales as $d \cdot 2^{-d}$ or possibly
$d \cdot \ln(d) \cdot 2^{-d}$ for large $d$; see Refs.  \onlinecite{To06d,Zh13b}.
While saturated RSA packings are nearly but not exactly hyperuniform,\cite{To06d}
as $d$ increases, the degree of hyperuniformity increases
and pair correlations markedly diminish,\cite{Zh13b} consistent with the 
decorrelation principle,\cite{To06b} which is described in Sec. \ref{high-d}.


\subsection{Maximally Random Jammed States}
\label{mrj-d}
  
Using the LS algorithm, Skoge {\it et al.} \cite{Sk06} generated and characterized  
MRJ packings in four, five and six dimensions. In particular, they estimated the MRJ packing fractions,
finding $\phi_{\mbox{\scriptsize MRJ}} \simeq 0.46$, $0.31$ and $0.20$ for $d=4$, $5$ and $6$, respectively.  To a good
approximation, the MRJ packing fraction obeys the scaling form $\phi_{\mbox{\scriptsize MRJ}}= c_12^{-d}+c_2 d \cdot2^{-d}$,
where $c_1=-2.72$ and $c_2=2.56$, which appears to be consistent
with high-dimensional asymptotic limit, albeit with different coefficients.
The dominant large-$d$ density scaling $d\cdot2^{-d}$ is supported by theoretical
studies.\cite{To02c,To06b,Ji10,Ka14}
Skoge {\it et al.} \cite{Sk06} also determined the MRJ  pair correlation function $g_{2}(r)$ and structure factor $S(k)$ for
these states and found that short-range ordering appreciably decreases with
increasing dimension, consistent with the 
{\it decorrelation principle}.\cite{To06b} This implies that, in the limit $d \to \infty$, $g_2(r)$ tends
to unity for all $r$ outside the hard-core, except
for a Dirac-delta function at contact due to the jamming constraint.\cite{To06b} As for $d=3$ (where $\phi_{\mbox{\scriptsize MRJ}}
\simeq 0.64$), the MRJ packings were found to be
isostatic, hyperuniform, and have a power-law divergence in $g_{2}(r)$ at contact; 
; $g_2(r) \sim 1/(r-D)^\alpha$ with $\alpha\approx 0.4$ as $r$ tends to $D^+$. Across dimensions, the
cumulative number of neighbors was shown to equal the {\it kissing} (contact) number of the
conjectured densest packing close to where $g_{2}(r)$ has its first
minimum.  Disordered jammed packings were also simulated and studied
in dimensions 7-10; see Ref. \onlinecite{Ch12}.

\subsection{Maximally Dense Sphere Packings}

The sphere packing problem seeks to answer the following
question:~\cite{Co93} Among all packings of congruent spheres in $\mathbb{R}^d$,
what is the maximal packing fraction $\phi_{\mbox{\scriptsize max}}$
and the corresponding arrangements of the spheres?
Until 2017, the  optimal solutions were known only for the first three space
dimensions.~\cite{Ha05} For $d=2$ and $d=3$, these are the triangular
lattice ($A_2$) with $\phi_{\mbox{\scriptsize max}}=\pi/\sqrt{12}=0.906899\ldots$
and checkerboard (fcc) lattice ($D_3$) and its stacking variants
with  $\phi_{\mbox{\scriptsize max}}=\pi/\sqrt{18}=0.74048\ldots$, respectively.
We now know the optimal solutions in two other space dimensions; namely, the $E_8$ and $\Lambda_{24}$ lattices are the densest packings
among all possible packings in $\mathbb{R}^8$ and $\mathbb{R}^{24}$, respectively;
see Refs. \onlinecite{Vi17} and \onlinecite{Co17}.
For $4 \le  d \le 9$, the densest known packings are (Bravais) lattice
packings.  \cite{Co93} The ``checkerboard" lattice $D_d$ is believed
to be optimal in $\mathbb{R}^4$ and $\mathbb{R}^5$. 
Interestingly, the non-lattice (periodic) packing $P_{10c}$
(with 40 spheres  per fundamental cell)
is the densest known packing in $\mathbb{R}^{10}$, which is the lowest
dimension in which the best known packing is not a (Bravais) lattice.

Table \ref{packings} lists the densest known or optimal sphere packings in  $\mathbb{R}^d$  for selected $d$.
For the first three space dimensions, the optimal sphere-packing solutions  (or their ``dual" solutions) are directly related to the best known solutions of the {\it number-variance} problem \cite{To03a,Za09}
as well as of two other well-known problems in discrete geometry: the {\it covering}
and {\it quantizer} problems, \cite{Co93,Sik08} but such relationships may or may not exist for $d\ge 4$, depending on the peculiarities
of the dimensions involved. \cite{To10d}

\begin{table}[bthp]
\caption{The densest known or optimal sphere packings in  $\mathbb{R}^d$  for selected $d$. 
For each packing, we provide the packing fraction $\phi$ and kissing number $Z$.
Except for the non-lattice packing $P_{10c}$ in $\mathbb{R}^{10}$, all
of the other densest known packings listed in this table are lattice packings:
$\mathbb{Z}$ is the integer lattice, $A_2$ is the triangular lattice,
$D_d$ is the checkerboard lattice (a generalization of the fcc
lattice), $E_d$ is one the root lattices, and $\Lambda_d$ is the laminated lattice.
For $d=8$ and $d=24$, it has recently been proved that
$E_8$ and $\Lambda_{24}$ are optimal among all packings, respectively; see Refs. \onlinecite{Vi17} and \onlinecite{Co17}.
The reader is referred to Ref. \onlinecite{Co93} for additional details. }
\label{packings}
\renewcommand{\baselinestretch}{1.2} \small 
\centering
\begin{tabular} {|c|c|c|c|}
\multicolumn{4}{c}{~} \\\hline
  ~$d$~ & ~Packing~& ~Packing Fraction, $\phi$~ & Kissing number, $Z$~\\ \hline
1 &  $\mathbb{Z}$ & $1$   &2\\

2 & $A_2$  & $\pi/\sqrt{12}=0.9068\ldots$ & 6\\

3 &  $D_3$  & $\pi/\sqrt{18}=0.7404\ldots$  &12 \\

4 &  $D_4$ &  $\pi^2/16= 0.6168\ldots$ & 24\\

5 &  $D_5$ & $2\pi^2/(30\sqrt{2})=0.4652\ldots$ &40 \\

6 &  $E_6$ &  $3\pi^2/(144\sqrt{3})=0.3729\ldots$ & 72\\
7 &  $E_7$ &  $\pi^3/105=0.2952\ldots$ & 126\\
8 &  $E_8$ &  $\pi^4/384=0.2536\ldots$ & 240 \\
9 &   $\Lambda_9$  &  $2\pi^4/(945\sqrt{2})=0.1457\ldots$ & 272\\
10&   $P_{10c}$ & $\pi^5/3072=  0.09961\ldots$ & 372\\
16&  $\Lambda_{16}$ & $\pi^8/645120= 0.01470\ldots$ & 4320\\
24&  $\Lambda_{24}$ & $\pi^{12}/479001600=0.001929\ldots$& 196360\\ \hline
\end{tabular}
\end{table}

The TJ linear-programming packing algorithm was adapted by Marcotte and Torquato \cite{Ma13c} to determine the densest lattice
packing (one particle per fundamental cell) in some high dimension. These authors applied it  for $2 \le d\le  19$
 and showed that it was able
to rapidly and reliably discover the densest known lattice
packings without a priori knowledge of their existence.
It was found to be appreciably faster than
previously known algorithms at that time.\cite{Gr11,An12}
The TJ algorithm was used to generate an ensemble of isostatic jammed hard-sphere lattices and study 
the associated pair statistic and force distributions.\cite{Ka13}
It was shown that  this special ensemble of lattice-sphere packings retain many of the crucial structural features of the classical hard-sphere model.

It is noteworthy that for sufficiently large $d$,
lattice packings are most likely not the densest
(see Fig.~\ref{holes}), but it becomes increasingly difficult to find explicit
dense packing constructions as $d$ increases.
Indeed, the problem of finding the shortest
lattice vector in a particular lattice packing  grows super-exponentially
with $d$ and is in the class of NP-hard problems.
\cite{At98}

\begin{figure}
\begin{center}
\includegraphics[height=1.2in, keepaspectratio]{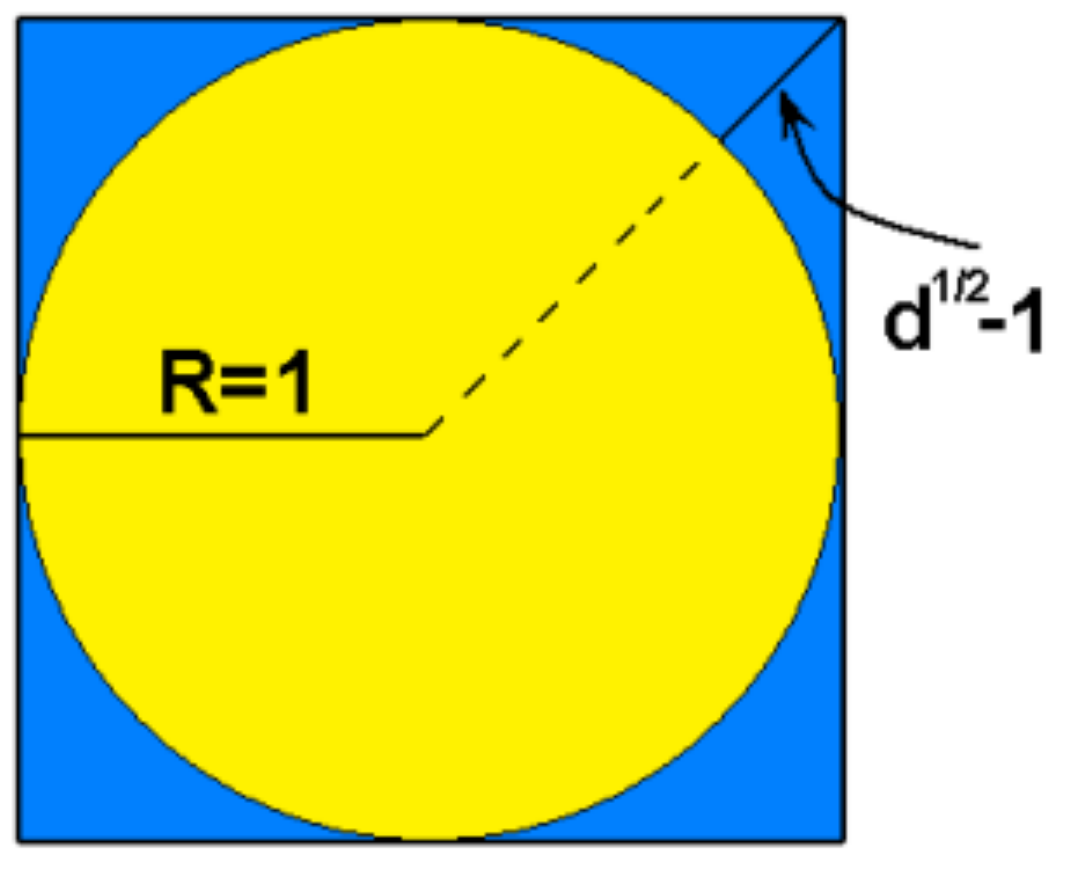}\hspace{0.25in}
\includegraphics[height=1.2in, keepaspectratio]{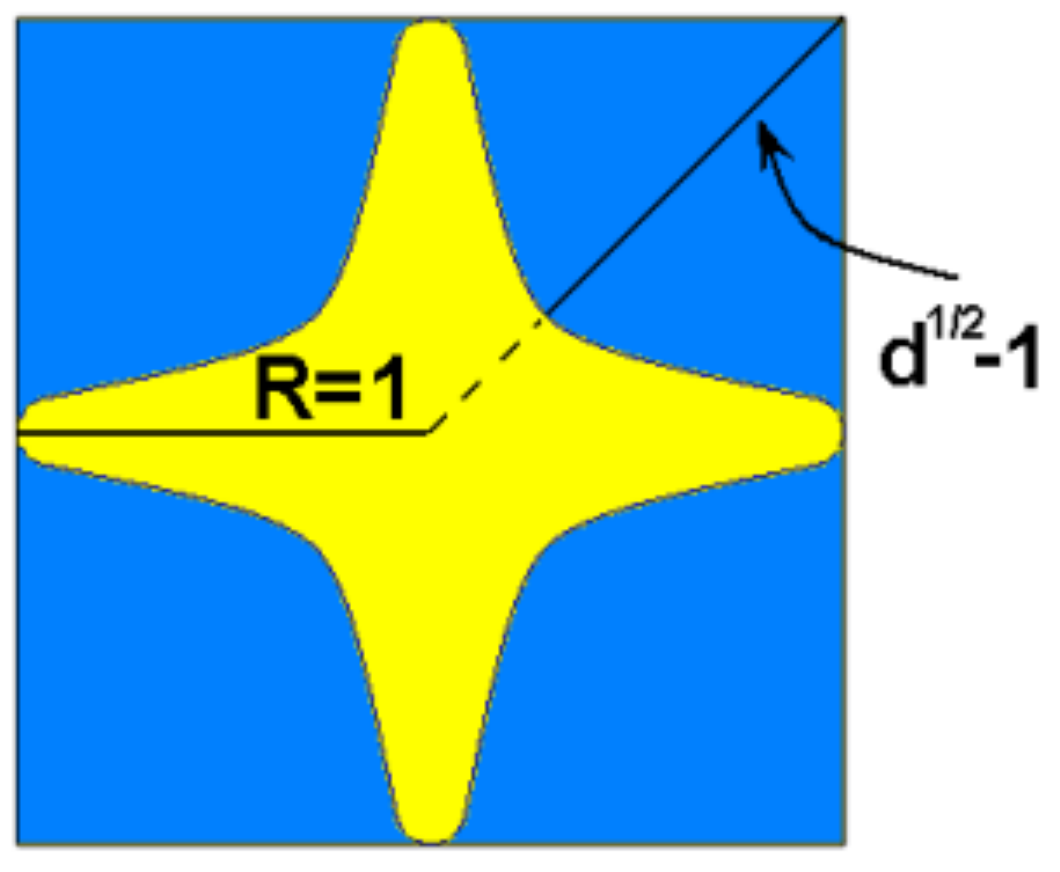}
\caption{(color online) Lattice packings in sufficiently
high dimensions are not dense  because
the ``holes" (space exterior to the spheres) eventually dominates 
the space $\mathbb{R}^d$ and hence become {\it unsaturated}.\cite{Co93}
For illustration purposes, we consider
the hypercubic lattice $\mathbb{Z}^d$. Left panel: A fundamental
cell of $\mathbb{Z}^d$ represented in two dimensions. The distance
between the point of intersection of the longest diagonal in the hypercube with the hypersphere
boundary and the vertex of the cube along this diagonal is given by
$\sqrt{d} -1$ for a sphere of unit radius. This means that $\mathbb{Z}^d$
already becomes unsaturated at $d=4$. Placing an additional sphere
in $\mathbb{Z}^4$ doubles the density of $\mathbb{Z}^4$ and yields
the four-dimensional checkerboard lattice packing $D_4$, which is believed
to be the optimal packing in $\mathbb{R}^4$. Right panel:
A schematic ``effective" distorted representation of the hypersphere within the hypercubic
fundamental cell for large $d$, illustrating that the volume content
of the hypersphere relative to the hypercube rapidly diminishes asymptotically. Indeed,
the packing fraction of $\mathbb{Z}^d$ is given by $\phi=\pi^{d/2}/(\Gamma(1+d/2) 2^d)$.
Indeed, the checkerboard lattice $D_d$ with packing fraction $\phi=\pi^{d/2}/(\Gamma(1+d/2) 2^{(d+2)/2})$
becomes suboptimal in relatively low dimensions because it also
becomes dominated by larger and larger holes as $d$ increases.}
\label{holes}
\end{center}
\end{figure}

For large $d$, the best that one can do theoretically
is to devise  upper and lower bounds on 
$\phi_{\mbox{\scriptsize max}}$. \cite{Co93}
The {\it nonconstructive} lower bound of Minkowski \cite{Mi05} established the
existence of reasonably dense lattice packings. He found
that the maximal packing fraction $\phi^L_{\mbox{\scriptsize max}}$ among all lattice packings 
for $d \ge 2$ satisfies
\begin{equation}
\phi^L_{\mbox{\scriptsize max}} \ge \frac{\zeta(d)}{2^{d-1}},
\label{mink}
\end{equation}
where $\zeta(d)=\sum_{k=1}^\infty k^{-d}$ is the Riemann zeta function.
Note that for large values of $d$,
the asymptotic behavior of the Minkowski lower bound is controlled by $2^{-d}$.
Since 1905, many extensions and generalizations of the lower bound (\ref{mink})
have been derived \cite{Da47,Ball92,Co93,Va09}, but none of these investigations have been able to improve
upon the dominant exponential term $2^{-d}$; they instead only improve
on the latter by a factor linear in $d$.

It is trivial to prove that the  packing fraction of a {\it saturated} packing of congruent spheres
in $\mathbb{R}^d$ satisfies~\cite{To06a}
\begin{equation}
\phi \ge \frac{1}{2^d}
\label{sat}
\end{equation}
for all $d$. This ``greedy" bound (\ref{sat})
 has the same dominant exponential term as Minkowski's bound (\ref{mink}).

Nontrivial upper bounds on $\phi_{\mbox{\scriptsize max}}$  in $\mathbb{R}^d$ 
for any $d$ have been derived.\cite{Bl29,Ro58,Ro64,Ka78,Co03}
The linear-programming (LP) upper bounds due to Cohn and Elkies 
\cite{Co03} provides the basic framework for proving the best known upper bounds on $\phi_{\mbox{\scriptsize max}}$
for  dimensions in the range $4 \le d  \le 36$. Recently, these LP bounds
have been used to prove that no packings in $\mathbb{R}^8$ and $\mathbb{R}^{24}$
have densities that can exceed those of the $E_8$ (Ref. \onlinecite{Vi17}) and $L_{24}$ 
(Ref. \onlinecite{Co17}) lattices, respectively.
Kabatiansky and Levenshtein \cite{Ka78} found the best asymptotic upper bound, which in the limit $d\rightarrow \infty$
yields $\phi_{\mbox{\scriptsize max}} \le 2^{-0.5990d}$,
proving that the maximal packing fraction tends to zero in the limit $d \rightarrow \infty$.
This rather counterintuitive high-dimensional property of sphere
packings  can be understood by recognizing that almost
all of the volume of a $d$-dimensional sphere for large $d$ is concentrated
near the sphere surface. 
 
\subsection{Are Disordered Packings the Densest in High Dimensions?} 
\label{high-d}

Since 1905, many extensions and generalizations of Minkowski's bound
have been derived \cite{Co93}, but none of them have  improved
upon the dominant exponential term $2^{-d}$. The existence of the  {\it unjammed} disordered 
ghost RSA packing~\cite{To06a} (Sec. \ref{ghost}) that rigorously achieves a packing fraction of $2^{-d}$
strongly suggests that Bravais-lattice packings  (which are almost surely unsaturated
in sufficiently high $d$) are far from optimal for large $d$. 

Torquato and Stillinger \cite{To06b} employed a plausible conjecture  that strongly
supports the counterintuitive possibility that the densest sphere packings
for sufficiently large $d$ may be disordered  or
at least possess fundamental cells whose size and structural
complexity increase with $d$. They did so using the so-called $g_2$-invariant optimization procedure 
that maximizes $\phi$
associated with a radial ``test" pair correlation function $g_2(r)$
to provide the putative exponential improvement on Minkowski's 100-year-old bound
on  $\phi_{\mbox{\scriptsize max}}$. Specifically, a {\it $g_2$-invariant process} \cite{To02c} is one in which 
the functional form of a ``test" pair correlation $g_2({\bf r})$ 
function remains invariant as density varies, for all ${\bf r}$, over the range of
packing fractions $0 \le \phi \le \phi_*$ subject to the  satisfaction of the nonnegativity
of the structure factor $S({\bf k})$ and $g_2({\bf r})$. 
When there exist sphere packings with a $g_2$
satisfying these conditions in the  interval $[0,\phi_*]$,
then one has the lower bound on the maximal packing fraction
given by
\begin{equation}
\phi_{\mbox{\scriptsize max}} \ge \phi_*\;.
\end{equation}
Torquato and Stillinger \cite{To06b} conjectured that a test
function $g_2({\bf r})$ is a pair-correlation function of a translationally 
invariant disordered sphere packing in $\mathbb{R}^d$ for
$0 \le \phi \le \phi_*$ for sufficiently large $d$ if and only if 
the nonnegativity conditions on $S({\bf k})$ and $g_2({\bf r})$.
The {\it decorrelation principle},\cite{To06b} among other results,\cite{Sc08,To10c,An16} provides
justification for the conjecture. This principle states that unconstrained
correlations in disordered sphere packings vanish asymptotically in high dimensions
and that the $g_n$ for any $n \ge 3$ can be inferred entirely (up to small
errors) from a knowledge of $\rho$ and $g_2$.  
This is vividly exhibited by the exactly solvable
ghost RSA packing process \cite{To06a} as well as by computer simulations
of high-dimensional MRJ \cite{Sk06} and  RSA \cite{To06c}  packings. Interestingly, this optimization problem is the {\it dual}
of the infinite-dimensional linear program (LP) devised by Cohn and Elkies \cite{Co02}
to obtain upper bounds on $\phi_{\mbox{\scriptsize max}}$.

Using a particular test pair correlation corresponding to a disordered sphere
packing, Torquato and Stillinger \cite{To06b} found a conjectural lower bound on $\phi_{\mbox{\scriptsize max}}$ 
that is controlled by $2^{-(0.77865\ldots) d}$, thus providing the first putative exponential
improvement on Minkowski's lower bound (\ref{mink}).
Scardicchio, Stillinger and Torquato \cite{Sc08} studied a wider class of test
functions (corresponding to disordered packings) that lead to precisely the same putative exponential improvement 
on Minkowski's lower bound and therefore the asymptotic form $2^{-(0.77865\ldots) d}$ 
is much more general and robust than previously
surmised.

Zachary and Torquato \cite{Za11e} studied, among other quantities, pair statistics
of  high-dimensional generalizations of the periodic 2D kagom{\' e} and 3D diamond crystal
structures.  They showed that the decorrelation principle is remarkably already exhibited in
these periodic crystals in low dimensions, suggesting that it applies for any sphere packing in high
dimensions, whether disordered or not. This conclusion was bolstered in a subsequent study by
Andreanov, Scardicchio and Torquato\cite{An16} who showed that strictly jammed lattice sphere packings
visibly decorrelate as $d$ increases in relatively low dimensions.


\subsection{Remarks on Packing Problems in Non-Euclidean Spaces}
\label{curved}

Particle packing problems in non-Euclidean (curved) spaces arises in a variety of fields, including
physics, \cite{Ren05,Bo06,Ka07b} biology, \cite{Ta30,Go67,Pr90,To02d,Za04} communications theory, \cite{Co93}
and geometry. \cite{Co93,Ra03,Ha04,Co07,Cohn11} While a comprehensive overview
of this topic is beyond the scope of this review, it is useful to highlight here
some of the developments for the interested reader.
We will limit the discussion to sphere packings on the positively curved
unit sphere $S^{d-1} \subset \mathbb{R}^d$. The reader
is referred to the review by Torquato and Stillinger \cite{To10c}
for some discussion of negatively curved hyperbolic space $\mathbb{H}^d$.

Recall that the {\it kissing}  number $Z$ is the number of spheres of unit radius
that simultaneously touch a unit sphere  $S^{d-1}$. \cite{Co93}  The kissing
number problem asks for the maximal kissing number $Z_{max}$ in $\mathbb{R}^d$.
The determination of the maximal kissing number
in $\mathbb{R}^3$ spurred a famous debate between Issac Newton and
David Gregory in 1694. The former correctly thought the answer 
was 12, but the latter wrongly believed that it was 13. The maximal kissing number 
$Z_{max}$ for $d >3$ is only known in dimensions
four, \cite{Mu08} eight,\cite{Lev79,Od79} and twenty four.\cite{Lev79,Od79}

A packing of congruent spherical caps on the unit sphere $S^{d-1}$ in $\mathbb{R}^d$
yields a {\it spherical code} consisting of the centers of the caps. \cite{Co93}
A spherical code is optimal if the  minimal
distance (smallest angular separation between distinct points in the
code) is as large as possible. The reader is referred
to the paper by Cohn and Kumar \cite{Co07} and references therein for some
developments in the mathematics literature. Interestingly, the jamming of spherical
codes in a variety of dimensions has been investigated. \cite{Cohn11} It is noteworthy
that some optimal spherical codes \cite{Ho10a} are related to the densest
local packing of spheres around a central sphere. \cite{Ho10b,Ho11a}

\section{Packings of Nonspherical Particles}
\label{nonspherical}

The packing characteristics of equilibrium phases and jammed states of packings of nonspherical particle are considerably richer
than their spherical counterparts.  \cite{On49,Ho70,Fr83,Bez91,Ti93,Frenk99,Be00,Wi03,Do04b,Do04d,Ma05,Do05a,Do05b,Bet05,Co06,Do07a,
Bol97,Ch08,To09b,To09c,Gl09,Ji10b,To10b,Ka10,
Ch10,Chaik10,Ba10,Ji11b,Ji11c,Ag11,Gr11,Ma11,To12b,He12,Da12a,Da12b,At12,Ni12,Ga12,
Ba13,Ga13,Ga14,Ch14a,Zo14,Ch14,Ch14a,Ci15,Do17,Va18}
This is due to the fact that nonsphericity of the particle shape introduces rotational degrees
of freedom not present in sphere packings.
Our primary interest is in simple 3D convex shapes, such as ellipsoids, superballs,
spherocylinders and polyhedra, although we remark on more complex shapes,
such as concave particles as well as congruent ring tori, which are multiply connected nonconvex bodies
of genus 1.  Organizing principles to characterize
and classify very dense, possibly jammed packings of nonspherical particles in terms
of shape symmetry of the particles \cite{To09a,To09b,To12b} are discussed in Sec.\ref{organ}.

\subsection{Simple Nonspherical Convex Shapes}

\subsubsection{Ellipsoids}
\label{ellip}

One simple generalization of the sphere is an ellipsoid,
the family of which is a continuous deformation of a sphere. A $d$-dimensional ellipsoid in $\mathbb{R}^d$  is a centrally
symmetric  body  occupying the region 
\begin{equation}
\left(\frac{x_1}{a_1}\right)^{2}+  \left(\frac{x_2}{a_2}\right)^{2}+\cdots +\left(\frac{x_d}{a_d}\right)^{2}\le 1,
\end{equation}
where $x_i$ $(i=1,2,\ldots,d)$ are Cartesian coordinates
and $a_i$ are the semi-axes of the ellipsoid. Thus,
we see that an ellipsoid is an {\it affine} (linear) transformation of the sphere.

\subsubsection{Superballs}
\label{super}

A $d$-dimensional \textit{superball} in $\mathbb{R}^d$  is a centrally
symmetric  body occupying
the region 
\begin{equation}
|x_1|^{2p}+|x_2|^{2p}+\cdots+|x_d|^{2p} \le 1,
\end{equation}
where $x_i$ $(i=1,\ldots,d)$ are Cartesian coordinates
and $p \ge 0$ is the \textit{deformation parameter} (not pressure,
as denoted in Sec.~\ref{phase}), which controls
the extent to which the particle shape has deformed from that of a
 $d$-dimensional sphere ($p=1$). Thus, superballs constitute  a large family of
both convex ($p\ge 1/2$) and concave ($0<p<1/2$) particles (see Fig.~\ref{superball}).
A ``superdisk",which is the designation in the 2D
case, possesses square
symmetry. As $p$ moves away from unity,
two families of superdisks with square symmetry can be obtained
depending on whether $p < 1$ or $p >1$. When $p<1/2$, the superdisk is concave; see Ref. \onlinecite{Ji08b}.

\begin{figure}
\begin{center}
\includegraphics[height=2cm, keepaspectratio]{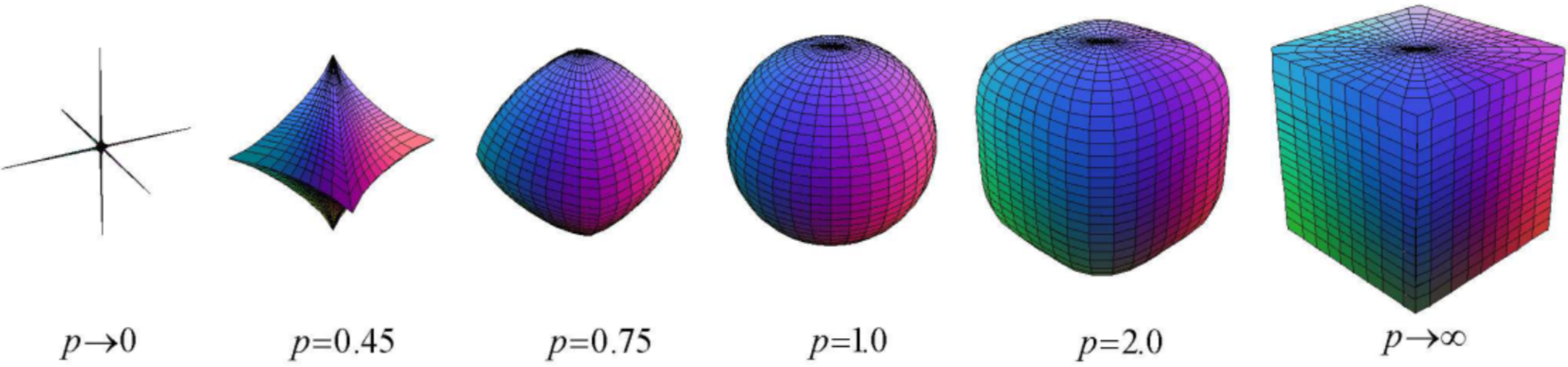}
\end{center}
\caption{(color online). Superballs with different values of the deformation
parameter $p$. We note that $p=0,1/2,1,$ and $\infty$ correspond to a 3D cross, regular octahedron,
sphere and cube, respectively.} \label{superball}
\end{figure}

\subsubsection{Spherocylinder}

A $d$-dimensional spherocylinder
in $\mathbb{R}^d$  consists of a cylinder of length $L$ and
radius $R$ capped at both ends by hemispheres of radius $R$ , and therefore is a centrally-symmetric
convex particle.  Its volume $V_{SC}$ for $d \ge 2$ is given by
\begin{equation}
V_{SC} =  \frac{\pi^{(d-1)/2} R^{d-1}}{\Gamma(1+(d-1)/2)} L + \frac{\pi^{d/2} R^d}{\Gamma(1+d/2)}.
\end{equation}
When $L=0$, a $d$-dimensional spherocylinder reduces to a $d$-dimensional sphere of radius $R$.
Figure \ref{fig_spherocylinder} shows 3D examples.

\begin{figure}[htbp]
\begin{center}
$\begin{array}{c}
\includegraphics[width=7.5cm,keepaspectratio]{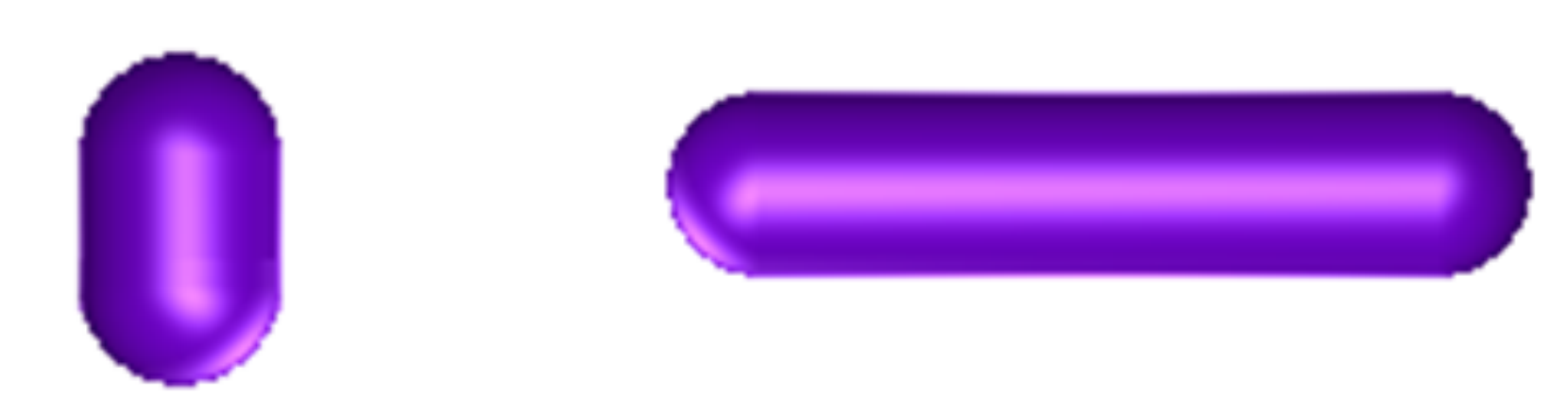} \\
\end{array}$
\end{center}
\caption{(Color online) Three-dimensional spherocylinders composed of cylinder with length $L$,
caped at both ends with hemispheres with radius $R$. Left panel: A
spherocylinder with aspect ratio $L/R = 1$. Right panel: A
spherocylinder with aspect ratio $L/R = 5$.}
\label{fig_spherocylinder}
\end{figure}

\subsubsection{Polyhedra}

\begin{figure}[bthp]
\begin{center}
\includegraphics[width=9.5cm,keepaspectratio]{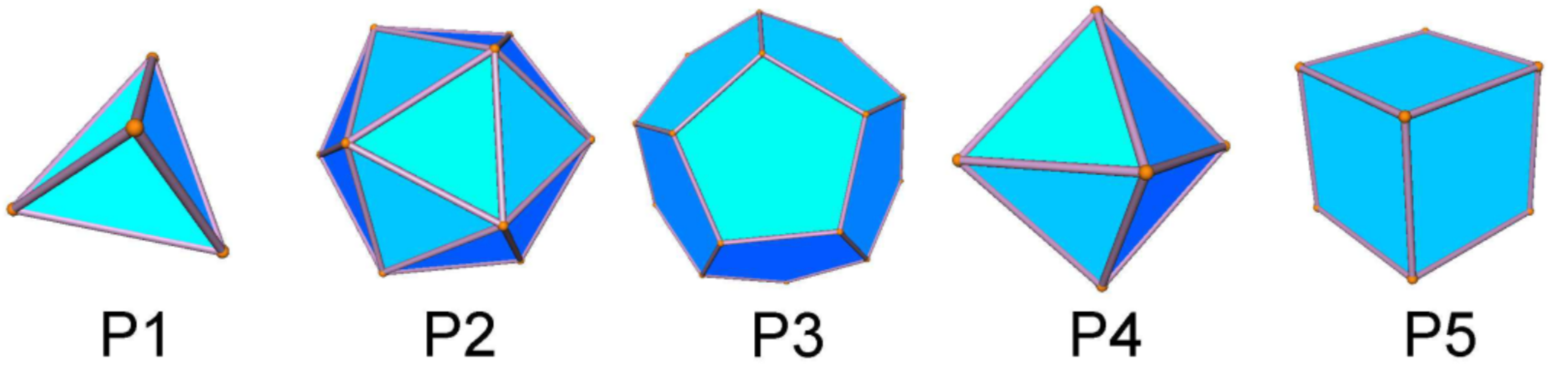} \\
\end{center}
\caption{(color online). The five Platonic solids: tetrahedron (P1), icosahedron (P2),
dodecahedron (P3), octahedron (P4), and cube (P5).}
\label{platonic}
\end{figure}

The Platonic solids  (mentioned in Plato's Timaeus)
are convex polyhedra with faces composed of congruent
convex regular polygons. There are exactly five such solids: the tetrahedron (P1), icosahedron (P2),
dodecahedron (P3), octahedron (P4), and cube (P5) (see Fig. \ref{platonic}) 
Note that viral capsids often have icosahedral symmetry; see,
for example, Ref. \onlinecite{Za04}.

\begin{figure}
\begin{center}
\includegraphics[width=9.5cm,keepaspectratio]{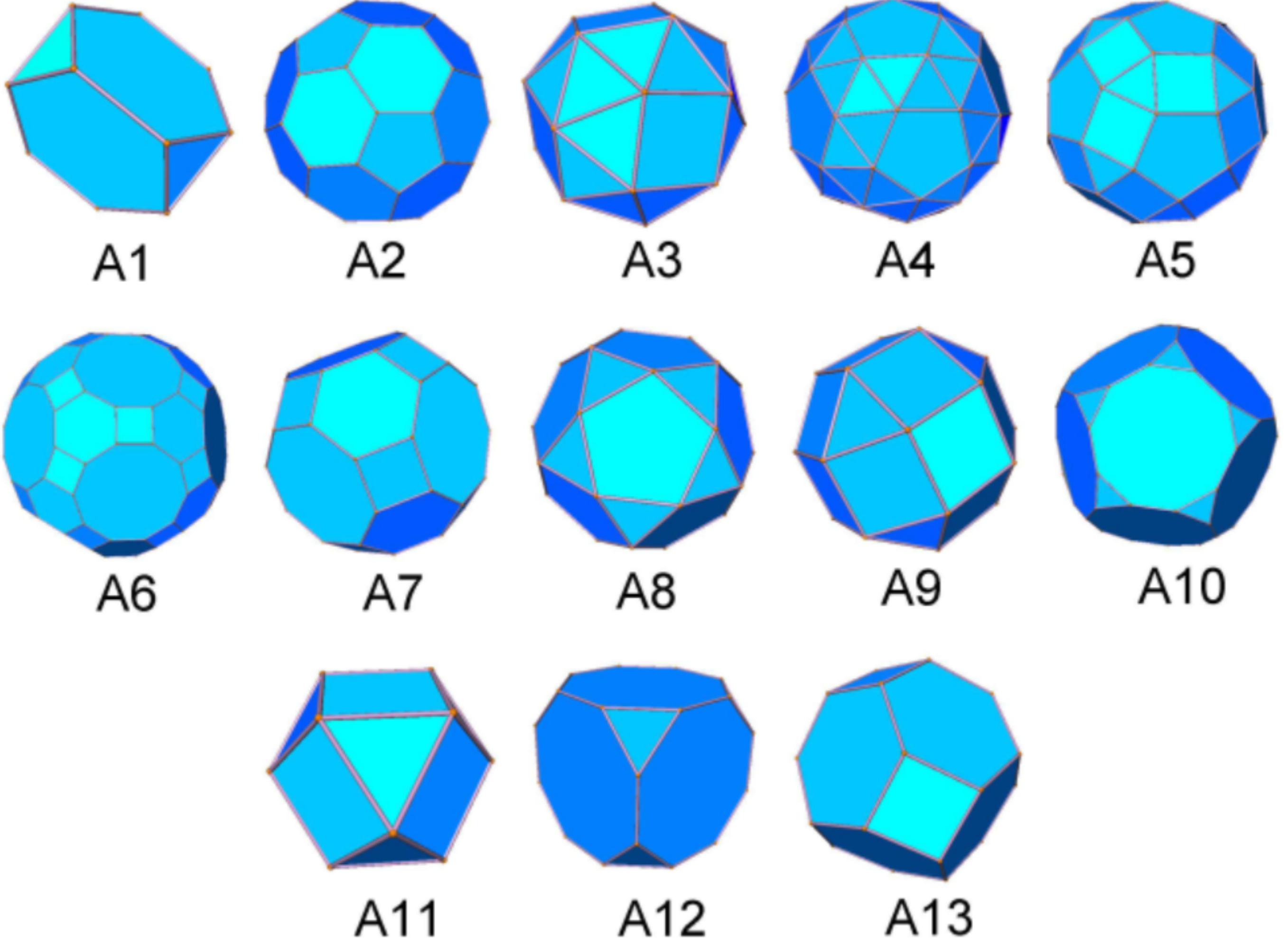} \\
\end{center}
\caption{(color online) The thirteen Archimedean solids: truncated tetrahedron (A1),
truncated icosahedron (A2),
snub cube (A3), snub dodecahedron (A4), rhombicosidodecahdron (A5),
truncated icosidodecahedron (A6),
truncated cuboctahedron (A7), icosidodecahedron (A8),  rhombicuboctahedron (A9),
truncated dodecahedron (A10), cuboctahedron (A11),
truncated cube (A12), and truncated octahedron (A13). } \label{archimedean}
\end{figure}

An Archimedean solid is a highly symmetric, semi-regular
convex polyhedron composed of two or more types of regular polygons meeting in
identical vertices. The thirteen Archimedean solids are depicted
in Fig. \ref{archimedean}. This typical enumeration does not count the chiral forms (not shown) of the snub cube (A3)
and snub dodecahedron (A4). The remaining 11 Archimedean solids
are non-chiral (i.e., each solid is superposable on its mirror image)
and the only non-centrally symmetric one  among these is the truncated tetrahedron.

It is noteworthy that the tetrahedron (P1) and
the truncated tetrahedron (A1) are the
only Platonic and non-chiral Archimedean solids, respectively,
that are not {\it centrally symmetric}. 
 We will see that the central symmetry
of the majority of the Platonic and Archimedean solids (P2 -- P5, A2 -- A13)
distinguish their dense packing arrangements from those of the
non-centrally symmetric ones (P1 and A1)
in a fundamental way.

\subsection{Equilibrium and Metastable Phase Behavior}

Hard nonspherical particles exhibit a richer phase diagram 
than that of hard spheres because the former can possess different
degrees of translational and orientational 
order; see Refs. \onlinecite{On49,Fr83,Ti93,Frenk99,Ba13,Ci15,Bol97,Ba10,Ga12,Bet05,Ag11,Ga13,Da12a,Da12b,Ji11b,Ch14a}.
Nonspherical-particle systems can form isotropic liquids, a variety of liquid-crystal phases, 
rotator crystals, and  solid crystals.  Particles in liquid phases
have neither translational nor orientational order.
Examples of liquid-crystal states include a  nematic phase in which the particles are aligned (i.e., with orientational order) while the system lacks any long-range translational
order and a  smectic phase in which the particles have
ordered orientations and possess translational order in one
direction. A rotator (or plastic) phase is one in which particles possess
translational order but can rotate freely. Solid crystals are
characterized by both translational and orientational order.

Ordering transitions in systems of hard nonspherical
particles are entropically driven, i.e., the stable phase is determined by a competition
between translational and orientational entropies.  
This principle was first established in the pioneering work of Onsager,\cite{On49}
where it was shown that needle-like shapes exhibit a liquid-nematic
phase transition at low densities because in the nematic phase
the drop in orientational entropy is offset by the increase 
in translational entropy, i.e., the available space for any needle increases
as the needle tends to align.

The stable phase formed by systems of hard-nonspherical
particles  is influenced by its symmetry and surface smoothness, which
in turn determines the overall system entropy. Here we briefly highlight 3D numerical studies
that use Monte Carlo methods and/or free-energy calculations to determine the phase diagrams. 
Spheroids exhibit not only fluid, solid crystal, and nematic phases 
for some aspect ratios, but rotator phases for nearly spherical
particle  shapes at intermediate
densities.\cite{Fr84,Ba13}  Hard ``lenses" (common volume to two overlapping
identical spheres) have  a qualitatively
similar phase diagram to hard oblate spheroids, but 
differences between them are more pronounced in the high-density crystal phase up to the
densest-known packings. \cite{Ci15} Spherocylinders exhibit  five different possible phases, depending on the packing fraction and 
aspect ratio: isotropic fluid, smectic, nematic, rotator, and solid crystal  phases.\cite{Bol97}
Apart from fluid and crystal phases, superballs 
can form rotator phases at intermediate densities.\cite{Ba10,Ga12}
Systems of tetragonal parallelepipeds can exhibit
liquid-crystalline and cubatic phases, depending
on the aspect ratio and packing fraction.\cite{Bet05}
Various convex space-filling polyhedra (including truncated octahedron,
rhombic dodecahedron, two types of prisms and cube)
possess unusual liquid-crystalline and rotator-crystalline phases at intermediate packing
fractions.\cite{Ag11} Truncated
cubes exhibit a rich diversity in crystal structures that depend
sensitively on the amount of truncation.\cite{Ga13} A study of a
wide class of polyhedra revealed that  entropy maximization favors mutual alignment of particles 
along their facets. \cite{Da12b}
By analytically constructing the densest known packings 
of congruent Archimedean truncated tetrahedra (which nearly fill
all of space), the melting properties of such systems were examined
by decompressing this densest packing and equilibrating them.\cite{Ji11b}
A study of the entire phase diagram of truncated tetrahedra showed that 
the system undergoes two first-order phase transitions  as the density increases: 
first a liquid-solid transition and then a solid-solid transition.\cite{Ch14a}

\subsection{Maximally Random Jammed States}

The fact that M\&M candies (spheroidal particles with aspect ratio $\alpha \approx 1.9$) were shown experimentally
to  randomly pack more densely than spheres ($\phi \approx 0.66$)
motivated the development of a modified LS algorithm  \cite{Do05a,Do05b}  to obtain frictionless MRJ-like packings with even higher densities
for other aspect ratios. This  included nearly-spherical ellipsoids 
with  $\phi \approx 0.74$,\cite{Do04b,Do07a} i.e., packing fractions
approaching those of the densest 3D sphere packings. Note that these other densest MRJ packings are realizable
experimentally. \cite{Ma05}

\begin{figure}[bthp]
\centerline{\includegraphics[height=2in,keepaspectratio,clip=]
{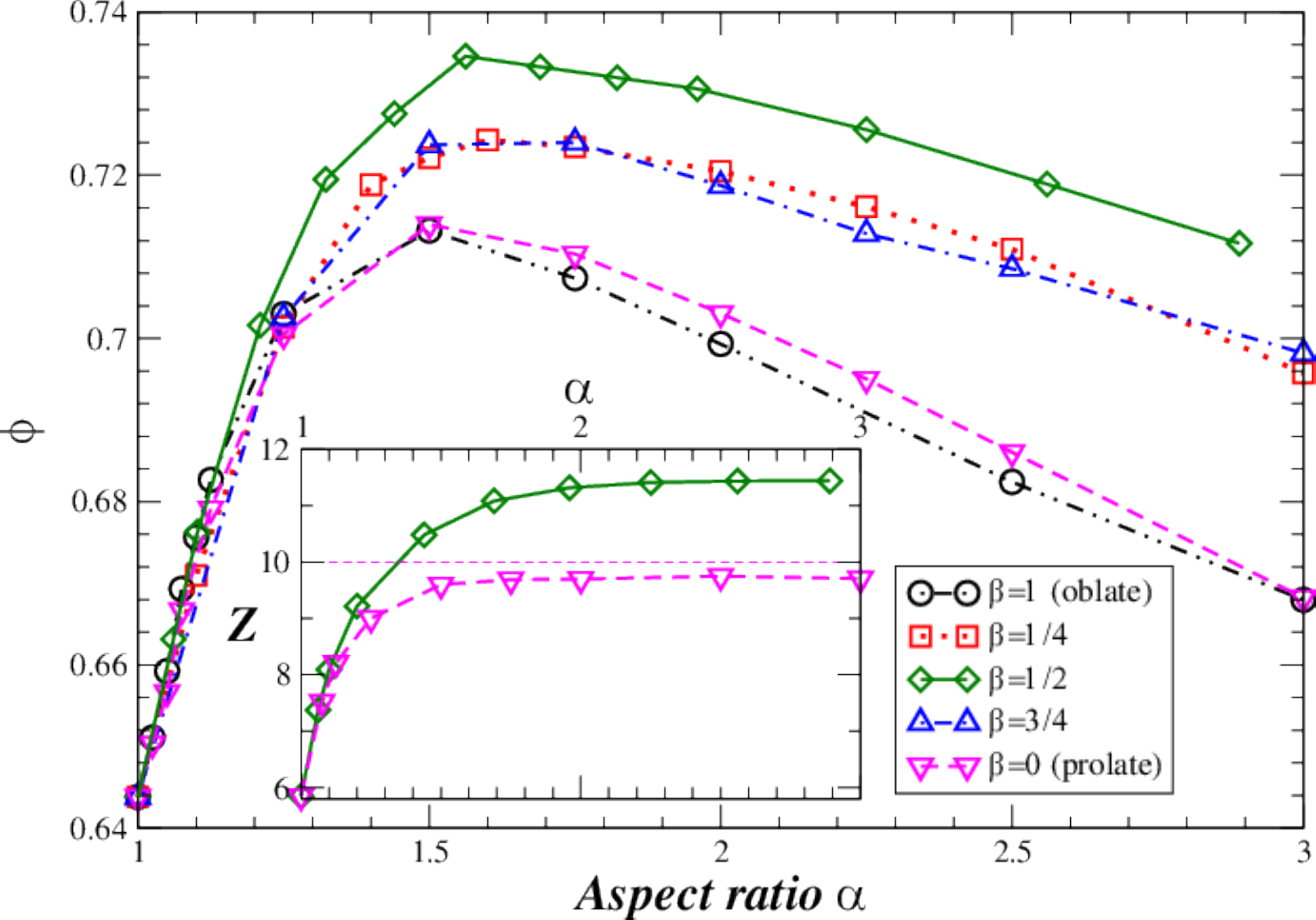}}
\caption{Packing fraction $\phi$  versus aspect ratio $\alpha$ for MRJ packings
of 10,000 ellipsoids as obtained in Ref. \onlinecite{Do07a}. The semi-axes here are $1, \alpha^\beta, \alpha$. 
The inset shows the mean contact number $\overline Z$ as  a function of $\alpha$.
Neither the spheroid (oblate or prolate) or general ellipsoids
cases attain their isostatic values of $\overline Z=10$ or $\overline Z=12$, respectively.}
\label{10000}
\end{figure}

Figure \ref{10000} shows separate plots of $\phi$ and  mean contact number $\overline Z$ as a function
of $\alpha$ as predicted by the more refined simulations of Donev {\it et al.}. \cite{Do07a} 
Each plot shows  a cusp (i.e., non-differentiable) minimum at the sphere point,
and $\phi$ versus aspect ratio $\alpha$ possesses a density maximum.
The existence of a cusp at the sphere point runs counter to the conventional expectation
that for ``generic" (disordered) jammed frictionless particles,  the total number
of (independent) constraints equals the total number of degrees of
freedom $d_f$, which has been
referred to as the {\it isostatic} conjecture. \cite{Al98} 
This conjecture implies a mean contact number $\overline Z=2d_f$, where
$d_{f}=2$ for disks, $d_{f}=3$
for ellipses, $d_{f}=3$ for spheres, $d_{f}=5$ for spheroids, and
$d_{f}=6$ for general ellipsoids.
Since $d_{f}$ increases discontinuously with the introduction of
rotational degrees of freedom as one makes the particles nonspherical,
the isostatic conjecture predicts that $\overline Z$ should have
a jump increase at aspect ratio $\alpha=1$ to a value of $\overline Z=12$ for a general
ellipsoid.  Such a discontinuity was not originally observed by
Donev {\it et al.} \cite{Do04b}, rather, they found that jammed ellipsoid packings
are \emph{hypostatic}  (or sub-isostatic), ${\overline Z}<2d_{f}$, near the sphere point,
and only become nearly  isostatic for large aspect ratios. 
In fact, the isostatic conjecture is only rigorously true 
for amorphous sphere packings after removal of rattlers; generic nonspherical-particle
packings should generally be hypostatic. \cite{Ro00,Do07a}
It has been rigorously shown that packings of nonspherical particles
are generally not jammed to
first order in the ``jamming gap" $\delta$ [see Sec. \ref{polytope}], but are jammed to second order in 
$\delta$ due to curvature deviations from the sphere.\cite{Do07a}
In striking contrast with MRJ sphere packings, the rattler concentrations 
of the MRJ ellipsoid packings appear practically to vanish
outside of some small neighborhood of the sphere point.\cite{Do07a}
The reader is referred to recent work on hypostatic jammed 2D packings of
noncircular  particles.\cite{Va18}

Jiao, Stillinger and Torquato \cite{Ji10b} have computed the packing fraction $\phi_{\mbox{\scriptsize MRJ}}$ of  MRJ packings
of binary superdisks in $\mathbb{R}^2$ and identical superballs in $\mathbb{R}^3$.
They found that $\phi_{\mbox{\scriptsize MRJ}}$
increases dramatically and nonanalytically 
as one moves away from the circular-disk or sphere point ($p=1$). Moreover,
these disordered packings were shown to be  {\it hypostatic}.
Hence, the local particle arrangements are necessarily nontrivially
correlated to achieve strict jamming and hence ``nongeneric,'' the degree of which
was quantified. MRJ packings of binary superdisks  and of ellipses are
effectively hyperuniform.\cite{Za11d}
Note that the geometric-structure approach was used to derive
a highly accurate formula for the packing fraction $\phi_{\mbox{\scriptsize MRJ}}$ of MRJ binary  packings of convex
superdisks\cite{Ti15}  that is valid for almost all size ratios, relative concentrations and deformation
parameter $p\ge 1/2$. For the special limit of monodisperse circular disks, this formula predicts
$\phi_{\mbox{\scriptsize MRJ}}=0.837$,
which  is in very good agreement with the recently numerically discovered MRJ isostatic state\cite{At14} with $\phi_{\mbox{\scriptsize MRJ}}=0.827$.

3D MRJ packings of the four non-tiling Platonic solids (tetrahedra,
octahedra, dodecahedra, and icosahedra) were generated using the ASC optimization scheme. \cite{Ji11c}
The MRJ packing fractions for tetrahedra, octahedra,
dodecahedra, and icosahedra are $0.763 \pm 0.005$, $0.697 \pm 0.005$, 
$0.716 \pm 0.002$, and $0.707 \pm 0.002$, respectively.
It was shown that as the number of facets of the particles increases, the translational order in the packings increases
while the orientational order decreases. Moreover, such MRJ packings were found to be hyperuniform
with a total correlation function $h({\bf r})$ that decays to zero asymptotically 
with same power-law as MRJ spheres, i.e, like $-1/r^4$. These results suggest that hyperuniform quasi-long-range
correlations are a universal feature of MRJ packings of frictionless particles of general shape. However, unlike
MRJ packings of ellipsoids, superballs, and superellipsoids, which are hypostatic, MRJ packings of the non-tiling
Platonic solids are isostatic.\cite{Ji11c}
In addition, 3D MRJ packings of truncated tetrahedra with an average 
packing fraction of 0.770 were also generated.~\cite{Ch14a}

\subsection{Maximally Dense Packings}

We focus here mainly on exact constructions of the densest known packings
of nonspherical particles in which the lattice vectors as well as particle positions
and orientations are expressible analytically. Where appropriate, we also
cite some work concerning packings obtained via computer simulations
and laboratory experiments.

 Rigorous upper bounds on the maximal packing fraction $\phi_{\mbox{\scriptsize max}}$
of packings of nonspherical particles of general shape in $\mathbb{R}^d$ can be used to assess
their packing efficiency. However, it has been highly challenging to
formulate upper bounds for non-tiling particle
packings that are nontrivially less than unity.
It has recently been shown  that
$\phi_{\mbox{\scriptsize max}}$ of a packing of congruent nonspherical particles of
volume $v_{P}$ in $\mathbb{R}^d$ is bounded from above according to
\begin{equation}
\phi_{\mbox{\scriptsize max}}\le \phi_{\mbox{\scriptsize max}}^U= \mbox{min}\left[\frac{v_{P}}{v_{S}}\;
\phi^{S}_{\mbox{\scriptsize max}},1\right], \label{bound}
\end{equation}
where $v_{S}$ is the volume of the largest sphere that can be inscribed
in the nonspherical particle and  $\phi^{S}_{\mbox{\scriptsize max}}$  
is the maximal packing fraction of a packing of $d$-dimensional of identical spheres 
\cite{To09b,To09c} (e.g., $\phi^{S}_{\mbox{\scriptsize max}}=\pi/\sqrt{12}$ for $d=2$
and $\phi^{S}_{\mbox{\scriptsize max}}=\pi/\sqrt{18}$ for $d=3$). The upper bound (\ref{bound}) 
will be relatively tight  provided that the {\it asphericity} $\gamma$
(equal to the ratio of the circumradius to the inradius)  of the
particle is not large.  Since bound (\ref{bound}) cannot generally be sharp (i.e.,
exact) for a non-tiling particle, any packing  whose density is close to the upper bound
(\ref{bound}) is nearly optimal, if not optimal.
Interestingly, a majority of the  centrally symmetric  Platonic and Archimedean solids have relatively small asphericities, 
explaining the corresponding small differences between $\phi_{\mbox{\scriptsize max}}^U$ and  $\phi_{\mbox{\scriptsize max}}^L$,
the packing fraction of the densest lattice packing.
\cite{Mi05,Be00,To09b,To09c} As discussed below, these densest lattice packings
are conjectured to be the densest among all packings.\cite{To09b,To09c}
Upper bound (\ref{bound}) will also be relatively tight for superballs (superdisks) for 
deformation parameters $p$ near the sphere (circle) point ($p=1$). 
Dostert {\it et al.} \cite{Do17} recently obtained upper bounds
on $\phi_{\mbox{\scriptsize max}}$ of {\it translative}
packings of superballs and 3D convex bodies with tetrahedral symmetry.

\subsubsection{Ellipsoids}

\begin{figure}[bthp]
\centerline{\includegraphics[height=2.6in,keepaspectratio,clip=]
{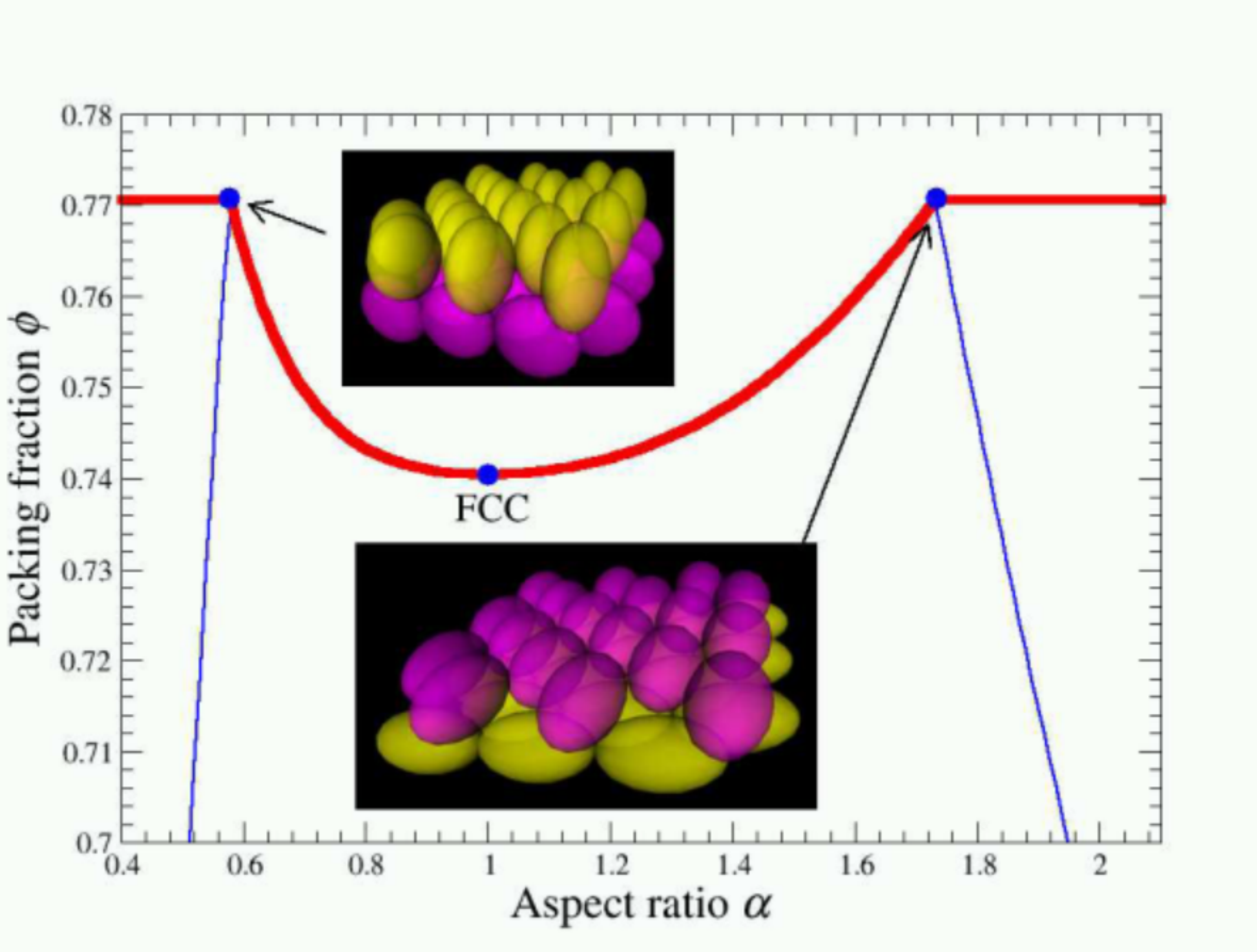}}
\caption{The packing fraction of the ``laminated" non-Bravais lattice packing of
ellipsoids (with a two-particle basis) as a function of
the aspect ratio $\alpha$. The point $\alpha=1$ corresponding
to the face-centered cubic lattice sphere packing is shown, along with the two sharp
maxima in the packing fraction for prolate ellipsoids with $\alpha=\sqrt{3}$
and oblate ellipsoids with $\alpha=1/\sqrt{3}$, as shown in
the insets. For both $\alpha < 1/\sqrt{3}$ and $\alpha >\sqrt{3}$, the
packing fraction drops off precipitously holding
the particle orientations fixed (blue
lines). The presently maximal achievable packing fraction
$\phi=0.7707\ldots$  is highlighted with
a thicker red line, and is constant for $\alpha \le 1/\sqrt{3}$ and $\alpha \ge \sqrt{3}$;  see Ref. \onlinecite{Do04d}.}
\label{ellipsoids}
\end{figure}

The fact that MRJ-like packings of nearly-spherical ellipsoids 
exist with  $\phi \approx 0.74$ (see Refs. \onlinecite{Do04b,Do07a}) suggested that there exist ordered ellipsoid packings
with appreciably higher densities. The densest known packings of identical 3D ellipsoids were  obtained
analytically by Donev {\it et al.};\cite{Do04d} see Fig. \ref{ellipsoids}.
These are exact constructions 
and represent a family of {\it non-Bravais} lattice packings of ellipsoids with
a packing fraction that always exceeds the that of the
corresponding densest Bravais lattice packing
($\phi =0.74048 \ldots$) with a maximal packing fraction of $\phi =0.7707 \ldots$,
for a wide range of aspect ratios ($\alpha \le 1/\sqrt{3}$ and $\alpha \ge \sqrt{3}$). In these densest known packings,  each ellipsoid has 14 contacting neighbors and there are two particles per fundamental cell.

While a convex ``lens" fits snugly within an oblate spheroid of the 
same aspect ratio, the densest known lens packings
are denser for general aspect ratios (except for a narrow
range of intermediate values) than their spheroid counterparts,
achieving the highest packing fraction of $\phi=\pi/4=0.7853\ldots$
in the ``flat-lens" limit. \cite{Ci15}

\subsubsection{Superballs}

Exact analytical constructions  for candidate maximally dense packings  of 2D superdisks
were recently proposed for all convex and concave shapes.\cite{Ji08b} These are achieved by two different
families of Bravais lattice packings such that $\phi_{\mbox{\scriptsize max}}$
is nonanalytic at the ``circular-disk" point ($p=1$) and increases significantly as
$p$ moves away from unity. 
The broken rotational symmetry of superdisks influences the packing
characteristics in a non-trivial way
that is distinctly different from ellipse  packings.
For ellipse packings, no improvement over the maximal
circle packing fraction ($\phi_{\mbox{\scriptsize max}}=\pi/\sqrt{12}=0.906899\ldots$) is possible, since the former is an affine transformation
of the latter. \cite{To10c} For superdisks, one can take advantage of 
the four-fold rotationally symmetric shape of the particle to obtain a  substantial
improvement on the maximal circle packing fraction.
By contrast, one needs to use higher-dimensional counterparts of ellipses ($d \ge 3$) in
order to improve on $\phi_{\mbox{\scriptsize max}}$ for spheres. Even for 3D ellipsoids, $\phi_{\mbox{\scriptsize max}}$ increases smoothly as the aspect ratios of the semi-axes vary from unity, \cite{Do04d} and hence has
no cusp at the sphere point. In fact, 3D ellipsoid packings
have a cusp-like behavior at the sphere point only when they are randomly
jammed. \cite{Do04b}

Jiao, Stillinger and Torquato  \cite{Ji09a} showed that increasing the dimensionality 
of a superball from two to three dimensions imbues the optimal  packings with  structural characteristics
that are richer than their 2D counterparts. 
They  obtained analytical constructions for
the densest known superball packings for all convex and concave cases, which  are certain families of Bravais lattice packings in which each particle has 12 contacting neighbors.
For superballs in the cubic regime ($p>1$), the
candidate optimal packings are achieved by two families of Bravais
lattice packings ($\mathbb{C}_0$ and $\mathbb{C}_1$ lattices) possessing two-fold  and three-fold rotational
symmetry, respectively.
For superballs in the octahedral regime ($0.5<p< 1$), there are also
two families of Bravais lattices ($\mathbb{O}_0$ and $\mathbb{O}_1$ lattices) obtainable from continuous
deformations of the fcc lattice, which are apparently optimal
in the vicinity of the sphere point and the octahedron point,
respectively. 

The exact maximal packing fraction $\phi_{\mbox{\scriptsize max}}$ as a function of deformation
parameter $p$ is plotted in Fig.~\ref{fig4}. 
As $p$ increases from unity, the initial increase of $\phi_{\mbox{\scriptsize max}}$
is linear in $(p-1)$ and subsequently $\phi_{\mbox{\scriptsize max}}$
 increases monotonically with $p$ until it reaches unity as the
particle shape tends to the cube. These characteristics stand in contrast to
those of the densest known ellipsoid packings (see Fig. \ref{ellipsoids})
whose packing fraction as a function of aspect ratios has zero
initial slope and is bounded from above by a value of
$0.7707\ldots$. \cite{Do04d} As $p$
decreases from unity, the initial increase of $\phi_{\mbox{\scriptsize max}}$ is
linear in $(1-p)$. Thus, $\phi_{\mbox{\scriptsize max}}$ is a nonanalytic function of
$p$ at $p=1$. \cite{Ji08b} However, 
the behavior of $\phi_{\mbox{\scriptsize max}}$ as the superball shape moves off the
sphere point  is distinctly different from that of optimal spheroid
packings, for which $\phi_{\mbox{\scriptsize max}}$ increases smoothly as the aspect
ratios of the semi-axes vary from unity and hence has no cusp at
the sphere point. \cite{Do04d} 
These distinctions between superball versus ellipsoid packings result from differences
in which rotational symmetries in these two packings are broken.\cite{Ji09a}
For the small range $0.79248 < p < 1$, Ni {\it et al.} ~\cite{Ni12} 
numerically found lattice packings that are very slightly 
denser than those of the theoretically predicted $\mathbb{O}_0$ lattices
of Ref. \onlinecite{Ji09a}. All of the results, support the Torquato-Jiao
conjecture that the densest packings of all convex superballs are their densest 
lattice packings.\cite{To09c}

\begin{figure} 
$\begin{array}{c}
\includegraphics[height = 4.5cm, keepaspectratio]{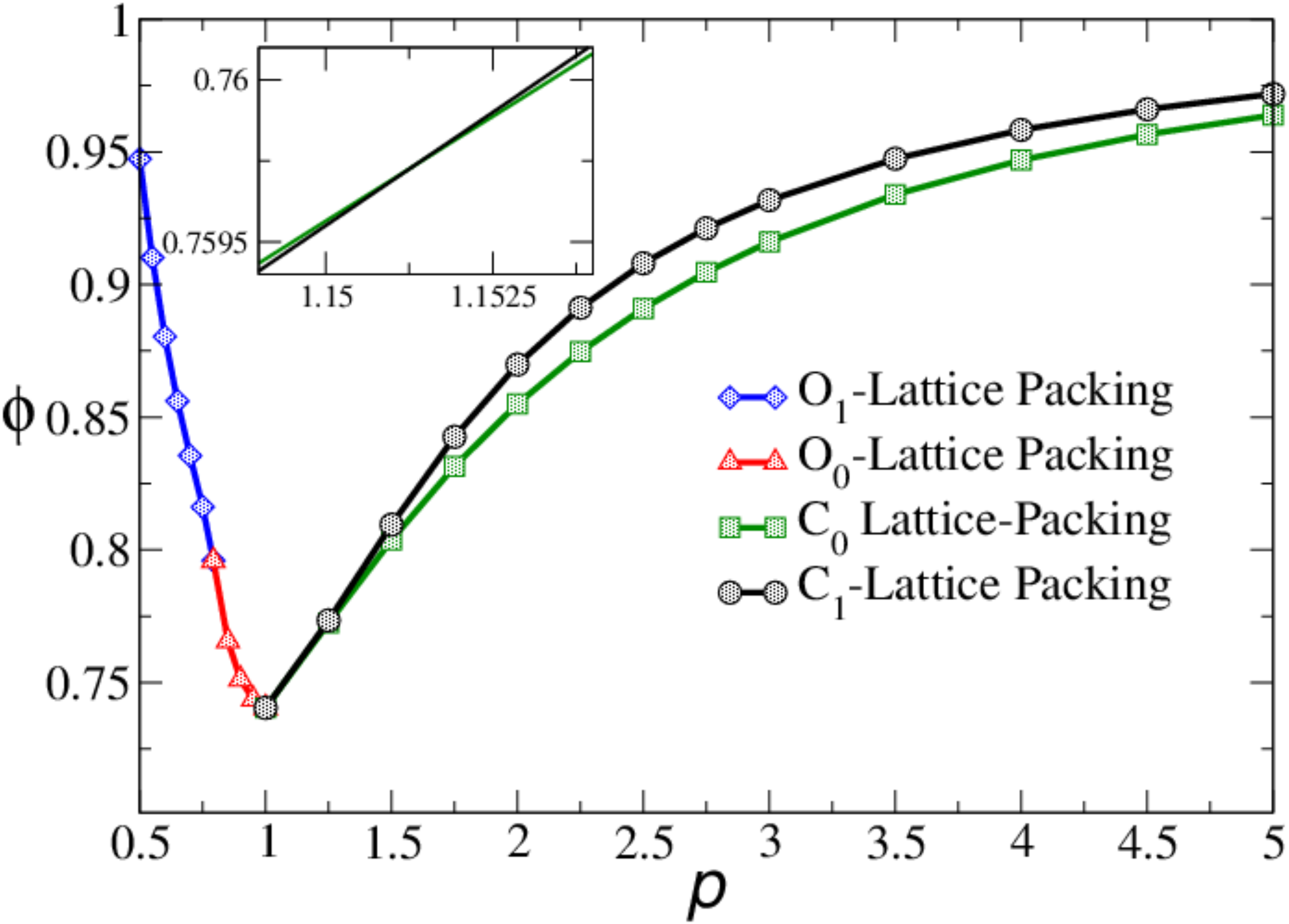}
\end{array}$
\caption{(color online). Packing fraction versus  deformation parameter $p$
for the packings of convex superballs. \cite{Ji09a} Insert: Around $p^*_c = 1.1509\ldots$,
the two curves are almost locally parallel to each other.} \label{fig4}
\end{figure}

\begin{figure}
$\begin{array}{c}
\includegraphics[height = 4.5cm, keepaspectratio]{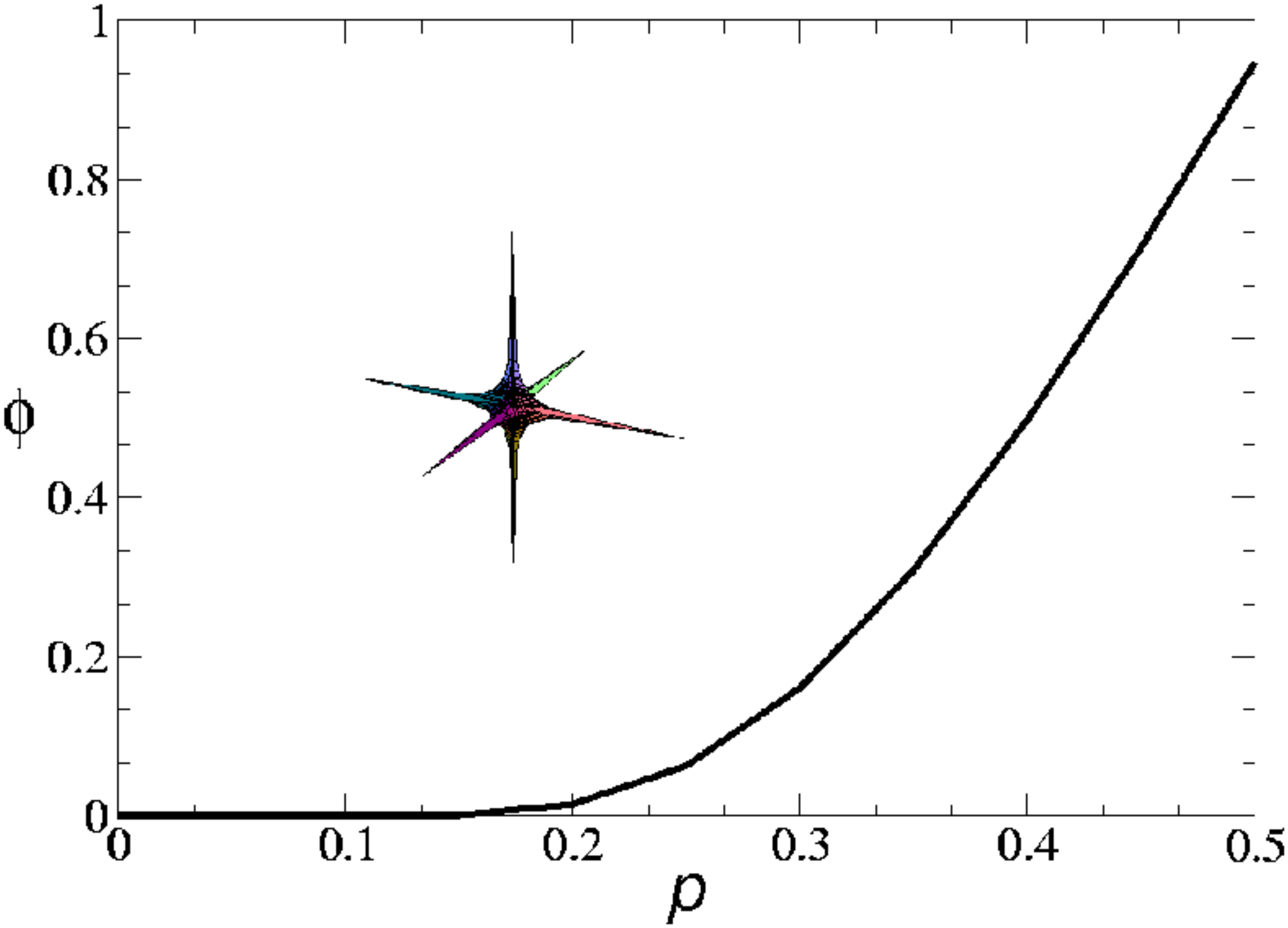}
\end{array}$
\caption{(color online). Packing fraction versus  deformation parameter $p$
for the lattice packings of concave superballs. \cite{Ji09a}  Insert: a concave
superball with $p = 0.1$, which will becomes a 3D
cross at the limit $p \rightarrow 0$.} \label{fig12}
\end{figure} 

\subsubsection{Spherocylinders}
\label{sph-cyl}

The role of curvature in determining dense packings of smoothly-shaped particles is still not very well
understood. While the densest packings of 3D ellipsoids are (non-Bravais)
lattice packings, \cite{Do04d} the densest packings of superballs
appear to lattice packings.\cite{Ji09a,To09b}
For 3D spherocylinders with $L>0$, the
optimal Bravais-lattice packing is very dense and might be one of the actual densest
packings. This is because the local principal
curvature of the cylindrical surface is zero along the spherocylinder axis (i.e., a ``flat" direction),
and thus, spherocylinders can have very dense lattice packings by an appropriate
alignment of  the spherocylinders along their axes.
The packing fraction of the densest Bravais-lattice packing is given by
\begin{equation}
\label{sphero_phi}
\phi =\frac{\pi}{\sqrt{12}}~\frac{L+\frac{4}{3}R}{L+\frac{2\sqrt{6}}{3}R},
\end{equation}
where $L$ is the length of the cylinder and $R$ is the radius of
the spherical caps. This lattice packing corresponds to stacking layers of aligned
spherocylinders in the same manner as fcc spheres
and hence  there are a uncountably-infinite
number of non-Bravais-lattice packings of spherocylinders (in correspondence to the Barlow stackings of  spheres\cite{To10c}) with the
same packing densities (\ref{sphero_phi}).
Thus, the set of dense nonlattice packings of
spherocylinders is overwhelmingly larger than that of the lattice packing. We will see that these dense packings are consistent
with Conjecture 3 described in Sec. \ref{organ}.

\subsubsection{Polyhedra}

About a decade ago, very little was known about the densest packings of
polyhedral particles. The difficulty
in obtaining dense packings of polyhedra is related to their complex rotational degrees of freedom and to the non-smooth nature of their shapes.
It was the investigation of Conway and Torquato of dense packings of tetrahedra 
\cite{Co06} that spurred the flurry of activity over several years to find
the densest packings of tetrahedra,\cite{To09b,To09c,Gl09,Chaik10,Ka10,To10a,Ch10,To10b} which in turn led 
to studies of the densest packings of other polyhedra and many other 
nonspherical convex and concave particles. \cite{Ji11b,To12b,Da12a,Da12b,Ga13,Ci15}

Torquato and Jiao \cite{To09b,To09c} employed the  ASC optimization scheme to determine
dense packings of the non-tiling Platonic solids and of all of the Archimedean solids.
For example, they were able  to find
the densest known packings of the octahedra, dodecahedra
and icosahedra (three non-tiling Platonic solids with central symmetry) with densities  $0.947\ldots$,
$0.904\ldots$, and $0.836\ldots$, respectively. Unlike the densest
tetrahedron packing, which must be a non-Bravais  lattice
packing, the densest packings of the other non-tiling Platonic
solids found by the algorithm are their previously known densest (Bravais)
lattice packings \cite{Mi05,Be00}; see Fig. \ref{nontiling}.  These simulation 
results as well as other theoretical considerations led them to general organizing principles
concerning the densest packings of a class of nonspherical particles,
which are described in Sec. \ref{organ}.

\begin{figure}
\centerline{
\includegraphics[  width=1.in,keepaspectratio]
{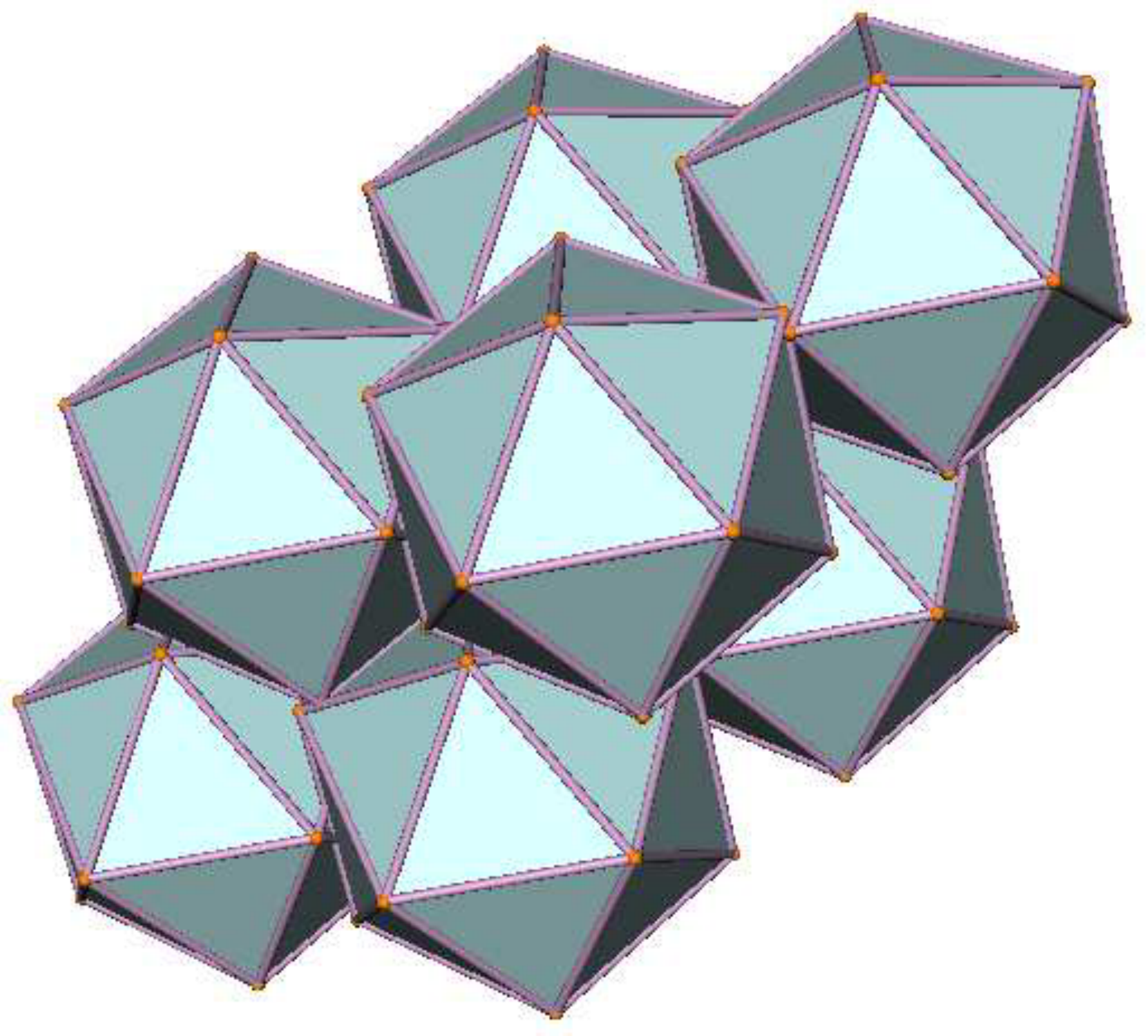}
\hspace{0.16in}
\includegraphics[  width=1.in,keepaspectratio]
{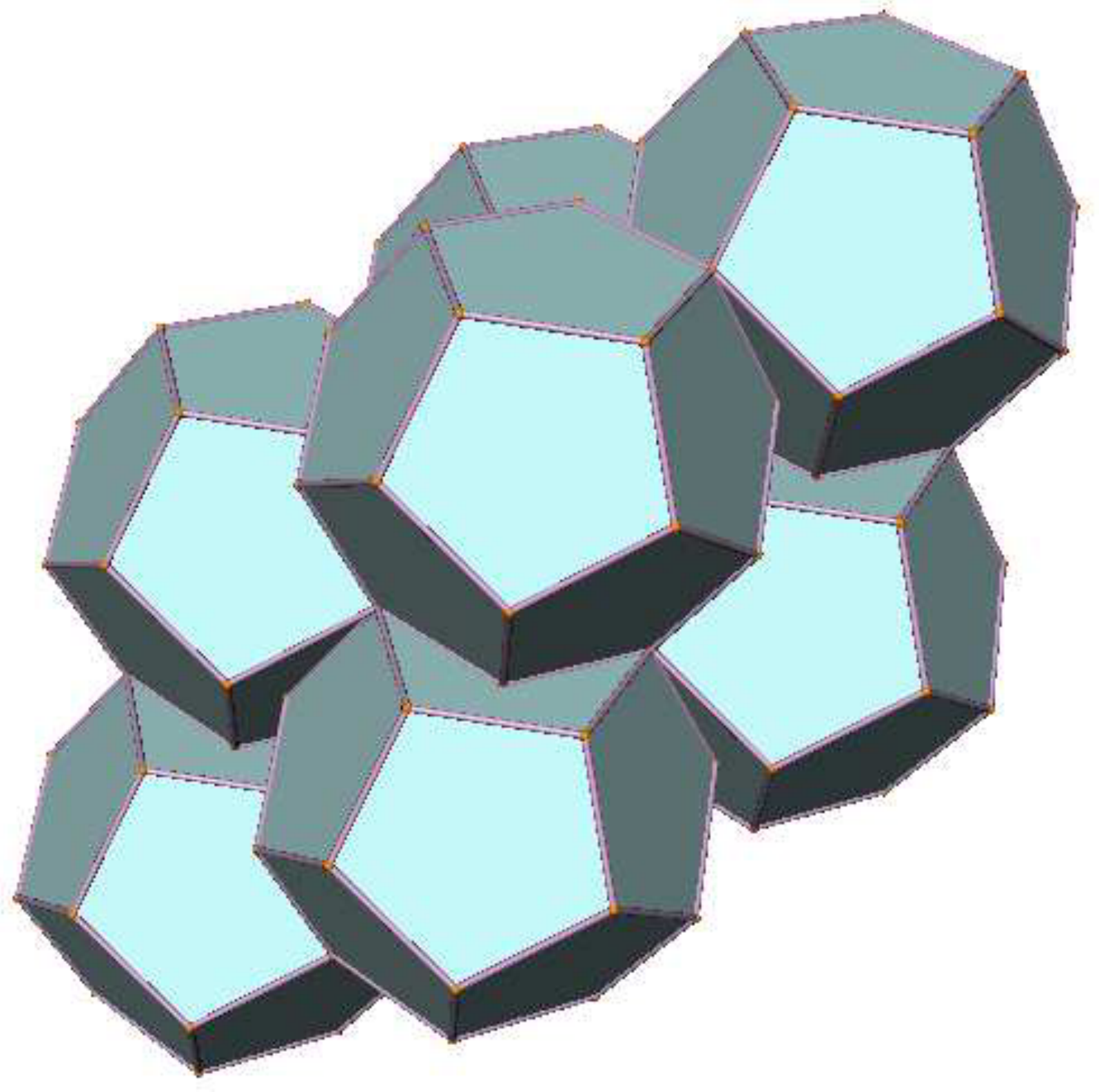}
\hspace{0.16in}
\includegraphics[  width=1.in,keepaspectratio]
{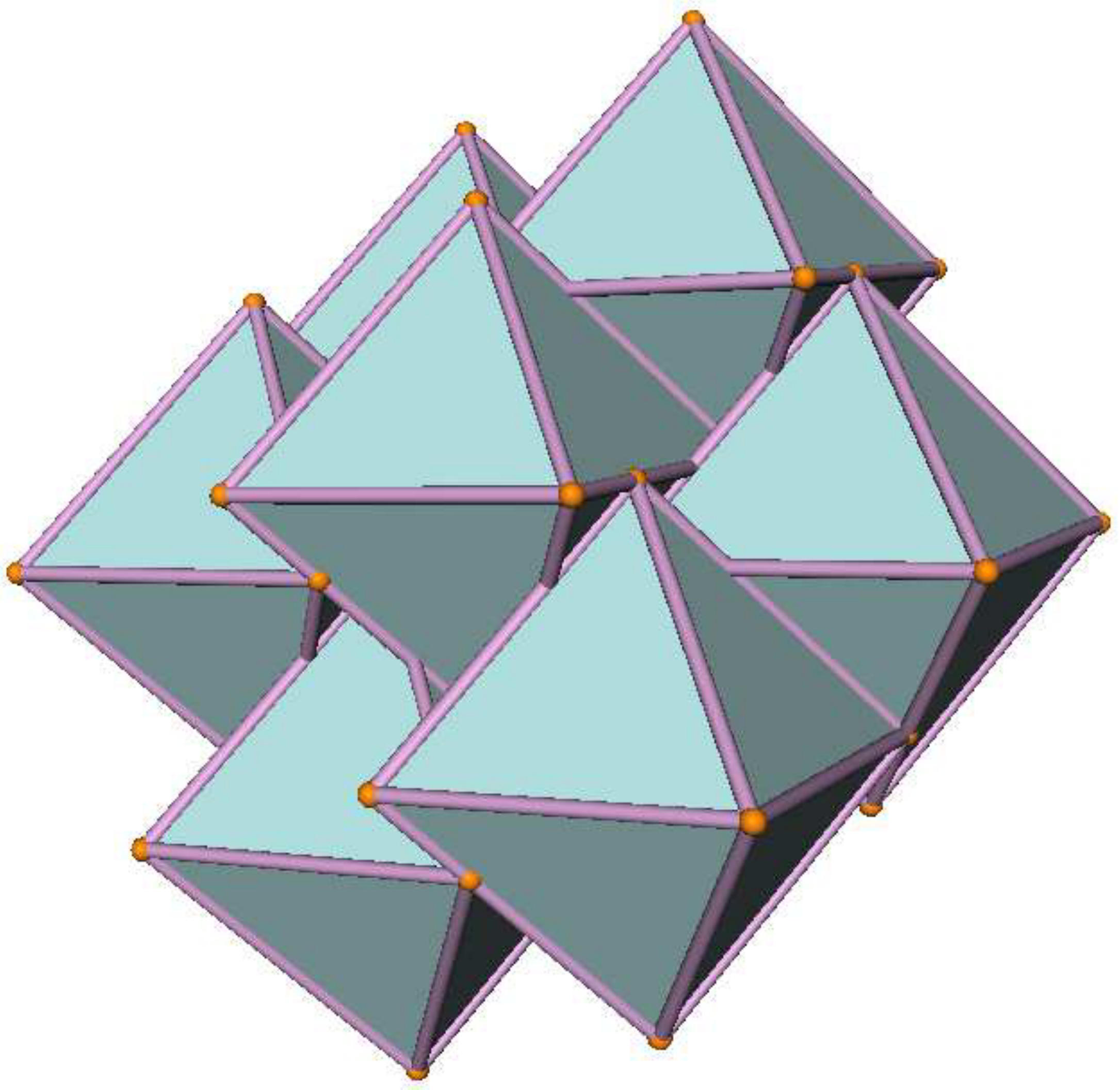}}
\caption{(color online) Portions of the densest lattice packings
of three of the centrally symmetric Platonic solids found by the 
ASC optimization scheme. \cite{To09b,To09c} Left panel: Icosahedron packing
with packing fraction $\phi=0.8363\ldots$. Middle panel:
Dodecahedron packing with packing fraction $\phi=0.9045\ldots$.
Right panel:  Octahedron packing
with packing fraction $\phi=0.9473\ldots$.} 
\label{nontiling}
\end{figure}

Conway and Torquato \cite{Co06} showed
that the maximally dense packing of regular tetrahedra  cannot be a Bravais lattice.
Among other non-Bravais lattice packings, they obtained a simple, {\it uniform}
packing of ``dimers" (two particles per fundamental cell) with packing fraction $\phi=2/3$. A {\it uniform} packing has a  symmetry (in this case a point inversion symmetry) 
that takes one tetrahedron to another. A {\it dimer} is composed of a pair of 
regular tetrahedra that exactly share a common face.
They also found a non-Bravais lattice (periodic) packings of regular tetrahedra with $\phi \approx 0.72$,
which doubled that the density of the corresponding densest Bravais-lattice packing 
($\phi=18/49=0.367 \ldots$), which was the record before 2006.
This work spurred many studies that improved on this density. \cite{To09b,To09c,Gl09,Chaik10,Ka10,To10a,Ch10,To10b}
Kallus {\it et al.} \cite{Ka10} found a remarkably simple ``uniform" packing of tetrahedra
with high symmetry consisting of only four particles per fundamental
cell (two ``dimers") with packing fraction $\phi={100}/{117}=0.854700\ldots$. 
 Torquato and Jiao \cite{To10a} subsequently
presented an analytical formulation to construct a three-parameter
family of  dense uniform dimer packings of tetrahedra again with four particles per fundamental cell.
Making an assumption about one of these parameters
resulted in a two-parameter family, including those with a packing fraction as high as
$\phi={12250}/{14319}=0.855506\ldots$.
Chen  {\it et al.} \cite{Ch10} used the full
three-parameter family to obtain the densest known dimer packings of tetrahedra with
the very slightly higher  packing fraction $\phi= {4000}/{4671} = 0.856347\ldots$
(see Fig.~\ref{tetra}).
Whether this is the optimal  packing  is an open question 
for reasons given by Torquato and Jiao.\cite{To10b}
The fact that the  first nontrivial upper bound
on the maximal packing fraction is infinitesimally smaller
that unity \cite{Gr11} ($\phi_{\mbox{\scriptsize max}} \le 1-2.6\times10^{-25}$)  is a
testament to the difficulty in accounting for the orientations of
the tetrahedra in a dense packing.

\begin{figure}
\centerline{
\includegraphics[  width=1.2in,keepaspectratio]
{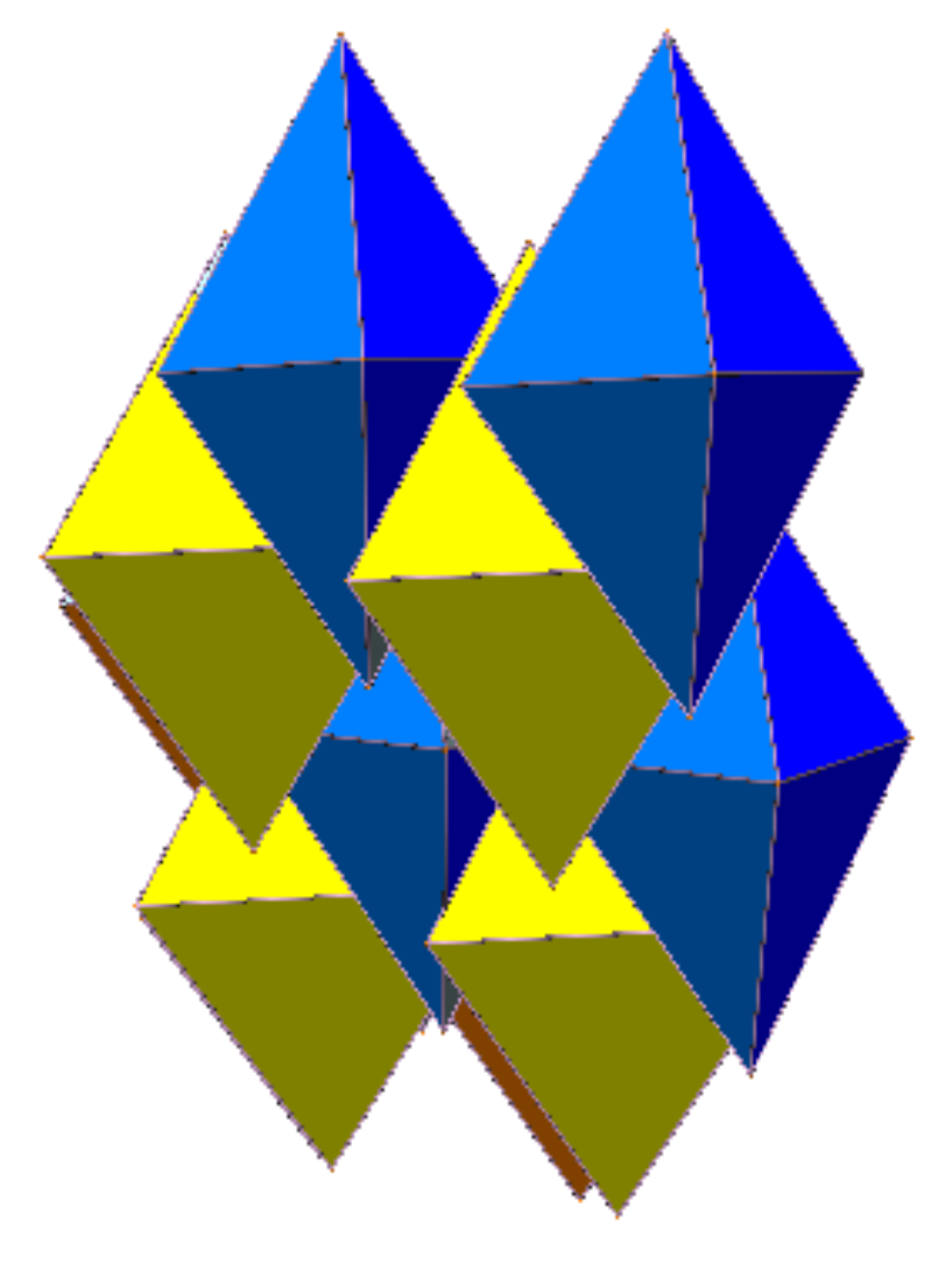}}
\caption{(color online) \footnotesize  A portion of  the
densest three-parameter family of tetrahedron packings with 4 particles per fundamental cell
and packing fraction $\phi = \frac{4000}{4671} = 0.856347\ldots$ obtained
by Chen {\it et al.}; see Ref. \onlinecite{To10b} for a general treatment.}
\label{tetra}
\end{figure}

\subsubsection{Organizing Principles for Convex and Concave Particles}
\label{organ}

Torquato and Jiao \cite{To09b,To09c} showed that substantial
face-to-face contacts between any of the centrally symmetric Platonic
and Archimedean solids allow for a  higher packing fraction.
They also demonstrated that central symmetry enables maximal 
face-to-face contacts when particles are {\it aligned}, which is
consistent with the densest packing being
the {\it optimal lattice packing}. 
The aforementioned simulation results, upper bound,
and theoretical considerations led to three conjectures
concerning the densest packings of polyhedra and other nonspherical
particles in $\mathbb{R}^3$; see Refs.  \onlinecite{To09b,To09c,To10b}. 

\noindent
{\bf Conjecture 1}: {\sl The densest packings of the centrally symmetric
Platonic and Archimedean solids are given by their corresponding optimal (Bravais) lattice packings.}
\smallskip

\noindent
{\bf Conjecture 2}: {\sl The densest packing of any convex, congruent polyhedron
without central symmetry generally is not a (Bravais) lattice packing, i.e., the
set of such polyhedra whose optimal packing
is not a lattice is overwhelmingly larger than the set whose optimal
packing is a lattice. 
}
\smallskip

\noindent
{\bf Conjecture 3}: {\sl The densest
packings of congruent, centrally symmetric particles that do not possess three
equivalent principle axes (e.g., ellipsoids) are generally not  (Bravais) lattice packings.}
\smallskip

Conjecture 1 is the analog of Kepler's sphere conjecture for 
the centrally symmetric Platonic and Archimedean solids. 
On the experimental side, it has been shown \cite{He12} that such silver polyhedral 
nanoparticles  self-assemble into the conjectured densest lattice packings of such shapes.
Torquato and Jiao \cite{To09c} have also commented on the validity
of Conjecture 1 to polytopes in four and higher dimensions.
The arguments leading to Conjecture 1 
also strongly suggest that the densest packings of superballs and other smoothly-shaped
centrally symmetric convex particles having surfaces without ``flat" directions are given by their corresponding optimal lattice packings.\cite{To09c}
Consistent with Conjecture 2,  the densest known packing of the non-centrally symmetric truncated tetrahedron 
is a non-lattice packing with packing fraction $\phi= 207/208 = 0.995192\ldots$, which is amazingly close 
to unity and strongly implies its optimality.\cite{Ji11b}
We note that non-Bravais-lattice packings of
elliptical cylinders (i.e., cylinders with an elliptical basal
face) that are denser than the corresponding optimal lattice
packings for any aspect ratio greater than unity have been constructed,\cite{Bez91,Be13} 
which is consistent with Conjecture 3. 
A corollary to Conjecture 3 is that the densest packings of congruent, centrally symmetric 
particles that  possess three equivalent principle axes (e.g., superballs) are generally  Bravais lattices. \cite{Ji09a}

Subsequently, Torquato and Jiao \cite{To12b} generalized the aforementioned
three conjectures in
order to guide one to ascertain the densest packings of other
convex nonspherical particles as well as {\it concave} shapes. These
generalized organizing principles are explicitly stated as 
the following four distinct propositions:
\smallskip

\noindent{\bf Proposition 1}: {\sl Dense packings of centrally
symmetric convex, congruent particles with three equivalent axes
are given by their corresponding densest lattice packings,
providing a tight density lower bound that may be optimal.}
\smallskip

\noindent{\bf Proposition 2}: {\sl Dense packings of convex,
congruent polyhedra without central symmetry are composed of
centrally symmetric compound units of the polyhedra with the
inversion-symmetric points lying on the densest lattice associated
with the compound units, providing a tight density lower bound
that may be optimal.}
\smallskip

\noindent{\bf Proposition 3}: {\sl  Dense packings of centrally
symmetric concave, congruent polyhedra are given by their
corresponding densest lattice packings, providing a tight density
lower bound that may be optimal. }
\smallskip

\noindent{\bf Proposition 4}: {\sl Dense packings of concave,
congruent polyhedra without central symmetry are composed of
centrally symmetric compound units of the polyhedra with the
inversion-symmetric points lying on the densest lattice associated
with the compound units, providing a tight density lower bound
that may be optimal.}
\smallskip

\noindent{Proposition 1 originally concerned polyhedra. It is generalized  here
to include any centrally symmetric convex, congruent particle with three equivalent axes 
to reflect the arguments put forth by Torquato and Jiao. \cite{To09b,To12b}}

All of the aforementioned organizing principles in the form of  conjectures and propositions have been tested in Ref. \onlinecite{To12b} against a  comprehensive set of both convex and concave particle shapes, including but not limited
to the Platonic and Archimedean solids,\cite{To09a,To09b,Ji11b} Catalan solids,\cite{De11} prisms
and antiprisms,\cite{De11}  Johnson solids,\cite{De11}   cylinders,\cite{Bez91,Be13} lenses \cite{Ci15}, truncated cubes\cite{Ga13} and various concave polyhedra.\cite{De11} 
These general organizing principles also enable one to construct analytically
the densest known packings of certain convex nonspherical
particles, including spherocylinders, and
square pyramids and rhombic pyramids.\cite{To12b}  Moreover, it was shown how to
apply these principles to infer the high-density equilibrium
crystalline phases of hard convex and concave particles.\cite{To12b}
We note that the densest known 2D packings of a large family of 2D
convex and concave
particles (e.g., crosses, curved triangles, moon-like
shapes) fully adhere to the aforementioned
organizing principles.\cite{At12}

Interestingly, the densest known packing of any identical
3D convex particle studied to date
has a density that exceeds that of the optimal sphere packing value $\phi^S_{max}=\pi/\sqrt{18}=0.7408\ldots$.
These results are consistent with a conjecture of Ulam,
who proposed to Martin Gardner \cite{Ga01}, without any justification,  that the optimal
packing fraction for congruent sphere packings is smaller than that for any
other convex body. There is currently no proof of Ulam's conjecture.

\subsubsection{Dense Packings of Tori}

We have seen that the preponderance of studies of dense packings of nonspherical
shapes in $\mathbb{R}^3$ have dealt with convex bodies that
are simply connected and thus topologically equivalent
to a sphere. However, much less is known about
dense packings of multiply connected solid bodies.
Gabbrielli {\it et al.}\cite{Ga14} investigated the packing behavior of congruent
{\it ring} tori in $\mathbb{R}^3$, which are multiply connected non-convex
bodies of genus one, as well as {\it horn} and {\it spindle} tori.
Guided by the aforementioned organizing principles, they analytically constructed a
family of dense periodic packings of individual tori and
found that the horn tori as well as certain spindle and
ring tori can achieve a packing density not only higher than that
of spheres (i.e., $\pi/\sqrt{18}=0.7404\ldots$) but also higher
than the densest known ellipsoid packings (i.e., $0.7707\ldots$).
Moreover, they  studied dense packings of clusters of {\it pair-linked}
ring tori (i.e., Hopf links), which can possess much higher
densities than corresponding packings consisting of unlinked
tori; see Fig. \ref{hopf} for a specific example.

 \begin{figure}[h]
$\begin{array}{c@{\hspace{0.25cm}}c@{\hspace{0.25cm}}c}
\includegraphics[width=2.65cm]{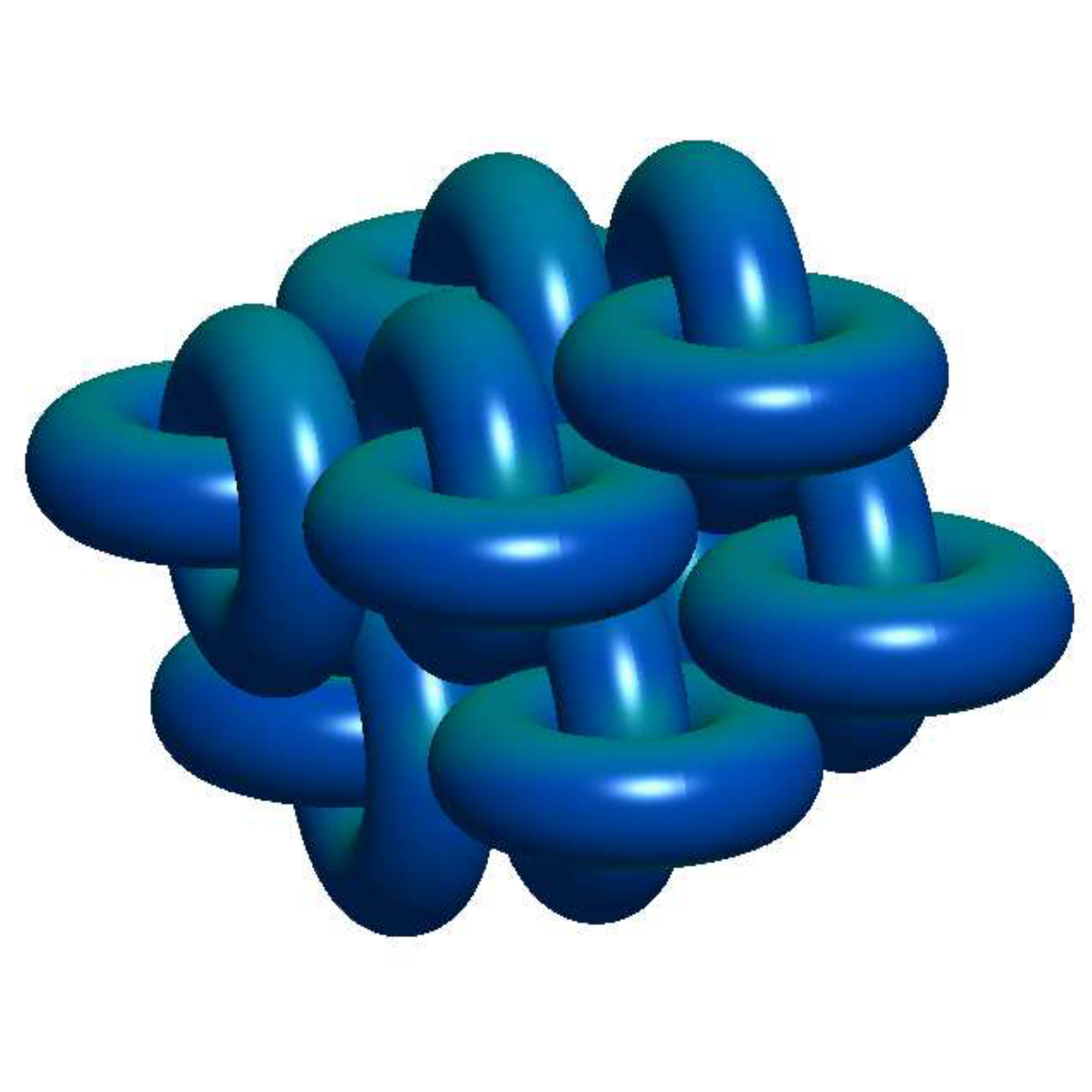} &
\includegraphics[width=2.65cm]{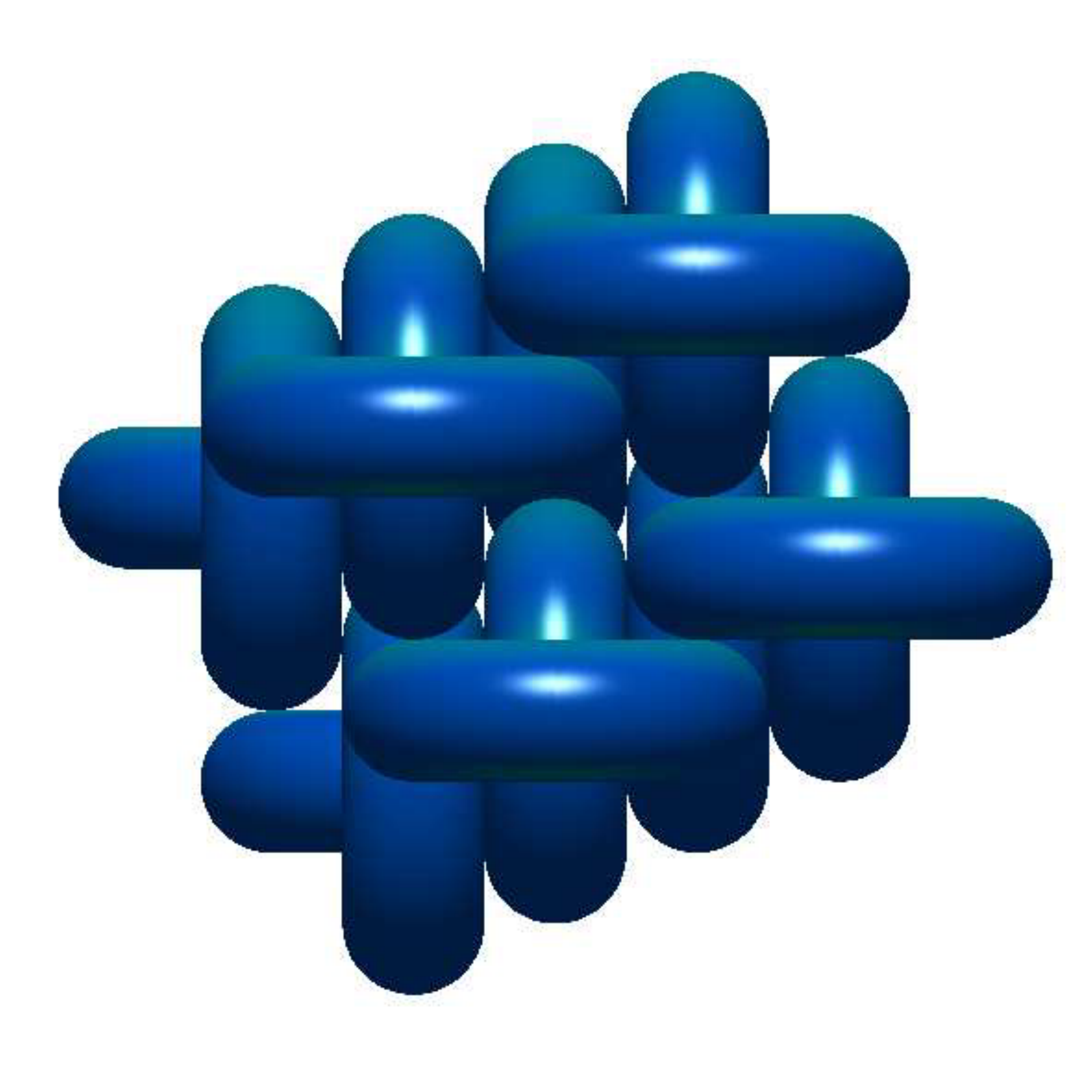} &
\includegraphics[width=2.65cm]{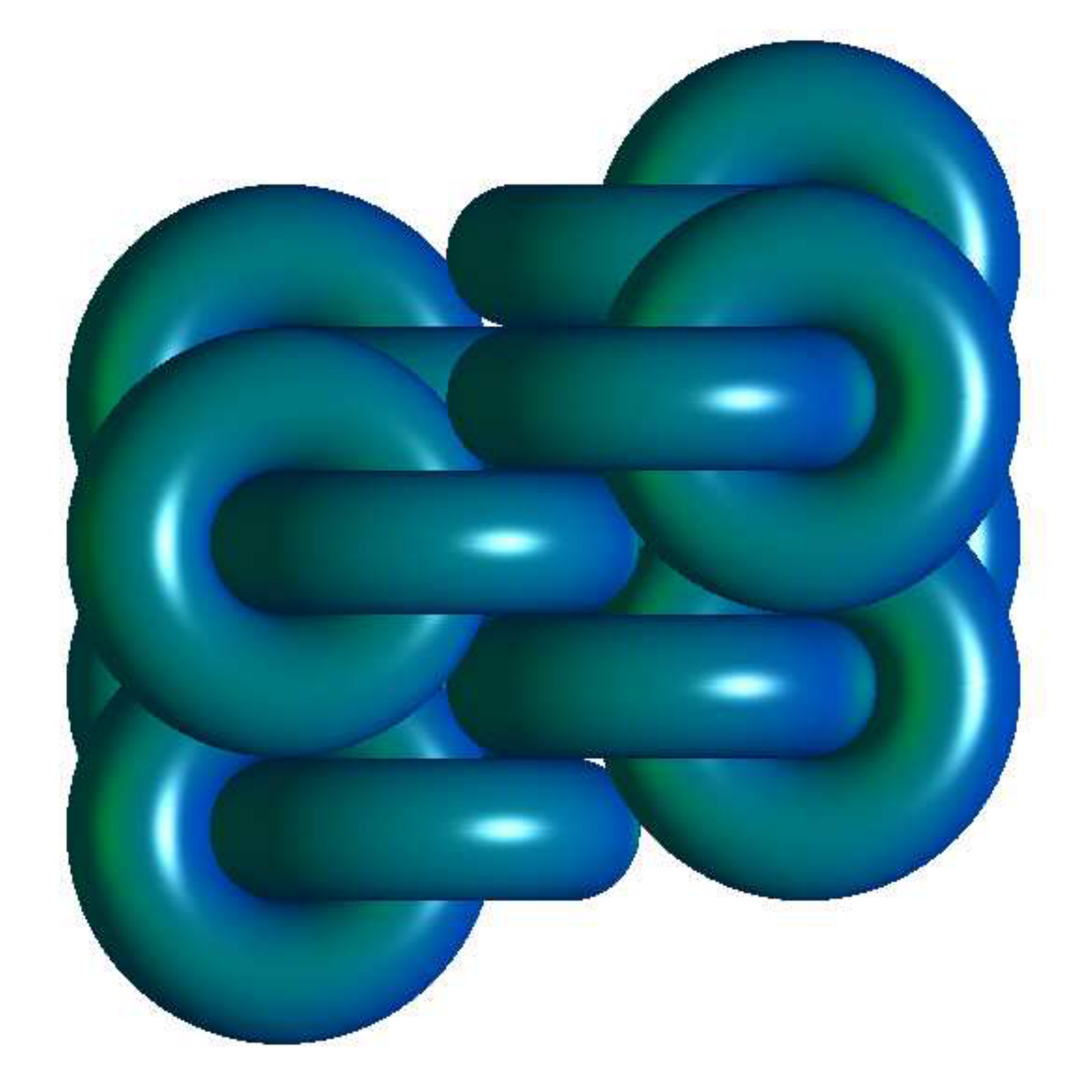} \\
\end{array}$
\caption{\label{hopf} The densest known packing of tori of radii
ratio 2 with a packing fraction $\phi\simeq 0.7445$. The images 
(taken from Ref. \onlinecite{Ga14}) show
four periodic units viewed from different angles, each containing four tori.}
\end{figure}

\section{Challenges and Open Questions}
\label{open}

Packing problems are fundamental and their solutions are often profound.
This perspective  provides only a glimpse into the richness of packing models
and their capacity to capture the salient structural and physical properties
of a wide class of equilibrium and nonequilibrium condensed phases of matter
that arise across the physical, mathematical and biological sciences
as well as technological applications.

Not surprisingly, there are many challenges and open questions.
Is it possible to prove a  first-order freezing transition in 3D hard-sphere systems?
Is the FCC lattice provably the entropically favored state as the maximal density
is approached along the stable crystal branch in this same system?
Can one devise numerical algorithms that produced large disordered strictly jammed isostatic
sphere packings that are rattler-free? Is the true MRJ state rattler-free 
in the thermodynamic limit? Can one prove the Torquato-Stillinger conjecture
that links strictly jammed sphere packings to hyperuniformity?
Can one identify incisive order metrics for packings of nonspherical  particles 
as well as  a wide class of many-particle systems that arise  in molecular, biological, cosmological,
and ecological systems? What are the appropriate generalizations of the jamming categories for packings of nonspherical particles? Is it possible to prove that the densest packings of the centrally
symmetric Platonic and Archimedean solids are given by
their corresponding optimal lattice packings?
Upon extending the geometric-structure approach to Euclidean dimensions
greater than three, do periodic packings with arbitrarily large unit cells 
or even disordered jammed packings ever provide the highest attainable densities?
Can one formulate a disordered sphere-packing model in $\mathbb{R}^d$ that can be rigorously shown
to have a packing fraction that exceeds $\phi=1/2^d$, the maximal value
achievable by the ghost RSA packing?
In view of the wide interest in packing problems across the sciences, it seems
reasonable to expect that substantial conceptual advances are forthcoming.

\begin{acknowledgments}

I am deeply grateful to Frank Stillinger,  Paul Chaikin, Aleksandar Donev,  Obioma Uche,  Antonello Scardicchio,   
Weining Man, Yang Jiao, Chase Zachary,  Adam Hopkins, {\'E}tienne Marcotte,
Joseph Corbo,   Steven Atkinson, Remi Dreyfus, Arjun Yodh, Ge Zhang, Duyu Chen, Jianxiang Tian, Michael Klatt, Enrqiue Lomba and Jean-Jacques Weis
with whom I have collaborated on topics described in this review article. I am very thankful to Jaeuk Kim, Duyu Chen, Zheng Ma, Yang Jiao and Ge Zhang for comments
that greatly improved this article.
 The author's work on packing models over the years has been supported by various grants from  the National
Science Foundation.
\end{acknowledgments}



%
\end{document}